\documentclass{aa}  

\usepackage{graphicx}
\usepackage{txfonts}
\usepackage{color}

\newcommand{\tablenotea}[1]{\parbox{8.7cm}{\indent \footnotesize{#1}}}
\newcommand{\tablenoteb}[1]{\parbox{18.3cm}{\indent \footnotesize{#1}}}
\newcommand{\acsca}{Acta Chem. Scand.}
\newcommand{\cjp}{Can. J. Phys.}
\newcommand{\cp}{Chem. Phys.}
\newcommand{\cpl}{Chem. Phys. Lett.}
\newcommand{\epsl}{Earth Planet. Sci. Lett.}
\newcommand{\epjd}{Eur. Phys. J. D}
\newcommand{\gcacta}{Geochim. Cosmochim. Acta}

\newcommand{\jaces}{J. Am. Ceram. Soc.}
\newcommand{\jacs}{J. Am. Chem. Soc.}
\newcommand{\jms}{J. Mol. Spectr.}
\newcommand{\jmst}{J. Mol. Struct.}
\newcommand{\jpc}{J. Phys. Chem.}
\newcommand{\jpca}{J. Phys. Chem. A}
\newcommand{\jpcrd}{J. Phys. Chem. Ref. Data}
\newcommand{\jpcm}{J. Phys. Condens. Matter}
\newcommand{\lps}{Lunar Planet. Sci.}
\newcommand{\mphys}{Mol. Phys.}
\newcommand{\nature}{Nature}
\newcommand{\natastro}{Nature Astron.}
\newcommand{\omet}{Organomettalics}
\newcommand{\ptrsa}{Philos. Trans. Royal Soc. A}
\newcommand{\pccp}{PCCP}
\newcommand{\prev}{Phys. Rev.}
\newcommand{\preva}{Phys. Rev. A}
\newcommand{\prevlet}{Phys. Rev. Lett.}
\newcommand{\rsi}{Rev. Sci. Instrum.}
\newcommand{\science}{Science}
\newcommand{\tca}{Theor. Chim. Acta}
\newcommand{\zfna}{Z. Naturforsch. A}

\begin{document}

\title{Chemical equilibrium in AGB atmospheres: Successes, failures, and prospects for small molecules, clusters, and condensates}

\titlerunning{Thermochemical equilibrium in AGB atmospheres}
\authorrunning{Ag\'undez et al.}

\author{M.~Ag\'undez\inst{1}, J.~I.~Mart\'inez\inst{2}, P.~L.~de Andres\inst{2}, J.~Cernicharo\inst{1}, \and J.~A.~Mart\'in-Gago\inst{2}}

\institute{
Instituto de F\'isica Fundamental, CSIC, C/ Serrano 123, 28006 Madrid, Spain \and
Instituto de Ciencia de Materiales de Madrid, CSIC, C/ Sor Juana In\'es de la Cruz 3, 28049 Cantoblanco, Spain
}

\date{Received; accepted}

 
\abstract
{Chemical equilibrium has proven extremely useful for predicting the chemical composition of AGB atmospheres. Here we use a recently developed code and an updated thermochemical database that includes gaseous and condensed species involving 34 elements to compute the chemical equilibrium composition of AGB atmospheres of M-, S-, and C-type stars. We include for the first time Ti$_x$C$_y$ clusters, with $x$ = 1-4 and $y$ = 1-4, and selected larger clusters ranging up to Ti$_{13}$C$_{22}$, for which thermochemical data are obtained from quantum-chemical calculations. Our main aims are to systematically survey the main reservoirs of each element in AGB atmospheres, review the successes and failures of chemical equilibrium by comparing it with the latest observational data, identify potentially detectable molecules that have not yet been observed, and diagnose the most likely gas-phase precursors of dust and determine which clusters might act as building blocks of dust grains. We find that in general, chemical equilibrium reproduces the observed abundances of parent molecules in circumstellar envelopes of AGB stars well. There are, however, severe discrepancies of several orders of magnitude for some parent molecules that are observed to be anomalously overabundant with respect to the predictions of chemical equilibrium. These are HCN, CS, NH$_3$, and SO$_2$ in M-type stars, H$_2$O and NH$_3$ in S-type stars, and the hydrides H$_2$O, NH$_3$, SiH$_4$, and PH$_3$ in C-type stars. Several molecules have not yet been observed in AGB atmospheres but are predicted with non-negligible abundances and are good candidates for detection with observatories such as ALMA. The most interesting ones are SiC$_5$, SiNH, SiCl, PS, HBO, and the metal-containing molecules MgS, CaS, CaOH, CaCl, CaF, ScO, ZrO, VO, FeS, CoH, and NiS. In agreement with previous studies, the first condensates predicted to appear in C-rich atmospheres are found to be carbon, TiC, and SiC, while Al$_2$O$_3$ is the first major condensate expected in O-rich outflows. According to our chemical equilibrium calculations, the gas-phase precursors of carbon dust are probably acetylene, atomic carbon, and/or C$_3$, while for silicon carbide dust, the most likely precursors are the molecules SiC$_2$ and Si$_2$C. In the case of titanium carbide dust, atomic Ti is the major reservoir of this element in the inner regions of AGB atmospheres, and therefore it is probably the main supplier of titanium during the formation of TiC dust. However, chemical equilibrium predicts that large titanium-carbon clusters such as Ti$_8$C$_{12}$ and Ti$_{13}$C$_{22}$ become the major reservoirs of titanium at the expense of atomic Ti in the region where condensation of TiC is expected to occur. This suggests that the assembly of large Ti$_x$C$_y$ clusters might be related to the formation of the first condensation nuclei of TiC. In the case of Al$_2$O$_3$ dust, chemical equilibrium indicates that atomic Al and the carriers of Al-O bonds AlOH, AlO, and Al$_2$O are the most likely gas-phase precursors.}


\maketitle

\section{Introduction}

During their late evolutionary stages, low- and intermediate-mass stars ($<$ 8 M$_{\odot}$) become red giants, increasing their radius by 2-3 orders of magnitude and decreasing their surface temperature to 2000-3000 K. At these temperatures, the material is essentially molecular. When these stars enter the so-called asymptotic giant branch (AGB) phase, they start to lose mass through nearly isotropic winds that give rise to circumstellar envelopes that are mainly composed of gaseous molecules and dust grains \citep{Hofner2018}.

Thermochemical equilibrium provides a simple but incredibly useful starting point to describe the chemical composition of matter in the atmospheres of AGB stars. For example, chemical equilibrium has provided an elegant explanation of the marked chemical differentiation between oxygen-rich and carbon-rich AGB stars based on the high bond energy of carbon monoxide \citep{Russell1934}. The high abundance of CO causes it to trap most of the limiting element and allows the element in excess to form either oxygen-bearing molecules when C/O $<$ 1 or carbon-bearing molecules when C/O $>$ 1. Moreover, the discovery of many molecules in envelopes around evolved stars, such as HCP, PO, AlOH, or TiO \citep{Agundez2007,Tenenbaum2007,Tenenbaum2010,Kaminski2013}, has largely been inspired by the predictions of chemical equilibrium calculations such as those of \cite{Tsuji1964,Tsuji1973}. During the past decades, however, observations have shown a significant number of discrepancies with the scenario depicted by chemical equilibrium such as the discovery of warm water vapor in carbon stars \citep{Decin2010}, which indicate that nonequilibrium processes are at work in AGB atmospheres.

Chemical equilibrium is also very useful for studying the types of dust that are formed in AGB ejecta. We know that AGB stars are the main sources of dust in the Galaxy \citep{Gehrz1989}, but identifying the chemical nature of the dust is difficult. Only a handful of solid materials have been identified so far in circumstellar envelopes of AGB stars (e.g., \citealt{Waters2011}), while some information is also available from the analysis of presolar material in meteorites \citep{Lodders2005}. Chemical equilibrium can provide the basic theoretical scenario with the types of condensates that are thermodynamically favored and their condensation temperatures, which determine the sequence in which they are expected to appear as matter flows from the AGB star and cools \citep{Sharp1995,Lodders1997,Lodders1999,Gail2013}.

Although the formation of dust in AGB outflows is a complex process that is likely governed by chemical kinetics, as indicated by the extensive theoretical work of Gail \& Sedlmayr (see, e.g., \citealt{Gail2013}), chemical equilibrium can provide clues on the sequence of clustering that initiates the formation of the first solid materials from a gas of atoms and small molecules. The identification of the most thermodynamically favored intermediate clusters is an important piece of information. Several works have studied from the point of view of chemical equilibrium the clustering process that initiates the formation of some of the condensates that are predicted to appear earlier in AGB winds, such as MgO \citep{Kohler1997}, SiC \citep{Yasuda2012,Gobrecht2017}, silicates \citep{Goumans2012,Goumans2013}, and Al$_2$O$_3$ \citep{Alvarez-Barcia2016,Gobrecht2016,Boulangier2019}. Today, the unprecedented angular resolution and sensitivity of observatories such as ALMA have the potential of identifying the building blocks of dust in the atmospheres of AGB stars, providing constraints on the clustering process based on their abundances and spatial distributions (see, e.g., \citealt{Kaminski2017,Decin2017,McCarthy2019}).

In this study, we revisit thermochemical equilibrium in AGB atmospheres with different C-to-O ratios (M, S, and C stars) using the latest thermochemical data to compare the predictions of chemical equilibrium with the current observational situation. Our main motivations are threefold. (1) We review the successes of chemical equilibrium in explaining the observed abundances of parent molecules in AGB envelopes and identify the main failures, all of which must be accounted for by any nonequilibrium scenario that is proposed for the atmospheres of AGB stars. (2) We identify potentially detectable molecules that have not yet been observed in AGB atmospheres. (3) We compute the condensation sequence of solid materials in the atmospheres of M, S, and C stars and evaluate the most likely gas-phase precursors of different condensates. We also determine which thermodynamically favorable clusters\footnote{In this work we use the term \textit{\textup{cluster}} to refer to large molecules, mostly large Ti$_x$C$_y$ and Si$_x$C$_y$ molecules.} might play a role as intermediate species in the clustering process. In particular, we have computed thermochemical properties for various Ti$_x$C$_y$ clusters to evaluate their abundances and role in the formation of titanium carbide dust in the atmospheres of C-type stars.

\section{Thermochemical equilibrium calculations}

\subsection{Method of computation}

The composition of a mixture of gases and condensates at thermochemical equilibrium is determined by the minimization of the Gibbs free energy of the system, and it only depends on three input parameters: pressure, temperature, and relative abundances of the elements. The calculations need to be fed with thermochemical data of the included species. Many programs based on different algorithms have been developed to compute chemical equilibrium in the atmospheres of cool stars, brown dwarfs, and planets. We can distinguish between two groups of methods: those based on equilibrium constants, and those that minimize the total Gibbs free energy of the system.

In the first group, the mathematical problem consists of a set of equations of conservation of each element, in which the partial pressure of each molecule is expressed in terms of the partial pressures of the constituent atoms by the equilibrium constant of atomization. In a first step, the system is solved only for the most abundant elements, and then the whole system including all trace elements is solved iteratively using the Newton-Raphson method or similar methods. The Newton-Raphson method, originally developed by \cite{Russell1934} for diatomic molecules and generalized by \cite{Brinkley1947}, was later applied by \cite{Tsuji1973} to atmospheres of cool stars. The method has been implemented with different refinements by \cite{Tejero1991} and by codes such as CONDOR \citep{Lodders1993}, GGChem \citep{Woitke2018}, and FastChem \citep{Stock2018}.

The second type of methods was introduced by \cite{White1958} and solves the problem of minimizing the total Gibbs energy of a mixture of species subject to the conservation of each element. This method is more general in that it makes no distinction between atoms, molecules, and condensates because all them are simply constituent species of the mixture. The method is widely used by different programs, for instance, by SOLGAS \citep{Eriksson1971}, NASA/CEA \citep{Gordon1994}, and more recently, TEA \citep{Blecic2016}.

\cite{Zeleznik1960} demonstrated that the methods of equilibrium constants and Gibbs minimization are computationally identical, and therefore the various existing programs are expected to converge to the same equilibrium composition regardless of the method used. Important differences can appear, however, when the included species are not the same or when the adopted thermochemical data are different. The precision of chemical equilibrium calculations is essentially limited by the completeness of the included species and by the availability of accurate thermochemical data.

Our chemical equilibrium code uses the Gibbs minimization method and is based on the algorithm implemented in the NASA/CEA program \citep{Gordon1994}. The code has been developed in recent years and has been applied to describe the chemical composition of hot-Jupiter atmospheres by \cite{Agundez2014a}.

\subsection{Thermochemical data}

To solve chemical equilibrium by minimizing the Gibbs free energy of a system, the basic thermodynamic quantity needed is the free energy of each species as a function of temperature $g^0(T)$. This quantity, also known as standard-state chemical potential, can be expressed as
\begin{equation}
g^0(T) = H^0(T) - T S^0(T),
\end{equation}
where $H^0(T)$ and $S^0(T)$ are the standard-state enthalpy and entropy, respectively, of the species, and standard-state refers to a standard pressure of 1 bar. These thermochemical properties are either given directly in compilations such as NIST-JANAF \citep{Chase1998}\footnote{\texttt{https://janaf.nist.gov/}} or are found parameterized as a function of temperature through NASA polynomial coefficients (see, e.g., \citealt{McBride2002}) in databases such as NASA/CEA \citep{McBride2002}\footnote{See \texttt{https://www.grc.nasa.gov/WWW/CEAWeb}} or the Third Millenium Thermochemical Database \citep{Goos}\footnote{\texttt{https://burcat.technion.ac.il/}}.

We here considered 919 gaseous species and 185 condensed species involving up to 34 elements. Thermochemical data were mostly taken from the library of NASA/CEA \citep{McBride2002} and from the Third Millenium Thermochemical Database \citep{Goos}. The NASA/CEA data are mostly based on the work carried out at the NASA Glenn Research Center until 2002 and in classical compilations such as \cite{Gurvich1989} and NIST-JANAF \citep{Chase1998}. The Third Millenium Thermochemical Database includes data from the NASA/CEA compilation, although it is larger and is continuously updated with data from dedicated ab initio calculations, the Active Thermochemical Tables \citep{Ruscic2014}, and from the recent literature. We note that the FeCl$_3$ data in the NASA/CEA library are incorrect because a formation enthalpy of $-1059$ kJ mol$^{-1}$ is assumed, while the value reported in the literature (e.g., NIST-JANAF; \citealt{Chase1998}) is $-253$ kJ mol$^{-1}$. The use of the incorrect data results in an overestimation of the FeCl$_3$ abundance, with important implications for the overall chlorine budget. For this species we therefore adopted the data from the Third Millenium Thermochemical Database.

In addition to these two large compilations, the NASA/CEA and Third Millenium Thermochemical Database, we also used thermochemical data from different literature sources either because the species was not included in the other two compilations or because more accurate data were available. For several molecules involving Li, Na, Mg, Ti, Fe, and Co, we took the data directly from the NIST-JANAF Thermochemical Tables \citep{Chase1998} because they are not included in the NASA/CEA library or in the Third Millenium Thermochemical Database. For PH$_3$, PH, PN, SH, S$_2$O, NS, and PS, we adopted the thermochemical data revised by \cite{Lodders1999REF,Lodders2004REF}. We also included some metal-containing diatomic molecules with thermochemical data from \cite{Barklem2016}, while for V$_2$O$_4$ , we used data from \cite{Balducci1983}. Thermochemical data for the two silicon-containing molecules SiCH and SiNH were taken from the Chemkin Thermodynamic Database \citep{Kee2000}. Data for Si$_x$C$_y$ clusters were taken from \cite{Deng2008}, while the thermochemical properties of Ti$_x$C$_y$ clusters were calculated in this work and are described in detail in Appendix~\ref{app:tixcy}. Concretely, we include all Ti$_x$C$_y$ clusters with $x$ = 1-4 and $y$ = 1-4 and the large stable clusters Ti$_3$C$_8$, Ti$_4$C$_8$, Ti$_6$C$_{13}$, Ti$_7$C$_{13}$, Ti$_8$C$_{12}$, Ti$_9$C$_{15}$, and Ti$_{13}$C$_{22}$. For the condensates CaTiO$_3$, NaAlSi$_3$O$_8$, KAlSi$_3$O$_8$, Fe$_3$C, CaMgSi$_2$O$_6$, CaAl$_2$Si$_2$O$_8$, and Ca$_2$Al$_2$SiO$_7$ we used thermochemical data from \cite{Robie1979}, while for CaAl$_4$O$_7$ and CaAl$_{12}$O${19}$, data were taken from \cite{Allibert1981} and \cite{Geiger1988}.

\subsection{Elemental composition}

\begin{table}
\caption{Thirty-four elements included and their abundances.} \label{table:elements}
\centering
\small
\begin{tabular}{lclclc}
\hline \hline
\multicolumn{1}{c}{Element} & \multicolumn{1}{c}{$\log \epsilon$~$^a$} & \multicolumn{1}{c}{Element} & \multicolumn{1}{c}{$\log \epsilon$~$^a$} & \multicolumn{1}{c}{Element} & \multicolumn{1}{c}{$\log \epsilon$~$^a$} \\\hline
H   & 12.00 & Al  & 6.45 & Mn & 5.43 \\
He & 10.93 & Si  & 7.51 & Fe  & 7.50 \\
Li   & $-$0.30~$^b$ & P   & 5.41 & Co  & 4.99 \\
Be  & 1.38 & S   & 7.12 & Ni   & 6.22 \\
B   & 2.70 & Cl  & 5.50 & Cu  & 4.19 \\
C   & 8.43~$^c$ & Ar  & 6.40 & Zn  & 4.56 \\
N   & 7.83 & K   & 5.03 & Rb  & 2.52~$^e$ \\
O   & 8.69 & Ca & 6.34 & Sr   & 2.87~$^e$ \\
F   & 4.48~$^d$ & Sc & 3.15 & Zr   & 2.58~$^e$ \\
Ne & 7.93 & Ti  & 4.95 & Ba  & 2.18~$^e$ \\
Na & 6.24 & V   & 3.93 \\
Mg & 7.60 & Cr  & 5.64 \\
\hline
\end{tabular}
\tablenotea{\\
$^a$~Abundance defined as $\log \epsilon$(X) = 12 + $\log$(X/H). Abundances are solar from \cite{Asplund2009} unless otherwise stated. $^b$~\cite{Abia1993}. $^c$~The abundance of C in S-type and carbon stars is increased over the solar value to have C/O  of 1.0 and 1.4, respectively. $^d$~\cite{Abia2015}. $^e$~The abundances of the $s$-process elements Rb, Sr, Zr, and Ba are increased over the solar values by 0.36, 1.01, 0.88, and 0.89 dex, respectively, in S-type stars \citep{Abia1998} and by 0.26, 0.46, 0.67, and 0.51 dex, respectively, in carbon stars \citep{Abia2002}.
}
\end{table}

Optical and infrared observations of AGB stars have found that the atmospheric elemental composition is nearly solar, with the exception of carbon and $s$-process elements, which are significantly enhanced in carbon stars because they are brought out to the surface by dredge-up processes. Determination of the abundances of C, N, and O in AGB stars indicates that these elements have essentially solar abundances, except for carbon, which in S- and C-type stars is enhanced, which results in C/O of $\sim1$ and $>1$, respectively \citep{Smith1985,Smith1986,Lambert1986}. In our calculations we consider C/O of 0.54 (solar), 1.00, and 1.40 for M-, S-, and C-type stars, respectively. Elements produced through neutron capture in the $s$-process such as Sr, Zr, and Ba are found to have moderate abundance enhancements in carbon stars \citep{Abia2002}. Other elements for which significant deviations from the solar abundances are expected in AGB stars are fluorine and lithium. In the case of fluorine, however, recent observational studies find only mild enhancements and point to abundances very close to the solar value \citep{Abia2015,Abia2019}. Although a few super-rich lithium stars ($\log \epsilon$ > 4) exist, the abundance of lithium in Galactic carbon stars is found to be below that in the Sun \citep{Abia1993}. The abundances adopted for the 34 elements we included in the chemical equilibrium calculations are given in Table~\ref{table:elements}.

\subsection{Radial profiles of temperature and pressure} \label{sec:pt-profile}

The winds associated with AGB stars cause them to have extended atmospheres, in which the gas cools and the density of particles drops as it moves away from the star. The temperatures and pressures in this extended atmosphere are critical for establishing the chemical equilibrium composition. For example, high temperatures favor an atomic composition, while low temperatures favor a molecular gas. It is therefore very important to have a realistic description of how the gas temperature and pressure vary with radius.

The situation becomes complicated by two facts. First, the atmospheres of AGB stars are not static, but are affected by dynamical processes that are ultimately driven by the pulsation of the star. Variability of the infrared flux has been observationally characterized for a long time, and it is interpreted as a consequence of the stellar pulsation, during which the size and effective temperature of the star experience important changes \citep{LeBertre1988,Suh2004}. Second, the low gravity of AGB stars causes the extended atmosphere to be affected by convective processes that lead to asymmetric structures, hot spots, and high-density clumps. This complex morphology is predicted by 3D hydrodynamical simulations \citep{Freytag2017} and is starting to be characterized in detail with high angular resolution observations at infrared and (sub-)millimeter wavelengths (e.g., \citealt{Khouri2016,Vlemmings2017,Fonfria2019}).

\begin{figure}
\centering
\includegraphics[angle=0,width=\columnwidth]{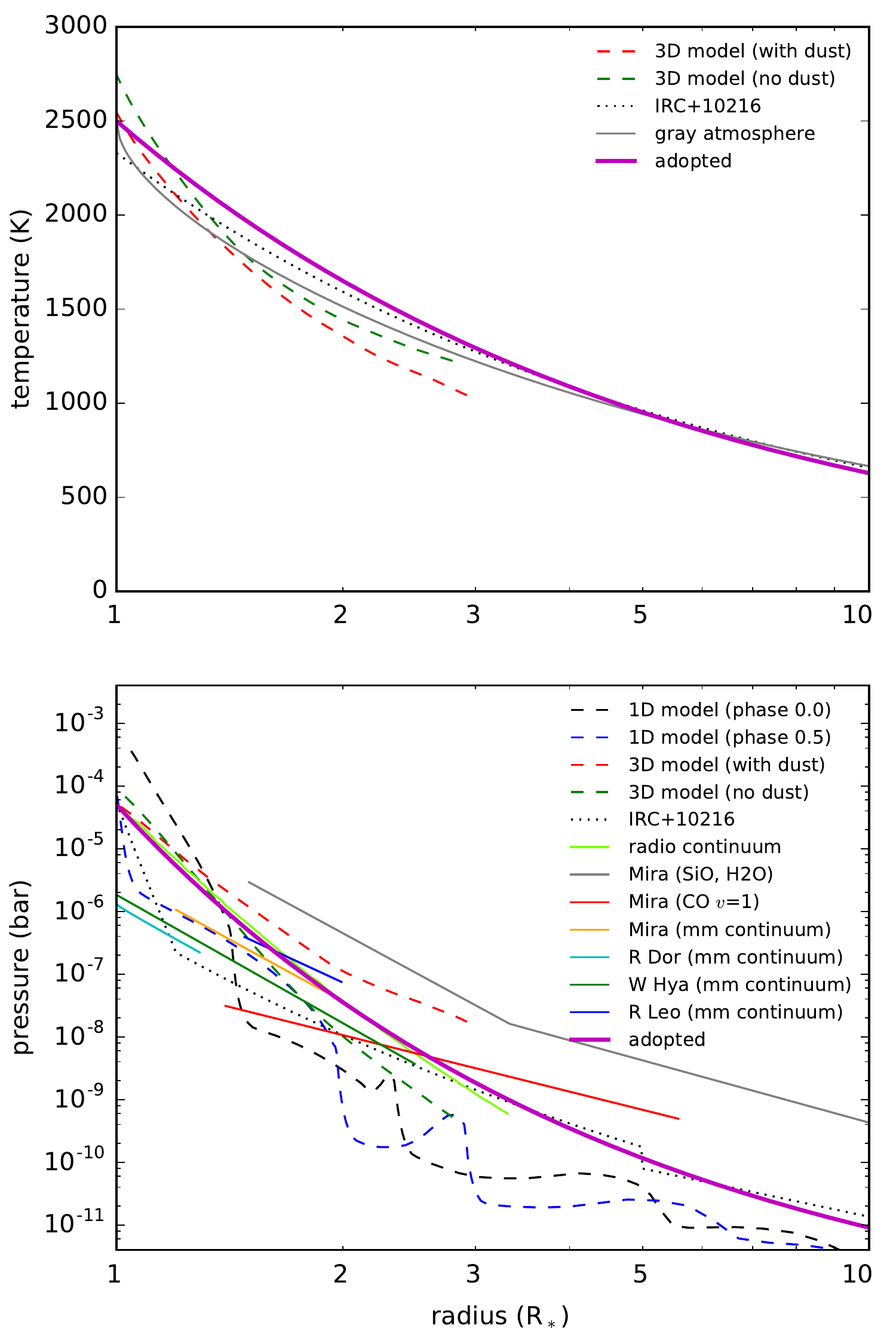}
\caption{Radial temperature (top panel) and pressure (bottom panel) profiles in the extended atmosphere of an AGB star. The black and blue dashed curves correspond to a 1D model of an AGB atmosphere at phases 0.0 and 0.5 (see Figure 1 of \citealt{Bladh2019}). The green and blue dashed curves show the profiles resulting from a 3D model of an AGB atmosphere with and without radiation pressure on dust (models st28gm06n06 and st28gm06n26 from \citealt{Freytag2017}, where profiles are averaged over spherical shells and time). The black dotted line shows the empirical profile derived for the carbon star IRC\,+10216 \citep{Agundez2012}. In the bottom panel we also show as thin solid lines several radial pressure profiles derived from high angular resolution observations of the radio continuum of several AGB stars \citep{Reid1997}, of Mira from ALMA observations of SiO and H$_2$O \citep{Wong2016} and of CO $v$ = 1 \citep{Khouri2018}, and of Mira, R\,Leo, W\,Hya, and R\,Dor from ALMA observations of the (sub)millimeter continuum \citep{Vlemmings2019}. The radial temperature and pressure profiles adopted in this study as representative of an AGB atmosphere are shown as thick magenta curves.} \label{fig:pt}
\end{figure}

Despite the complications related to the variation with time and the complex morphology, we adopted for our chemical equilibrium calculations a simple scenario that is representative of a generic AGB star in which the atmosphere is spherically symmetric and temperature and pressure vary smoothly with radius.
Effective temperatures of AGB stars are usually in the range 2000-3000 K \citep{Bergeat2001}. Here we adopt an effective temperature of 2500 K. The temperature gradient throughout the extended atmosphere can be usually well accounted for using a power law, that is, $T(r) \propto r^{-\alpha}$, with values of $\alpha$ in the range 0.5-1.0. For example, a gray atmosphere, in which $\alpha$ approaches 0.5 for $r > R_*$, has been adopted to model high angular resolution observations of continuum emission at radio and (sub)millimeter wavelengths \citep{Reid1997,Vlemmings2019}. Several works have modeled molecular lines arising from the inner envelope around the carbon star IRC\,+10216, finding values of $\alpha$ in the range 0.55-0.58 \citep{Fonfria2008,DeBeck2012,Agundez2012}. The 3D hydrodynamic models of \cite{Freytag2017} result in steeper radial temperature profiles close to the star, with a power-law index of 0.8-0.9 inside 2 $R_*$, and more shallow in the 2-3 $R_*$ region. For our chemical equilibrium calculations, we adopted a power law for the radial temperature profile with an index of 0.6 (see the thick magenta line in the upper panel of Fig.~\ref{fig:pt}), which results in a temperature profile similar to that derived for IRC\,+10216 and those resulting from 3D hydrodynamic models. In our adopted profile, the gas temperature decreases from 2500 K at the stellar surface down to $\sim630$ K at 10 $R_*$.

The radial pressure profile is expected to be given by hydrostatic equilibrium at the stellar surface, while in the outer parts of the circumstellar envelope, where the gas has reached the terminal expansion velocity, mass conservation implies that the gas density varies with radius as a power law, $n(r) \propto r^{-\beta}$, with $\beta$ = 2. The region in between these two parts, the extended atmosphere, is a complex environment where the gas is accelerated and the radial density profile is expected to be shallower than at hydrostatic equilibrium but steeper than the $r^{-2}$ power law. In general, this behavior is supported by models and observations, although estimated densities can easily differ by several orders of magnitude in different studies. For example, hydrodynamic models that explain the formation of AGB winds through a combination of stellar pulsation plus radiation pressure on dust grains can provide estimates of the gas density in the extended atmosphere \citep{Hofner2018}. These models can result in very different densities depending on the adopted parameters and the included processes, however. For example, we compare in the lower panel of Fig.~\ref{fig:pt} the various dashed curves, which correspond to a 1D model by \cite{Bladh2019} at two different phases and to two 3D models from \cite{Freytag2017}. High angular resolution observations from radio to infrared wavelengths can provide constraints on the densities in the extended atmosphere of AGB stars (see the thin solid curves in the lower panel of Fig.~\ref{fig:pt}). Infrared observations of R\,Dor, W\,Hya, and IK\,Tau indicate $\beta$ values between 2.7 and 4.5 in regions that extend to a few stellar radii \citep{Khouri2016,Ohnaka2017,Adam2019}. From ALMA (sub)millimeter continuum observations of the low-mass loss-rate objects Mira ($o$\,Cet), R\,Dor, W\,Hya, and R\,Leo, \cite{Vlemmings2019} derived values of $\beta$ in the range 5-6 for the 1-3 $R_*$ region. An even steeper radial density profile is obtained for the same 1-3 $R_*$ region from high angular resolution observations of radio continuum emission from various AGB stars \citep{Reid1997} and from 3D hydrodynamic models \citep{Freytag2017}.

As an illustration of the differences found in the literature, we show in the lower panel of Fig.~\ref{fig:pt} three radial density profiles derived from ALMA data of the star Mira, using SiO and H$_2$O \citep{Wong2016}, CO $v$ = 1 $J$ = 3-2 \citep{Khouri2018}, and (sub)millimeter continuum \citep{Vlemmings2019}. Although the derived slopes are similar, the absolute densities differ by as much as two orders of magnitude. The most striking feature is that when SiO and H$_2$O data are used, the densities needed to properly excite the observed lines are significantly higher than those derived from vibrationally excited CO or (sub)millimeter continuum. It is clear that further observational studies are needed to determine the best density tracers and to converge in the density estimates.

It seems that a single power law cannot adequately reproduce the variation of density through the whole extended atmosphere. It is likely that the radial density profile becomes progressively flatter away from the star until a power law with $\beta$ = 2 is reached outside the acceleration region. The radial pressure profile adopted for the chemical equilibrium calculations catches this idea and is shown as the thick magenta line in the lower panel of Fig.~\ref{fig:pt}. With this prescription, the pressure at the stellar surface is $5\times10^{-5}$ bar, which agrees with typical values from hydrodynamical models. Then, pressure decreases to a few $10^{-8}$ bar at 2 $R_*$, which is in between the values derived from high angular resolution observations, and it finally becomes $\sim10^{-11}$ bar at 10 $R_*$, which is in the range of values expected for a high mass-loss rate of $\sim10^{-5}$ M$_{\odot}$ yr$^{-1}$, as is the case of IRC\,+10216.

\section{Parent molecules: successes and failures of chemical equilibrium}

\begin{table*}
\caption{Abundances (relative to H$_2$) of parent molecules other than H$_2$ and CO derived from observations of M, S, and C stars.} \label{table:observations}
\centering
\small
\begin{tabular}{lrl@{\hspace{1.0cm}}rl@{\hspace{1.0cm}}rl}
\hline \hline
Molecule & \multicolumn{2}{c}{M stars} & \multicolumn{2}{c}{S stars} & \multicolumn{2}{c}{C stars} \\
\hline
\textcolor{blue}{\textbf{H$_2$O}}         & $(0.3-4)\times10^{-4}$ & 4 stars $^{(1)}$ & \textcolor{blue}{$(1.2-1.5)\times10^{-5}$} & \textcolor{blue}{2 stars $^{(2,3)}$} & \textcolor{blue}{$(0.1-5)\times10^{-6}$} & \textcolor{blue}{2 stars $^{(4,5)}$} \\
C$_2$H$_2$ & -- & & -- & & $(0.75-8)\times10^{-5}$ & IRC\,+10216 $^{(6)}$ \\
\textcolor{blue}{\textbf{HCN}}              & \textcolor{blue}{$(0.18-5)\times10^{-7}$} & \textcolor{blue}{25 stars $^{(7)}$} & $(0.06-4.5)\times10^{-6}$ & 18 stars $^{(7)}$ & $(0.17-8)\times10^{-5}$ & 26 stars $^{(7)}$ \\
CO$_2$         & $3\times10^{-7}$ & SW\,Vir $^{(8)}$ & -- & & -- & \\
CH$_4$         & -- & & -- & & $3.5\times10^{-6}$ & IRC\,+10216 $^{(9)}$ \\

\textcolor{blue}{\textbf{NH$_3$}}         & \textcolor{blue}{$(0.25-1)\times10^{-6}$} & \textcolor{blue}{3 stars $^{(10)}$} & \textcolor{blue}{$1.7\times10^{-5}$} & \textcolor{blue}{W\,Aql $^{(3)}$} & \textcolor{blue}{$6\times10^{-8}$} & \textcolor{blue}{IRC\,+10216 $^{(11)}$} \\
C$_2$H$_4$ & -- & & -- & & $6.9\times10^{-8}$ & IRC\,+10216 $^{(12)}$ \\
\hline
SO                & $(0.12-6.0)\times10^{-6}$ & 30 stars $^{(13)}$ & -- & & -- & \\
\textcolor{blue}{\textbf{CS}}                 & \textcolor{blue}{$(0.014-1.1)\times10^{-7}$} & \textcolor{blue}{30 stars $^{(13)}$} & $(0.1-8.2)\times10^{-6}$ & 6 stars $^{(14)}$ & $(0.027-2.1)\times10^{-5}$ & 25 stars $^{(15)}$ \\
H$_2$S        & $(0.05-3)\times10^{-5}$ & 5 stars $^{(16)}$ & -- & & $4\times10^{-9}$ & IRC\,+10216 $^{(17)}$ \\
SH                & -- & & $2\times10^{-7}$ & R\,And $^{(18)}$ & -- & \\
\textcolor{blue}{\textbf{SO$_2$}}        & \textcolor{blue}{$(0.04-7.4)\times10^{-6}$} & \textcolor{blue}{30 stars $^{(13)}$} & -- & & -- & \\
\hline
SiO                & $(0.02-5.4)\times10^{-5}$ & 45 stars $^{(19)}$ & $(0.04-6.8)\times10^{-5}$ & 25 stars $^{(20)}$ & $(0.003-1)\times10^{-5}$ & 25 stars $^{(15)}$ \\
SiS                & $(0.0062-1.9)\times10^{-6}$ & 30 stars $^{(13)}$ & $(0.18-1.5)\times10^{-6}$ & 5 stars $^{(14)}$ & $(0.096-1.1)\times10^{-5}$ & 25 stars $^{(15)}$ \\
SiC$_2$        & -- & & -- & & $(0.037-3.7)\times10^{-5}$ & 25 stars $^{(21)}$ \\
Si$_2$C        & -- & & -- & & $(0.4-2)\times10^{-7}$ & IRC\,+10216 $^{(22)}$ \\
\textcolor{blue}{\textbf{SiH$_4$}}        & -- & & -- & & \textcolor{blue}{$2.2\times10^{-7}$} & \textcolor{blue}{IRC\,+10216 $^{(9)}$} \\
\hline
HCP              & -- & & -- & & $2.5\times10^{-8}$ & IRC\,+10216 $^{(17,23)}$ \\
PO                & $(0.55-1)\times10^{-7}$ & 3 stars $^{(24)}$ & -- & & -- & \\
PN                & $(1-2)\times10^{-8}$ & 3 stars $^{(24)}$ & -- & & $3\times10^{-10}$ & IRC\,+10216 $^{(25)}$ \\
\textcolor{blue}{\textbf{PH$_3$}}        & -- & & -- & & \textcolor{blue}{$1\times10^{-8}$} & \textcolor{blue}{IRC\,+10216 $^{(26)}$} \\
\hline
HF                & -- & & -- & & $8\times10^{-9}$ & IRC\,+10216 $^{(27)}$ \\
HCl               & -- & & $1.5\times10^{-8}$ & R\,And $^{(18)}$ & $1\times10^{-7}$ & IRC\,+10216 $^{(27)}$ \\
\hline
NaCl             & $4\times10^{-9}$ & IK\,Tau $^{(28)}$ & -- & & $1.8\times10^{-9}$ & IRC\,+10216 $^{(17,29)}$ \\
KCl               & -- & & -- & & $5\times10^{-10}$ & IRC\,+10216 $^{(17)}$ \\
AlCl              & $(0.09-2.5)\times10^{-8}$ & 2 stars $^{(30)}$ & -- & & $7\times10^{-8}$ & IRC\,+10216 $^{(17)}$ \\
AlF               & -- & & -- & & $1\times10^{-8}$ & IRC\,+10216 $^{(17)}$ \\
AlO               & $(0.1-9.5)\times10^{-8}$ & 3 stars $^{(30,31)}$ & -- & & -- & \\
AlOH            & $(1.4-4.4)\times10^{-9}$ & 2 stars $^{(30)}$ & -- & & -- & \\
TiO               & $(0.1-1.0)\times10^{-7}$ & $o$\,Cet $^{(32)}$ & -- & & -- & \\
TiO$_2$       & $(0.1-1.0)\times10^{-7}$ & $o$\,Cet $^{(32)}$ & -- & & -- & \\
NaCN           & -- & & -- & & $3\times10^{-9}$ & IRC\,+10216 $^{(17,33)}$ \\
KCN             & -- & & -- & & $6\times10^{-10}$ & IRC\,+10216 $^{(34)}$ \\
\hline
\end{tabular}
\tablenoteb{Note: The cases in which observed abundances are much higher than expected from chemical equilibrium are highlighted in blue.\\
References: (1) \cite{Maercker2016}. (2) \cite{Schoier2011}. (3) \cite{Danilovich2014}. (4) \cite{Decin2010}. (5) \cite{Neufeld2010}. (6) \cite{Fonfria2008}. (7) \cite{Schoier2013}. (8) \cite{Tsuji1997}. (9) \cite{Keady1993}. (10) \cite{Wong2018}. (11) \cite{Schmidt2016}. (12) \cite{Fonfria2017}. (13) \cite{Massalkhi2020}. (14) \cite{Danilovich2018}. (15) \cite{Massalkhi2019}. (16) \cite{Danilovich2017}. (17) \cite{Agundez2012}. (18) \cite{Yamamura2000}. (19) \cite{Gonzalez-Delgado2003}. (20) \cite{Ramstedt2009}, V386\,Cep removed. (21) \cite{Massalkhi2018}. (22) \cite{Cernicharo2015}. (23) \cite{Agundez2007}. (24) \cite{Ziurys2018}. (25) \cite{Milam2008}. (26) \cite{Agundez2014b}. (27) \cite{Agundez2011}. (28) \cite{Milam2007}. (29) \cite{Quintana-Lacaci2016}. (30) \cite{Decin2017}. (31) \cite{Kaminski2016}. (32) \cite{Kaminski2017}. (33) \cite{Quintana-Lacaci2017}. (34) \cite{Pulliam2010}.}
\end{table*}

\begin{figure*}
\centering
\includegraphics[angle=0,width=\textwidth]{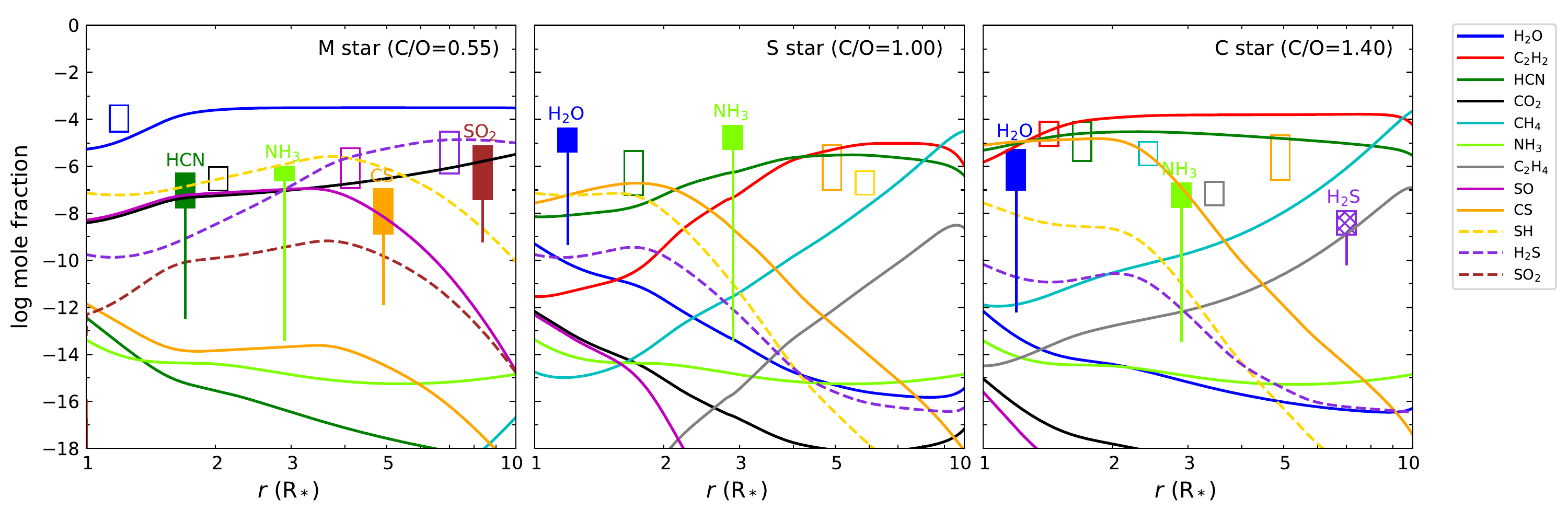} \includegraphics[angle=0,width=\textwidth]{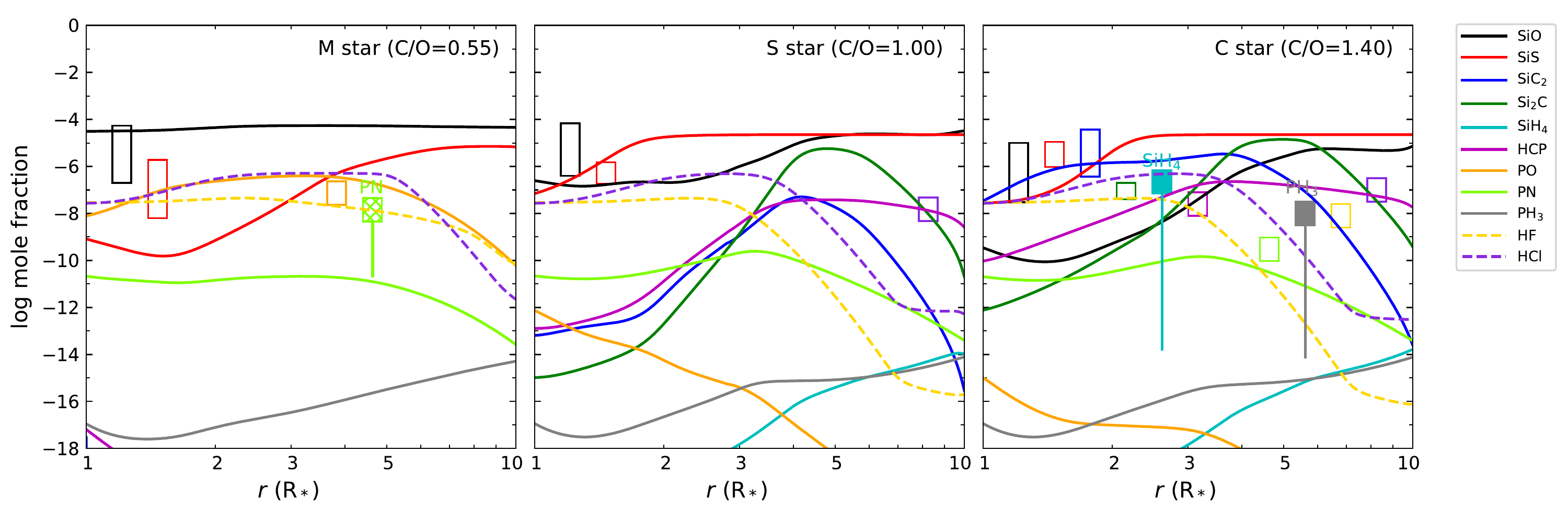} \includegraphics[angle=0,width=\textwidth]{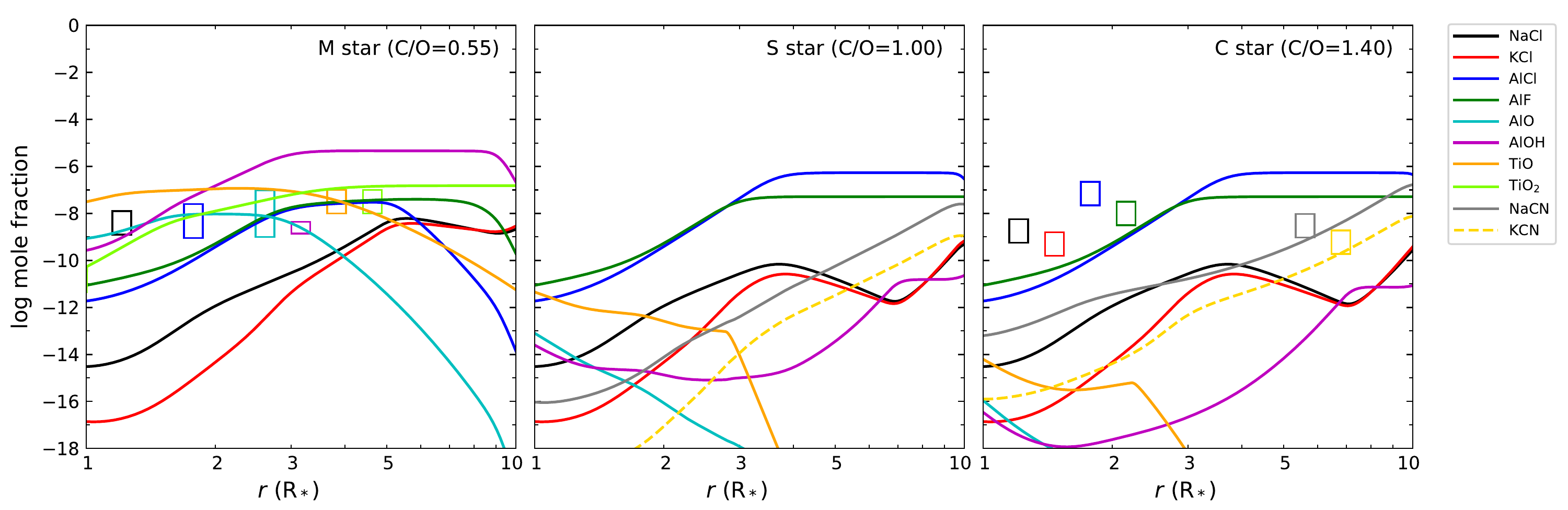}
\caption{Calculated chemical equilibrium abundances of parent molecules in M-, S-, and C-type AGB atmospheres are shown as a function of radius and are compared with abundances derived from observations. Chemical equilibrium calculations include only gaseous species. Observed abundances are indicated by rectangles, whose vertical extent corresponds to the range of observed abundances given in Table~\ref{table:observations}. Rectangles are located at different radii to facilitate visualization. Empty rectangles correspond to cases in which observed abundances agree with any of the abundances calculated by chemical equilibrium in the 1-10 $R_*$ range (usually the maximum abundance). Filled rectangles are used to indicate cases with a severe disagreement (by several orders of magnitude) between observed and calculated abundances, while we use hatched rectangles to indicate a significant disagreement (by more than one order of magnitude). The level of disagreement between the observed and maximum calculated abundance is indicated by a vertical line.} \label{fig:abun_obs}
\end{figure*}

\subsection{Successes}

Thermochemical equilibrium has been remarkably successful at explaining the molecular composition of circumstellar envelopes around AGB stars (e.g., \citealt{Tsuji1964,Tsuji1973}). A major success is that chemical equilibrium has provided the theoretical framework for understanding the chemical differentiation between envelopes around M-, S-, and C-type AGB stars according to the elemental C/O at the star surface. In this scenario, CO is the most abundant molecule after H$_2$ and it locks most of the carbon or oxygen depending on whether the C/O is lower or higher than one. This basic fact is at the heart of the most widely used method for determining mass-loss rates from AGB stars from observations of circumstellar emission in rotational lines of CO \citep{Hofner2018}.

Predictions of chemical equilibrium for the budget of major elements have in the main been confirmed by observations. In Table~\ref{table:observations} we list the parent molecules that have been observed in envelopes around AGB stars of M-, S-, and C-type and the derived abundance ranges. We refer to molecules that are formed in the inner regions of AGB envelopes as parent molecules, as opposed to daughter molecules that are formed in the external layers of the envelope. The parent character of the molecule has been confirmed for most by observation of high-energy lines or through interferometric maps. For a few, the information of their spatial distribution is not conclusive, although formation in the inner envelope is the most likely origin. The observed abundances are compared in Fig.~\ref{fig:abun_obs} with the results from the chemical equilibrium calculations performed in this study for a standard AGB atmosphere, that is, using the elemental composition given in Table~\ref{table:elements} and the pressure-temperature profile discussed in Sec.~\ref{sec:pt-profile}. Only gaseous species are included in the calculations presented in Fig.~\ref{fig:abun_obs}.

Calculated abundances are expressed here as mole fractions, while observed abundances are usually given in the literature relative to H$_2$ (where it is implicitly assumed that most hydrogen is molecular). These two quantities are identical throughout most of the atmosphere. Only in the hot innermost regions, where atomic hydrogen may become more abundant than H$_2$ (inner to $\sim2$ $R_*$ for our adopted radial profiles of pressure and temperature), can the two abundance measures differ by as much as a factor of two. For our purposes, this is not very important because calculated and observed abundances are compared at an order-of-magnitude level.

Chemical equilibrium calculations (e.g., \citealt{Tsuji1964,Tsuji1973}) make clear predictions for the main reservoirs of C, N, and O in AGB atmospheres. The main carrier of oxygen (except for CO) in envelopes around oxygen-rich AGB stars is predicted to be H$_2$O. This has been verified by observations \citep{Gonzalez-Alfonso1999,Maercker2016}. In carbon-rich AGB atmospheres, the main carriers of carbon (except for CO) are predicted to be C$_2$H$_2$ and HCN, which is also supported observationally \citep{Fonfria2008,Schoier2013}. Molecular nitrogen is predicted to be the main carrier of nitrogen regardless of the C/O, although this has never been confirmed by observations because it is difficult to detect N$_2$. Hydrocarbons such as CH$_4$ and C$_2$H$_4$ are calculated to be quite abundant at 10 $R_*$ in C-type atmospheres (see Fig.~\ref{fig:abun_obs}) and have been observed in IRC\,+10216 with abundances of about the predicted ones \citep{Keady1993,Fonfria2017}. Furthermore, carbon dioxide is calculated with a mole fraction in the range $10^{-8}$-$10^{-6}$ in M-type atmospheres (see Fig.~\ref{fig:abun_obs}), and it is observed with an abundance relative to H$_2$ of $3\times10^{-7}$ in SW\,Vir \citep{Tsuji1997}.

Sulfur is predicted to be largely in the form of molecules such as CS and SiS in C-rich atmospheres, SiS in S-type stars, and H$_2$S in O-rich atmospheres (see Fig.~\ref{fig:abun_obs}), which essentially agrees with observations \citep{Danilovich2017,Danilovich2018,Massalkhi2019}. Other S-bearing molecules that are predicted to be abundant in M-type atmospheres are SO and SiS, and these species are indeed observed with relatively high abundances in some O-rich envelopes \citep{Bujarrabal1994,Schoier2007,Danilovich2016,Massalkhi2020}. The radical SH is predicted to be relatively abundant in the hot inner regions of AGB atmospheres and has been observed through infrared observations in the atmosphere of the S-type star R\,And \citep{Yamamura2000}.

Silicon monoxide (SiO) is predicted to be the most abundant Si-bearing molecule in the entire 1-10 $R_*$ range in the atmospheres of M stars. In S-type atmospheres, the calculated abundance of SiO decreases by two orders of magnitude in the 1-5 $R_*$ but retains a very high abundance beyond, and the same occurs in C-rich atmospheres, although in this case, the abundance drop in the 1-5 $R_*$ is even more pronounced (see Fig.~\ref{fig:abun_obs}; see also \citealt{Agundez2006}). Observations indicate that the abundance of SiO does not differ significantly between envelopes around M-, S-, and C-type stars, although in all them the SiO abundance decreases with increasing mass-loss rate \citep{Gonzalez-Delgado2003,Schoier2006,Ramstedt2009,Massalkhi2019,Massalkhi2020}. This decline in the SiO abundance with increasing envelope density is not a consequence of chemical equilibrium \citep{Massalkhi2019}, but has been interpreted as evidence that SiO disappears from the gas phase at high densities to be incorporated into dust grains \citep{Gonzalez-Delgado2003,Schoier2006,Ramstedt2009,Massalkhi2019,Massalkhi2020}. It therefore appears that the gradual abundance decline calculated for SiO in the 1-5 $R_*$ region from stellar type in the sense M $\rightarrow$ S $\rightarrow$ C does not have a direct consequence in the SiO abundance that is injected into the expanding wind. However, this behavior predicted by chemical equilibrium probably explains why SiO masers are observed in M-type stars but not toward carbon stars (e.g., \citealt{Pardo2004}). Except for these details, chemical equilibrium and observations agree in the fact that SiO is one of the most abundant carriers of silicon in the atmospheres of M-, S-, and C-type stars. Calculations and observations also agree for SiS in that it is an abundant molecule regardless of the C/O. However, observations indicate a differentiation between C- and O-rich envelopes, with SiS being on average one order of magnitude more abundant in carbon-rich sources \citep{Schoier2007,Danilovich2018,Massalkhi2019,Massalkhi2020}. Moreover, in some oxygen-rich envelopes, the fractional abundance of SiS relative to H$_2$ is as low as $\sim$10$^{-8}$, which is well below the predictions of chemical equilibrium \citep{Danilovich2019,Massalkhi2020}.

Two silicon-bearing molecules, SiC$_2$ and Si$_2$C, become quite abundant in C-rich atmospheres according to chemical
equilibrium \citep{Tejero1991,Takano1992,Yasuda2012}, and somewhat less abundant in S-type atmospheres (see Fig.~\ref{fig:abun_obs}). The two molecules are observed to be abundant in the carbon star IRC\,+10216, as reported by \cite{Cernicharo2010,Cernicharo2015}, although \cite{Massalkhi2018} found that the abundance of SiC$_2$ is not uniform in C-rich envelopes, but that, similarly to the case of SiO, it decreases with increasing mass-loss rate. As \cite{Massalkhi2018} pointed out, this behavior does not arise from chemical equilibrium, but indicates that SiC$_2$ is a potential gas-phase precursor of SiC dust.

For phosphorus, chemical equilibrium predicts that HCP is a main carrier in C-type atmospheres, while PO dominates to a large extent in M-type stars \citep{Agundez2007,Milam2008}. The two molecules have been detected in the corresponding environments, which confirms this point \citep{Agundez2007,Ziurys2018}. Calculations also predict a relative abundance of about $10^{-10}$ for PN in C-rich atmospheres. This value agrees with the abundance derived in the C-star envelope IRC\,+10216 \citep{Milam2008}.

The halogen elements fluorine and chlorine are predicted to be largely in the form of HF and HCl in the inner regions of AGB atmospheres, regardless of the C/O (see Fig.~\ref{fig:abun_obs}). The fact that most F is predicted to be in the form of HF has been used to derive fluorine abundances in carbon stars by observing the $v$ = 1-0 vibrational band of HF \citep{Abia2010}. An independent measurement of the abundance of HF was provided by the detection of the $J$ = 1-0 rotational transition in the C-rich envelope IRC\,+10216 \citep{Agundez2011}. The abundance derived in this study was found to be $\sim10$~\% of the value expected if fluorine were mostly in the form of HF with a solar abundance, which was interpreted in terms of depletion onto dust grains. \cite{Agundez2011} also reported observations of low-$J$ transitions of HCl in IRC\,+10216 and derived an abundance for HCl of 15~\% of the solar abundance of chlorine, while \cite{Yamamura2000} derived an even lower abundance for HCl in the S-type star R\,And from observations of ro-vibrational lines. The missing chlorine might be depleted onto dust grains or in atomic form. Given the variation in chemical equilibrium abundances of HF and HCl with radius (see Fig.~\ref{fig:abun_obs}) and the uncertainties in the abundances derived from observations, we can consider that calculations and observations agree that HF and HCl are important carriers of fluorine and chlorine, respectively, in AGB atmospheres. However, observations of these two molecules in more sources is needed to understand the chemistry of halogens in AGB atmospheres better.

Unlike in interstellar clouds, where metals largely form part of dust grains, a wide variety of metal-bearing molecules are observed in the envelopes around AGB stars. Many of them are in fact formed in the hot stellar atmosphere, where they are relatively abundant according to chemical equilibrium, and are later incorporated into the expanding envelope. Early observations of metal-containing molecules revealed the metal halides NaCl, KCl, AlCl, and AlF \citep{Cernicharo1987}, while the metal cyanides NaCN and KCN were detected later \citep{Turner1994,Pulliam2010}. All these molecules were discovered in the C-star envelope IRC\,+10216, and some of them have been observed exclusively in this source. Other metal cyanides such as MgNC or CaNC are observed in IRC\,+10216 \citep{Kawaguchi1993,Guelin1993,Cernicharo2019a}, although they are formed in the outer envelope and thus are not parent molecules. The parent character of NaCN has been confirmed through interferometric observations \citep{Quintana-Lacaci2017}, while in the case of KCN, the parent character is merely suggested by chemical equilibrium, which predicts a similar behavior as for NaCN in C-star atmospheres (see Fig.~\ref{fig:abun_obs}). In general, the abundances derived for NaCl, KCl, AlCl, AlF, NaCN, and KCN in IRC\,+10216 \citep{Pulliam2010,Agundez2012} are consistent with the expectations from chemical equilibrium calculations of C-rich AGB atmospheres. The metal halides NaCl and AlCl have also been observed in a few O-rich envelopes \citep{Milam2007,Decin2017} with abundances that agree with chemical equilibrium calculations.

Some metal oxides have long been observed at optical and near-infrared wavelengths in the spectra of AGB stars. For example, TiO and VO are detected toward M- and S-type stars \citep{Joyce1998}, and the oxides of $s$-process elements ZrO, YO, and LaO are conspicuous in S-type stars \citep{Keenan1954}. More recently, metal oxides such as AlO, AlOH, TiO, and TiO$_2$ have been detected toward M-type AGB stars through their rotational spectrum, which allows constraining their abundances \citep{Kaminski2016,Kaminski2017,Decin2017}. In general, the abundances derived for AlO, AlOH, TiO, and TiO$_2$ agree well with the values calculated by chemical equilibrium in O-rich AGB atmospheres (see Fig.~\ref{fig:abun_obs}). These molecules are clearly parent molecules, as confirmed by interferometric observations, and studying them is interesting because some of them may act as gas-phase precursors in the formation of dust around M stars.

\subsection{Failures} \label{sec:failures}

In recent years,  increasingly more observational discoveries of parent molecules in AGB envelopes have severely disagreed with the predictions of chemical equilibrium. Several molecules are observed in envelopes around AGB stars of certain chemical type with abundances well above (by several orders of magnitude) the expectations from chemical equilibrium calculations. It is interesting to note that the disagreement with chemical equilibrium always means anomalously overabundant molecules. That is, no molecule is predicted to be abundant that is observed to be much less abundant. These anomalously overabundant molecules are highlighted in blue in Table~\ref{table:observations} and are also indicated by filled rectangles in Fig.~\ref{fig:abun_obs}, with a vertical line that shows the level of disagreement between observations and chemical equilibrium.

One of the most noticeable failures of chemical equilibrium concerns H$_2$O. This molecule is predicted and observed to be very abundant in O-rich envelopes, but in S- and C-type stars, water is predicted to have a negligible abundance and is expected not to be observable, although it is detected with a relatively high abundance (see Fig.~\ref{fig:abun_obs}). The history of the problem of water started with the detection of the low-energy rotational line at 557~GHz in the carbon star IRC\,+10216 with the space telescopes Submillimeter Wave Astronomy Satellite (SWAS) and \textit{Odin} \citep{Melnick2001,Hasegawa2006}. Different scenarios were offered to explain this unexpected discovery, and they included the sublimation of cometary ices from a putative Kuiper belt analog \citep{Melnick2001,Ford2001}, Fischer-Tropsch catalysis on iron grains \citep{Willacy2004}, and the formation in the outer envelope through the radiative association of H$_2$ and O \citep{Agundez2006}. The subsequent detection with \textit{Herschel} of many high-energy rotational lines of H$_2$O in IRC\,+10216 ruled out the three previous scenarios and constrained the formation region of water to the very inner regions of the envelope, within 5-10 $R_*$ \citep{Decin2010}. The presence of water in C-rich envelopes was found to be a common phenomenon and not restricted to IRC\,+10216 \citep{Neufeld2011}. Moreover, the problem of water in carbon stars has been extended to S-type AGB stars with the detection of abundant H$_2$O in some sources \citep{Schoier2011,Danilovich2014}. As illustrated in Fig.~\ref{fig:abun_obs}, the problem of water is a problem of more than four to five orders of magnitude.

Ammonia is also a remarkable example of chemical equilibrium breakdown in AGB atmospheres. The chemical equilibrium abundance of NH$_3$ is vanishingly small regardless of the C/O (see Fig.~\ref{fig:abun_obs}). Nonetheless, infrared observations of ro-vibrational lines and \textit{Herschel} observations of the low-energy rotational line at 572 GHz have outlined a scenario in which NH$_3$ is fairly abundant, with abundances relative to H$_2$ in the range 10$^{-7}$-10$^{-5}$ in envelopes around M-, S-, and C-type AGB stars \citep{Danilovich2014,Schmidt2016,Wong2018}. The formation radius of NH$_3$ is constrained to be in the range 5-20 $R_*$ in the C-star envelope IRC\,+10216 \citep{Keady1993,Schmidt2016}. For ammonia, the disagreement with chemical equilibrium is a problem of more than six orders of magnitude.

The carbon-bearing molecules HCN and CS are together with C$_2$H$_2$ the main carriers of carbon in C-type atmospheres. HCN and CS are observed with high abundances in C-rich envelopes, which agrees with expectations from chemical equilibrium. In the atmospheres of S-type stars, the two molecules are also predicted to be relatively abundant, although with somewhat lower abundances than in C-type atmospheres. This has also been verified observationally (see Fig.~\ref{fig:abun_obs}). HCN and CS show a clear chemical differentiation depending on the C/O because observed abundances in oxygen-rich envelopes are systematically lower than in C-rich sources by about two orders of magnitude \citep{Bujarrabal1994,Schoier2013,Danilovich2018,Massalkhi2019}. Although the observed abundances of HCN and CS in O-rich envelopes are below those in C-rich envelopes, they are still much higher than expected from chemical equilibrium, which for C/O $<$ 1 predicts negligible abundances for these two molecules. According to the interferometric observations by \cite{Decin2018a}, HCN is formed at radii smaller than about 6 $R_*$ in R\,Dor, while in the case of CS, the abundance profiles derived for IK\,Tau, W\,Hya, and R\,Dor indicate a formation radius of a few stellar radii \citep{Danilovich2019}. The difference between observed abundances and the values calculated by chemical equilibrium is large, more than four orders of magnitude for CS and more than five orders of magnitude for HCN (see Fig.~\ref{fig:abun_obs}).

Sulfur dioxide (SO$_2$) has been observed in oxygen-rich AGB envelopes for a long time \citep{Omont1993}. It was originally thought to be formed in the outer layers of the envelope, but infrared observations of the $\nu_3$ band at 7.3 $\mu$m by \cite{Yamamura1999} and observations of high-energy rotational lines \citep{Danilovich2016,Velilla-Prieto2017} indicated that in fact SO$_2$ originates in the inner regions of the envelope with abundances of about 10$^{-6}$ relative to H$_2$. The result is striking because the chemical equilibrium abundance predicted for SO$_2$ in M-type atmospheres is lower by at least three to four orders of magnitude (see Fig.~\ref{fig:abun_obs}). \cite{Velilla-Prieto2017} reported a formation radius for SO$_2$ in IK\,Tau in the range 1-8 $R_*$, although interferometric observations would be desirable.

In addition to H$_2$O and NH$_3$, there are two other hydrides, silane (SiH$_4$) and phosphine (PH$_3$), that are observed in C-rich envelopes with abundances well above the prediction of chemical equilibrium. Silane was detected in the carbon star IRC\,+10216 through infrared observations of ro-vibrational lines \citep{Keady1993}, while phosphine has been observed toward the same carbon star from observations of low-energy rotational transitions \citep{Agundez2008,Agundez2014b}. The formation radii of SiH$_4$ and PH$_3$ in IRC\,+10216 are not well constrained, although \cite{Keady1993} favored a distribution in which SiH$_4$ is present from $\sim40$ $R_*$. Thus, silane can be considered a late parent species because of its large formation radius. The discrepancy between the abundances derived from observations and the values calculated with chemical equilibrium is $>$ 6 orders of magnitude for SiH$_4$ and PH$_3$. It is curious to note that all the anomalously overabundant molecules in C-type AGB stars are hydrides. The same is not true for M-type AGB stars.

\begin{figure*}
\centering
\includegraphics[angle=0,width=\textwidth]{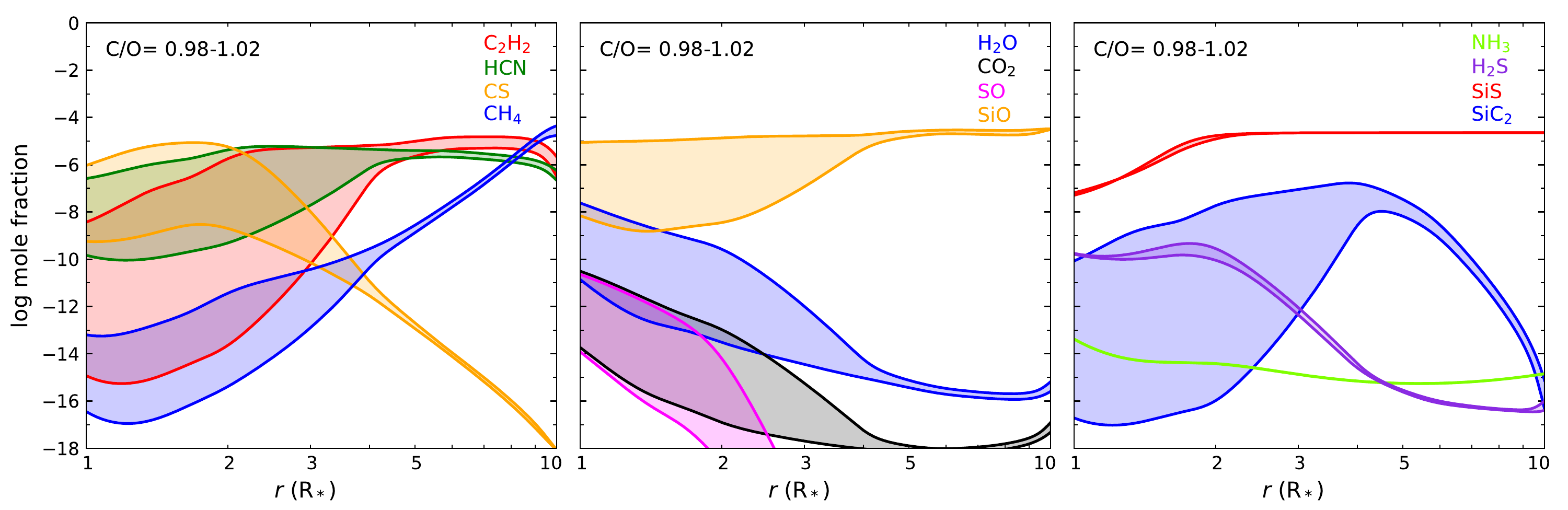}
\caption{Calculated chemical equilibrium abundances of selected parent molecules as a function of radius for an S-type atmosphere, where the shadowed regions correspond to calculated abundances adopting C/O in the range 0.98-1.02. Chemical equilibrium calculations include only gaseous species.} \label{fig:abun_c_o_ratio}
\end{figure*}

Several molecules show a large discrepancy between the abundances derived from observations and calculated by chemical equilibrium, although it is not as severe as for the molecules discussed above. We refer to PN in O-rich stars and H$_2$S in C-rich stars, which are indicated by hatched rectangles in Fig.~\ref{fig:abun_obs}. For PN in O-rich AGB atmospheres, the disagreement between the observed abundances, (1-2)$\times$10$^{-8}$ \citep{Ziurys2018}, and the calculated maximum chemical equilibrium abundance is almost three orders of magnitude. However, uncertainties on the observational and theoretical sides mean that the true level of disagreement is unclear. For example, while \cite{Ziurys2018} derived a PN abundance of 10$^{-8}$ relative to H$_2$ in IK\,Tau, \cite{DeBeck2013} and \cite{Velilla-Prieto2017} derived higher abundances, (3-7)$\times10^{-7}$, in this source. When we give preference to these latter abundances, the level of disagreement would be even higher. On the other hand, the formation enthalpy of PN is rather uncertain (see \citealt{Lodders1999REF}), which directly translates into the calculated chemical equilibrium abundance. In this study we adopted the thermochemical data for PN from \cite{Lodders1999REF}, who gives preference to a formation enthalpy at 298.15 K of 171.5 kJ mol$^{-1}$, while other compilations such as JANAF use lower values that would result in higher chemical equilibrium abundances for PN. This would reduce the level of disagreement. In the case of H$_2$S in C-rich AGB stars, the calculated maximum chemical equilibrium abundance is $7\times10^{-11}$, while the value derived from observations is $\sim50$ times higher. In this case, the observed abundance is based on the detection of only one line in only one source (see \citealt{Agundez2012}), and thus it has to be viewed with some caution.
In summary, the main failures of chemical equilibrium to account for the observed abundances of parent molecules in circumstellar envelopes are NH$_3$, HCN, CS, SO$_2$, and possibly PN in M-type stars, H$_2$O and NH$_3$ in S-type stars, and the hydrides H$_2$O, NH$_3$, SiH$_4$, PH$_3$, and perhaps H$_2$S as well in C-type stars. The large discrepancies between the abundances derived from observations and those calculated with chemical equilibrium necessarily imply that nonequilibrium chemical processes must be at work in AGB atmospheres. Any invoked nonequilibrium scenario must account for all these anomalously overabundant molecules, but must also reproduce the remaining molecular abundances that are reasonably well explained by chemical equilibrium. No scenario currently provides a fully satisfactory agreement with observations, although two mechanisms that can drive the chemical composition out of equilibrium have been proposed.

The first scenario involves shocks that are produced in the extended atmosphere as a consequence of the pulsation of the AGB star. A model based on this scenario was originally developed to study the inner wind of the carbon star IRC\,+10216 \citep{Willacy1998} and the oxygen-rich Mira star IK\,Tau \citep{Duari1999}, and was later generalized by \cite{Cherchneff2006} to AGB winds with different C/O. The main successes of these models were that they resulted in high abundances for HCN and CS in O-rich winds, although neither was H$_2$O efficiently produced in C-rich winds nor NH$_3$ independently of the C/O. Subsequent models in which the chemical network was modified resulted in a relatively high abundance of H$_2$O in C-rich AGB winds \citep{Cherchneff2011,Cherchneff2012}. In this new chemical network, however, rate constants of reverse reactions do not obey detailed balance, which may affect the predicted abundance of water. Similar models performed by \cite{Marigo2016} found that in O-rich atmospheres, HCN is only formed with abundances of about the observed ones when nearly isothermal shocks are considered.

A different scenario that consists of photochemistry driven by interstellar ultraviolet photons that would penetrate the inner envelope through the clumpy envelope was proposed by \cite{Decin2010} to explain the formation of water in the inner envelope of IRC\,+10216. The scenario was later generalized to AGB envelopes with different C-to-O ratios by \cite{Agundez2010}. In these models, warm photochemistry is able to efficiently form H$_2$O, NH$_3$, and H$_2$S in the inner regions of C-rich envelopes, while NH$_3$, HCN, and CS are synthesized in the inner layers of O-rich envelopes. Similar models based on a different formalism were carried out by \cite{VandeSande2018}, who found similar qualitative results. \cite{VandeSande2019} then found that ultraviolet photons from the AGB star could also lead to some photochemistry for sufficiently high stellar temperatures and degree of clumpiness.

Although the two scenarios are promising in that they result in an enhancement of some of the anomalously overabundant parent molecules observed in AGB envelopes, the main problem with them is that they are quite parametric, that is, they depend on parameters such as the shock strength or the degree of clumpiness. These parameters are poorly constrained from observations.

\subsection{S-type atmospheres: sensitivity to C/O } \label{sec:s_stars}

\begin{figure*}
\centering
\includegraphics[angle=0,width=\textwidth]{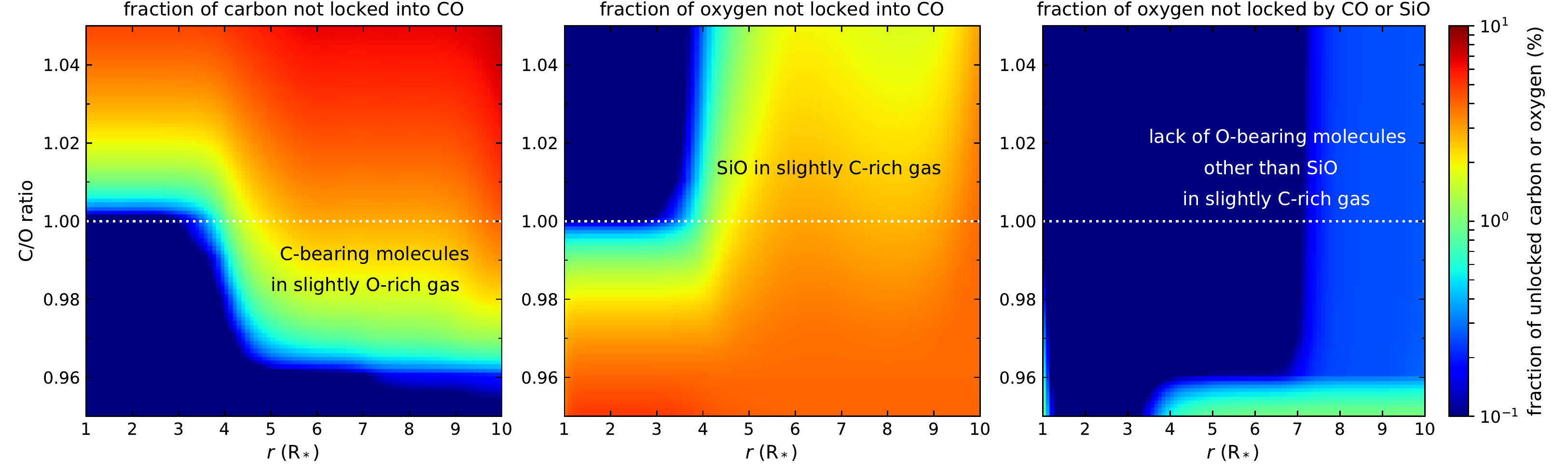}
\caption{Left and middle panels: Fractions of C and O that is not locked into CO as a function of radius and C/O, as calculated by gas-phase chemical equilibrium. To illustrate that the vast majority of oxygen that is not locked into CO is trapped by SiO for C-to-O ratios around one, we show in the right panel the fraction of oxygen that is not locked by CO or SiO. Chemical equilibrium calculations include only gaseous species.} \label{fig:c_o_excess}
\end{figure*}

The chemical equilibrium calculations for an S-type atmosphere shown in Fig.~\ref{fig:abun_obs} assume that the elemental C/O is exactly one. However, the C/O of S stars may range from slightly oxygen-rich to slightly carbon-rich. Because C/O $<$ 1 and C/O $>$ 1 imply a marked chemical differentiation, we here explore how sensitive molecular abundances are to slight changes in the C/O. This provides an idea of the diversity of the atmospheric composition between S stars with slightly different C-to-O ratios. In Fig.~\ref{fig:abun_c_o_ratio} we show the abundances of some important parent molecules as a function of radius for C-to-O ratios in the range 0.98-1.02. These calculations include only gaseous species. Over this narrow range of C-to-O ratios, some molecules experience large abundance variations (e.g., C$_2$H$_2$ and SiC$_2$), while other species remain almost insensitive to C/O (e.g., NH$_3$, H$_2$S, and SiS). The sensitivity to the C/O is more pronounced in the inner atmosphere, while at radii larger than $\sim5$~$R_*$, molecules tend to show little abundance variation with C/O. If the abundances of the parent molecules are set by the chemical equilibrium values in the hot inner regions, we therefore expect large abundance variations from source to source that reflect the diversity of C-to-O ratios from slightly below one to slightly above one. This would be most noticeable for molecules such as C$_2$H$_2$, HCN, CS, SiO, and SiC$_2$, for which the predicted abundances range from moderately high to very low, depending on whether the C/O is higher or lower than one. The severe disagreement that is found between chemical equilibrium and observations for H$_2$O and NH$_3$ in S-type stars (see the discussion in Sec.~\ref{sec:failures}) persists when the C/O is allowed to vary slightly around one.

In general, carbon-bearing molecules reach higher abundances than oxygen-bearing molecules for C-to-O ratios around one. For example, C-bearing molecules such as HCN, C$_2$H$_2$, and CH$_4$ maintain moderately high abundances over certain radii in slightly oxygen-rich conditions, while the only O-bearing molecule that is present with a non-negligible abundance under slightly carbon-rich conditions is SiO. Other O-bearing molecules such as H$_2$O and SO need C-to-O ratios well lower than one to reach moderately high abundances. The reason for this behavior is illustrated in Fig.~\ref{fig:c_o_excess}. For radii larger than $\sim4$~$R_*$, when the gas is slightly O-rich (C/O in the range 0.96-1.00), a few percent of the carbon is not locked by CO and goes to C-bearing molecules such as C$_2$H$_2$, HCN, and CH$_4$. When we now focus on slightly C-rich conditions, we see that for radii larger than $\sim4$~$R_*$, a few percent of oxygen is not trapped by CO. However, in this case, the oxygen that is not locked by CO is mostly in the form of SiO. SiO even competes with CO for the oxygen over a wide range of C-to-O ratios, and this is the origin of the relatively high abundance of SiO in C-rich AGB atmospheres. Except for SiO, no other O-bearing molecule is predicted with a significant abundance under slightly C-rich conditions. In summary, C-bearing molecules compete more efficiently with CO for the carbon than O-bearing molecules (other than SiO) for the oxygen. The consequence is that the chemical equilibrium composition of S-type atmospheres resembles that of carbon stars more closely than that  of M-type stars.

\section{Potentially detectable molecules} \label{sec:detectable}

The gas-phase budget of the different elements that are included in the chemical equilibrium calculations in M-, S-, and C-type atmospheres is discussed in detail in Appendix~\ref{app:budget}. The most abundant molecular reservoirs of the nonmetal elements (C, O, N, Si, S, P, F, Cl, and B) are typically observed in AGB envelopes with abundances in agreement with predictions from chemical equilibrium. The only cases where major molecular reservoirs are not detected correspond to nonpolar molecules, such as N$_2$ , or molecules containing elements with a very low abundance, such as B. Several molecules that are predicted to be relatively abundant have not yet been detected, however, and observations can therefore still be used to test the predictions of chemical equilibrium.

\begin{table*}
\caption{Potentially detectable molecules in AGB atmospheres.} \label{table:detectable}
\centering
\small
\begin{tabular}{l@{\hspace{0.00cm}}l@{\hspace{0.15cm}}rl@{\hspace{0.15cm}}l@{\hspace{0.20cm}}l@{\hspace{0.05cm}}l}
\hline \hline
\multicolumn{1}{l}{Molecule} & \multicolumn{2}{c}{Prediction~$^a$} & \multicolumn{2}{c}{Electric dipole moment (D)} & \multicolumn{2}{c}{Rotational spectrum measured$?$} \\
\hline
CH                & $\sim3\times10^{-7}$ & C stars & 1.46 & \tiny{\citep{Phelps1966}} & Yes & \tiny{\citep{Martin-Drumel2011}} \\
CH$_2$        & $\sim1\times10^{-8}$ & C stars & 0.591 & \tiny{\citep{Woon2009}} & Yes & \tiny{\citep{Brunken2005}} \\
NH                & $\sim5\times10^{-10}$ & M, S, C stars & 1.389 & \tiny{\citep{Scarl1974}} & Yes & \tiny{\citep{Klaus1997,Flores-Mijangos2004,Lewen2004}} \\
SiH               & $\sim1\times10^{-8}$ & S, C stars & 0.097 & \tiny{\citep{Yurchenko2018}} & Yes$^b$ & \tiny{\citep{Betrencourt1986,Ram1998}. See CDMS$^c$.}  \\
PH               & $\sim1\times10^{-10}$ & M, S, C stars & 0.396 & \tiny{(CDMS)$^c$} & Yes & \tiny{\citep{Goto1993,Klisch1998}} \\
O$_2$          & $\sim5\times10^{-8}$ & M stars & 0~$^b$ & & Yes & \tiny{\citep{Drouin2010}} \\
Si$_3$C      & $\sim1\times10^{-7}$ & S, C stars & $\sim0.1$ & \tiny{\citep{Stanton2005}} & No & \\
Si$_5$C      & $\sim3\times10^{-6}$ & S, C stars & 0.27 & \tiny{B3LYP/cc-pVTZ$^e$} & No & \\
SiNH           & $\sim1\times10^{-8}$ & S, C stars & 0.34 & \tiny{\citep{McCarthy2015}} & Yes & \tiny{\citep{Bogey1991,McCarthy2015}} \\
SiF              & $\sim1\times10^{-10}$ & S, C stars & 1.126 & \tiny{\citep{Karna1987}} & Yes & \tiny{\citep{Tanimoto1983}} \\
SiCl             & $\sim1\times10^{-9}$ & S, C stars & 1.53 & \tiny{\citep{Gosavi1986}} & Yes & \tiny{\citep{Tanimoto1984}} \\
PS               & $\sim1\times10^{-7}$ & M stars & 0.565 & \tiny{\citep{Muller2013}} & Yes & \tiny{\citep{Ohishi1988}} \\
HPO$_2$    & $\sim1\times10^{-7}$ & M stars & 1.61 & \tiny{B3LYP/cc-pVTZ$^e$} & Yes$^b$ & \tiny{\citep{Osullivan2006}} \\
BO              & $\sim5\times10^{-10}$ & M stars & 2.62 & \tiny{(CDMS)$^c$} & Yes & \tiny{\citep{Tanimoto1986}} \\
HBO           & $\sim1\times10^{-9}$ & M, S, C stars & 2.69 & \tiny{\citep{DeYonker2005}} & Yes & \tiny{\citep{Kawashima1987}} \\
HBO$_2$   & $\sim1\times10^{-9}$ & M stars & 2.78 & \tiny{\citep{Dewar1988}} & No & \\
BF              & $\sim1\times10^{-9}$ & S, C stars & 0.50 &  \tiny{\citep{Dewar1988}} & No & \\
\hline
AlH             & $\sim1\times10^{-8}$ & S, C stars & 0.30 & \tiny{\citep{Matos1988}} & Yes & \tiny{\citep{Halfen2004,Halfen2016}} \\
                   & $\sim1\times10^{-9}$ & M stars & & & & \\
AlS                  & $\sim1\times10^{-10}$ & M stars & 3.63 & \tiny{\citep{Guichemerre2000}} & Yes & \tiny{\citep{Takano1991,Breier2018}} \\
AlOF               & $\sim1\times10^{-8}$ & M stars & 1.62 &  \tiny{B3LYP/cc-pVTZ$^e$} & No & \\
MgS           & $\sim4\times10^{-9}$ & M stars & 6.88 & \tiny{\citep{Fowler1991}} & Yes & \tiny{\citep{Takano1989,Walker1997}} \\
MgO           & $\sim1\times10^{-10}$ & M stars & 6.2 & \tiny{\citep{Busener1987}} & Yes & \tiny{\citep{Torring1986}} \\
CaS                & $\sim4\times10^{-8}$ & M stars & 10.47 & \tiny{(CDMS)$^c$} & Yes & \tiny{\citep{Takano1989}} \\
CaOH             & $\sim1\times10^{-8}$ & M stars & 1.465& \tiny{\citep{Steimle1992}} & Yes & \tiny{\citep{Ziurys1992,Scurlock1993}} \\
CaCl               & $\sim1\times10^{-8}$ & M stars & 4.257 & \tiny{\citep{Ernst1984}} & Yes & \tiny{\citep{Moller1982,Ernst1983}} \\
                       & $\sim3\times10^{-10}$ & S, C stars & & \\
CaF$_2$        & $\sim1\times10^{-8}$ & M stars & 3.340 & \tiny{\citep{Szentpaly1990}} & No & \\
CaF                & $\sim5\times10^{-9}$ & M stars & 3.07 & \tiny{\citep{Childs1984}} & Yes & \tiny{\citep{Anderson1994}} \\
BaO              & $\sim1\times10^{-10}$ & M stars & 7.955 & \tiny{\citep{Wharton1962}} & Yes & \tiny{\citep{Tiemann1974,Hocking1978,Blom1992}} \\BaS              & $\sim1\times10^{-10}$ & M stars & 10.86 & \tiny{\citep{Melendres1969}} & Yes & \tiny{\citep{Tiemann1976,Helms1980,Janczyk2006}} \\
ScO                & $\sim1\times10^{-9}$ & M stars & 4.55 & \tiny{\citep{Shirley1990}} & Yes & \tiny{\citep{Halfen2017}} \\
ScO$_2$        & $\sim1\times10^{-9}$ & M stars & 5.29 & \tiny{\citep{Gutsev2000}} & No & \\
Ti$_8$C$_{12}$ & $\sim1\times10^{-8}$ & S, C stars & 0.72 & \tiny{B3LYP/cc-pVTZ + PP for Ti$^e$} & No & \\
ZrO              & $\sim1\times10^{-9}$ & M, S, C stars & 2.551 & \tiny{\citep{Suenram1990}} & Yes & \tiny{\citep{Suenram1990,Beaton1999}} \\
ZrO$_2$      & $\sim1\times10^{-9}$ & M stars & 7.80 & \tiny{\citep{Brugh1999}} & Yes & \tiny{\citep{Brugh1999}} \\
VO                 & $\sim5\times10^{-9}$ & M stars & 3.355 & \tiny{\citep{Suenram1991}} & Yes & \tiny{\citep{Suenram1991,Flory2008}} \\
VO$_2$         & $\sim1\times10^{-8}$ & M stars & 5.40 & \tiny{\citep{Gutsev2000}} & No & \\
CrS               & $\sim1\times10^{-9}$ & M stars & 4.91 & \tiny{\citep{Bauschlicher1995}} & Yes & \tiny{\citep{PulliamZiurys2010}} \\
CrO               & $\sim1\times10^{-10}$ & M stars & 3.88 & \tiny{\citep{Steimle1989}} & Yes & \tiny{\citep{Sheridan2002}} \\
CrCl              & $\sim1\times10^{-10}$ & M stars & 6.42 & \tiny{\citep{Harrison1999}} & Yes & \tiny{\citep{Oike1998,Katoh2004}} \\
MnH              & $\sim1\times10^{-10}$ & M, S, C stars & 10.65 & \tiny{\citep{Koseki2006}} & Yes & \tiny{\citep{Halfen2008}} \\
FeS              & $\sim4\times10^{-8}$ & M stars & 6.46 & \tiny{\citep{Sharkas2017}} & Yes & \tiny{\citep{Takano2004}} \\
FeO              & $\sim3\times10^{-10}$ & M stars & 4.50 & \tiny{\citep{Steimle2004}} & Yes & \tiny{\citep{Krockertskothen1987,Allen1996}} \\
CoH             & $\sim1\times10^{-7}$ & M, S, C stars & 1.88 & \tiny{\citep{Wang2009}} & No & \\
NiS              & $\sim4\times10^{-7}$ & M stars & 5.38 & \tiny{\citep{Sharkas2017}} & Yes & \tiny{\citep{Yamamoto2007}} \\
\hline
\end{tabular}
\tablenoteb{\\
Notes:\\
$^a$~Maximum calculated mole fraction in the 1-10 R$_*$ range and type of AGB star in which the molecule is predicted to be abundant.\\
$^b$~Rotational spectrum has not been directly measured, but it can be predicted from rotationally resolved vibrational or electronic spectra.\\
$^c$~Cologne Database for Molecular Spectroscopy \citep{Muller2005}: \texttt{https://cdms.astro.uni-koeln.de}\\
$^d$~Rotational transitions radiatively allowed by the nonzero magnetic dipole moment.\\
$^e$~From quantum-chemical calculations carried out in this study.\\
}
\end{table*}

In Table~\ref{table:detectable} we present a list of molecules that have not yet been observed in AGB atmospheres but are predicted with non-negligible abundances, and thus are potentially observable. We generally include molecules for which the maximum calculated mole fraction over the 1-10 $R_*$ range is $\geq$ 10$^{-10}$. These chemical equilibrium calculations include only gaseous species. We also list the electric dipole moment of each molecule and indicate whether the rotational spectrum has been measured in the laboratory. Some of these molecules are good targets for detection through high angular resolution and sensitive observations using observatories such as ALMA. Some molecules are more favorable for detection than others. Factors that play against detection are a low abundance, a low dipole moment, a complex rotational spectrum, which results in spectral dilution, and a low spatial extent restricted to the photosphere and near surroundings. The latter may occur for some radicals that are only abundant in the very inner atmosphere and may be converted into more stable molecules at larger radii by nonequilibrium chemistry, and for metal-bearing molecules that can be severely depleted from the gas phase at relatively short radii because they are incorporated into condensates. Sensitive high angular resolution observations able to probe the very inner atmosphere are the best way to observe these molecules. For some molecules, the probabilities of detection are uncertain because they only reach high abundances at large radii, close to 10 $R_*$, where chemical equilibrium is less likely to hold because the temperatures and pressures are lower.

\subsection{Nonmetal molecules}

The hydride radicals CH, CH$_2$, and NH reach maximum abundance in the photosphere and show a marked abundance fall-off with increasing radius. The probabilities of detecting these species depend on whether the photospheric abundance can be maintained throughout the extended atmosphere and be injected into the expanding wind, or if these radicals are chemically processed and converted into more stable molecules such as methane and ammonia. If their presence is restricted to the innermost atmosphere, observations with a high angular resolution or in the infrared domain might allow probing them. The hydrides SiH and PH are also abundant in the photosphere, but have a more extended distribution than the above three radicals, which makes it more likely that they survive the travel through the extended atmosphere. Their detection is complicated because their dipole moment is low, however.

Molecular oxygen is listed in Table~\ref{table:detectable}, although its detection appears difficult. It has a relatively low mole fraction ($\sim5\times10^{-8}$) over a narrow region in the very inner atmosphere, and more importantly, the rotational transitions have very low intrinsic line strengths because there is no electric dipole moment. Detection of O$_2$ in AGB envelopes must probably await sensitive observations by future space telescopes.

Two of the silicon-carbon clusters, except for the already known SiC$_2$ and Si$_2$C (see Table~\ref{table:observations}),  Si$_3$C and Si$_5$C, are predicted to form with high abundances in S- and C-type atmospheres (see Appendix~\ref{app:si} and the bottom panel of Fig.~\ref{fig:n}). The thermochemical data for these species are taken from \cite{Deng2008}. The low-dipole moments calculated for Si$_3$C and Si$_5$C hinder their detection. The detection of these two molecules must await the characterization of their rotational spectrum in the laboratory. The silicon-carbon cluster Si$_2$C$_2$ is also predicted with a non-negligible abundance in S- and C-type atmospheres. However, this molecule has a nonpolar rhombic structure in its ground state \citep{Lammertsma1988,Presilla-Marquez1995,Rintelman2001,Deng2008} and thus cannot be detected through its rotational spectrum.

The molecule iminosilylene (SiNH) is predicted to be relatively abundant (mole fraction of up to 10$^{-8}$) in atmospheres around S- and C-type stars. Thermochemical data for this species are taken from the Chemkin Thermodynamic Database \citep{Kee2000}, which assigns a formation enthalpy at 298.15 K of 160.6 kJ mol$^{-1}$ based on quantum calculations. An astronomical search is feasible because the rotational spectrum has been measured in the laboratory \citep{McCarthy2015}, although the calculated dipole moment is low (0.34 D; \citealt{McCarthy2015}). The silicon monohalides SiF and SiCl are also potentially detectable in AGB atmospheres. They are predicted with mole fractions up to 10$^{-10}$ and 10$^{-9}$, respectively, in S- and C-type atmospheres. Although the predicted abundances are not very high, the dipole moments in excess of 1 D (see Table~\ref{table:detectable}) may help in the detection.

Of the P-bearing molecules that are not yet observed in AGB atmospheres, PS has the highest probabilities of being detected. This molecule has been predicted to be the most abundant P-bearing molecule in O-rich atmospheres by \cite{Tsuji1973}, although searches for it have not been successful \citep{Ohishi1988}. According to our calculations, PS is predicted in oxygen-rich atmospheres with an abundance as high as that of PO (which has already been observed), with the main difference that PO locks most of the phosphorus in the 1.5-4.5 $R_*$ region, while PS is the main reservoir of P somewhat farther away (in the 4.5-7 $R_*$), although still in the region of influence of chemical equilibrium. The dipole moment is not very high (0.565 D; \citealt{Muller2013}), but the high predicted abundance should permit a detection. The P-bearing molecule HPO$_2$ is also predicted with a high abundance in O-rich atmospheres, although only in the outer regions ($>$ 8 $R_*$), where chemical equilibrium is less likely to hold. The most stable isomer of HPO$_2$ ($cis$ HOPO) has been characterized spectroscopically in the laboratory \citep{Osullivan2006}, although the pure rotational spectrum has not been directly measured.

Several B-containing molecules are predicted to lock most of the boron, depending on the region of the atmosphere and the C/O. Concretely, BO, HBO, and HBO$_2$ are predicted to be main reservoirs of boron in O-rich atmospheres, while in S- and C-type atmospheres, HBO and BF are the main molecular carriers of this element. The main handicap in detecting them is the low elemental abundance of boron ($5\times10^{-10}$ relative to H). Because these molecules are main reservoirs of boron and because their dipole moment is high (except for BF), however, observation of some of them is feasible, which would place some constraints on the so far unexplored chemistry of boron.

\subsection{Metal-bearing molecules}

The observation of metal-containing molecules is complicated by several facts. First, many of the metals are present at a level of trace. Second, in many cases, the main gas-phase reservoir of the metal is atomic, and molecules are predicted at a lower level. Third, metals have a refractory character and thus high condensation temperatures, which cause them to easily leave the gas phase to form condensates. In spite of these difficulties, a variety of metal-bearing molecules have been detected with abundances in agreement with expectations from chemical equilibrium. Currently, constraints on the molecular budget of metals in the atmospheres of AGB stars are restricted to Na, K, Al, and Ti (see Table~\ref{table:observations}), although there is still some margin for a further characterization of the molecular reservoirs of other metals.

In addition the Al-bearing molecules known in AGB atmospheres (AlCl, AlF, AlO, and AlOH), some others could be detected. Aluminum monohydride (AlH) is predicted with a non-negligible abundance in M-, S-, and C-type atmospheres. This molecule has been detected in the O-rich star $o$\,Cet through optical observations \citep{Kaminski2016}, although no abundance was derived. The rotational spectrum of AlH is known from the laboratory \citep{Halfen2004,Halfen2016}, and thus it can be searched for at (sub-)millimeter wavelengths. However, the low dipole moment (0.30 D; \citealt{Matos1988}) and the fact that the two lowest rotational transitions cannot be observed from the ground because they lie close to water atmospheric lines play against its detection. The molecules AlS and AlOF are also potentially detectable in O-rich atmospheres because they are predicted with non-negligible abundances (mole fractions up to 10$^{-10}$ and 10$^{-8}$, respectively). Other Al-bearing molecules are predicted to be quite abundant, such as Al$_2$O, Al(OH)$_3$ in M stars and Al$_2$C$_2$ in C stars (see the bottom panel of Fig.~\ref{fig:b}). However, their structures are predicted to be highly symmetric with zero or very low dipole moment \citep{Turney2005,Wang2007,Cannon2000,Naumkin2008,Dong2010}, which means that it is very unlikely if not impossible to detect them through their rotational spectrum.

Magnesium is predicted to be essentially in the form of neutral atoms in AGB atmospheres. Several Mg-bearing molecules have been observed in the C-rich AGB envelope IRC\,+10216, however, such as MgNC, MgCN, HMgNC, MgC$_3$N, MgC$_2$H, and MgC$_4$H \citep{Kawaguchi1993,Guelin1993,Ziurys1995,Cabezas2013,Agundez2014c,Cernicharo2019b}, although they are formed in the outer envelope and thus are not parent molecules. The most promising candidate of the  Mg-bearing parent molecules for detection are MgS and MgO, which are predicted with non-negligible abundances in O-rich atmospheres and have fairly high dipole moments \citep{Fowler1991,Busener1987}.

Neutral atoms are also the main reservoir in AGB atmospheres for calcium. However, some molecules are predicted to trap a fraction of Ca in the 3-10 $R_*$ range, reaching non-negligible abundances. This is the case of CaS, which is predicted to have a mole fraction as high as $4\times10^{-8}$ in O-rich atmospheres. Its very high dipole moment (10.47 D; see Table~\ref{table:detectable}) makes it a very interesting candidate for detection. The hydroxides CaOH and Ca(OH)$_2$ are also predicted to be abundant in the outer parts of O-rich atmospheres. While CaOH has a dipole moment of 1.465 D \citep{Steimle1992}, Ca(OH)$_2$ is predicted to be highly linear with a very low or zero dipole moment \citep{Wang2005,Vasiliu2010}. The calcium halides CaCl, CaCl$_2$, CaF, and CaF$_2$ are also predicted with non-negligible abundances, especially in oxygen-rich atmospheres. Of these, CaCl, CaF, and CaF$_2$ have high dipole moments, but CaCl$_2$ is predicted to be linear and thus nonpolar \citep{Vasiliu2010}. \cite{Ziurys1994} searched without success for CaF toward the C-rich AGB star IRC\,+10216. Our chemical equilibrium calculations (see the third panel in Fig.~\ref{fig:be}) indicate that CaF only reaches non-negligible abundances in O-rich atmospheres.

The trace metals Ba, Sc, Zr, and V form some molecules, mainly oxides, that might be detectable through their rotational spectrum in oxygen-rich atmospheres because they are predicted to have mole fractions in the range 10$^{-10}$-10$^{-8}$ and have fairly high dipole moments. These molecules are BaO, BaS, ScO, ScO$_2$, ZrO, ZrO$_2$, VO, and VO$_2$. Some of them, such as ZrO and VO, have long been known to be present in the atmospheres of S- and M-type AGB stars from observations at optical and near-infrared wavelengths \citep{Keenan1952,Keenan1954,Joyce1998}, and various absorption bands in the spectra of S-type stars have been assigned to BaO by \cite{Dubois1977}. Moreover, ScO has been observed in the optical spectrum of V1309\,Sco, a remnant of a stellar merger whose conditions resemble those of AGB outflows \citep{Kaminski2015}. However, these oxides (BaO, ScO, ZrO, and VO) have not yet been detected through their rotational spectrum, which would allow us to derive their abundances. Other molecules, such as BaCl$_2$, Sc$_2$O, Sc$_2$O$_2$, and ZrCl$_2$ , only reach mole fractions of about 10$^{-10}$ at large radii ($\sim10$ $R_*$) in certain AGB atmospheres, although in this region it is more uncertain that chemical equilibrium prevails. The vanadium-oxygen cluster V$_4$O$_{10}$ is predicted to be the main carrier of V in oxygen-rich atmospheres beyond $\sim$7 $R_*$ (see the bottom panel in Fig.~\ref{fig:sc}). However, this highly symmetric cluster is predicted to be nonpolar according to quantum-chemical calculations at the B3LYP/cc-pVTZ + PP for V level and thus cannot be observed through its rotational spectrum. The large titanium-carbon cluster Ti$_8$C$_{12}$ is predicted to lock most of the titanium in S- and C-type atmospheres beyond 2-3 $R_*$ (see the second panel in Fig.~\ref{fig:sc}). However, its large size is a handicap for detecting it through its rotational spectrum, which is likely to be crowded by lines so that spectral dilution is a serious problem.

Three Cr-bearing molecules (CrS, CrO, and CrCl) are potentially detectable in AGB stars. Although chromium is mostly in atomic form, these molecules are predicted with mole fractions in the range 10$^{-10}$-10$^{-9}$ in M-type atmospheres, and all of them have quite high dipole moments. CrO has been observed in the optical toward the stellar-merger remnant V1309\,Sco, an object where other oxides such as TiO, VO, ScO, and AlO have also been found \citep{Kaminski2015}. These observations support that CrO can plausibly be detected at radio wavelengths in O-rich AGB stars. 

The metal transition hydrides MnH and CoH are also in the list of potentially detectable molecules. Chemical equilibrium predicts that MnH is present with a uniform mole fraction of $\sim10^{-10}$ in AGB atmospheres, regardless of the C/O  (see the second panel in Fig.~\ref{fig:cr}). The high dipole moment of MnH (10.65 D; \citealt{Koseki2006}) can help to detect it. The dipole moment is lower for CoH (1.88 D; \citealt{Wang2009}), but this hydride is predicted to be the main carrier of Co in AGB atmospheres of any chemical type and is therefore expected to be present with a fairly large abundance. Although there is certainly a problem of incompleteness in the Co-bearing molecules included in the calculations, which might affect the predicted abundance of CoH (see Sec.~\ref{app:transition-metals}), this molecule is a very interesting target for future searches in AGB atmospheres when the rotational spectrum is measured in the laboratory.

Iron and nickel are, as many other transition metals, predicted to be mostly in the form of neutral atoms. However, in O-rich atmospheres, the sulfides FeS and NiS reach relatively high mole fractions (up to $\sim4\times10^{-8}$ and $\sim4\times10^{-7}$, respectively), which together with the high dipole moments of these molecules make them attractive candidates for detection. Iron monoxide (FeO) is also calculated with a non-negligible mole fraction (up to $\sim3\times10^{-10}$) in M-type atmospheres, and thus could be detectable given its high dipole moment. A claim of detection of FeO in the oxygen-rich AGB star R\,Dor has recently been made using ALMA \citep{Decin2018b}. The inferred abundance relative to H$_2$ is a few times 10$^{-8}$, about two orders of magnitude above the predictions of chemical equilibrium. Further observations are required to unambiguously establish the presence of FeO and derive its abundance. The molecule Fe(OH)$_2$ is predicted to have a rising abundance with increasing radius, reaching a mole fraction of $\sim10^{-8}$ at 10 $R_*$ (see the third panel in Fig.~\ref{fig:cr}). However, this molecule is predicted to have a linear O-Fe-O structure, and thus probably has a very low dipole moment \citep{Wang2006}.

\section{Condensates} \label{sec:condensates}

\subsection{Observational constraints}

It is well known that solid dust grains are formed in AGB atmospheres, and that the later ejection of this material into the interstellar medium constitutes the main source of interstellar dust in the Galaxy \citep{Gehrz1989}. Infrared observations have allowed us to identify a few solid compounds in AGB envelopes, although the identification of some of them is still under discussion (see reviews by \citealt{Molster2010} and \citealt{Waters2011}). The observational situation of condensates identified in AGB envelopes is summarized in Table~\ref{table:condensates}.

\begin{table}
\caption{Condensates identified in AGB dust.} \label{table:condensates}
\centering
\small
\begin{tabular}{lll}
\hline \hline
\multicolumn{1}{l}{Condensate} & \multicolumn{2}{l}{Identification} \\
\hline
\multicolumn{3}{c}{M stars} \\
\hline
Amorphous silicate  & IR & \cite{Woolf1969} \\
Crystalline silicates  & & \\
~~~~Olivine (Mg$_{(2-2x)}$Fe$_{2x}$SiO$_4$)  & IR & \cite{Waters1996} \\
~~~~Pyroxene (Mg$_{(1-x)}$Fe$_{x}$SiO$_3$)  & IR & \cite{Waters1996} \\
Alumina (Al$_2$O$_3$)                                        & IR & \cite{Onaka1989} \\
                                                                              & PM & \cite{Nittler1994} \\
                                                                              & & \cite{Hutcheon1994} \\
Mg$_{(1-x)}$Fe$_{x}$O with $x$=0.9                  & IR & \cite{Posch2002} \\
Spinel (MgAl$_2$O$_4$)                                    & IR $?$ & \cite{Posch1999} \\
                                                                            & PM & \cite{Nittler1997} \\
Hibonite (CaAl$_{12}$O$_{19}$)                        & PM & \cite{Choi1999} \\
Gehlenite (Ca$_2$Al$_2$SiO$_7$)                    & IR $?$ & \cite{Heras2005} \\
Fe                                                                        & IR $?$ & \cite{Kemper2002} \\
\hline
\multicolumn{3}{c}{C stars} \\
\hline
Amorphous carbon & IR & \cite{Martin1987} \\
Graphite                 & PM & \cite{Amari1990} \\
SiC                          & IR & \cite{Treffers1974} \\
                                & PM & \cite{Bernatowicz1987} \\
MgS                        & IR & \cite{Goebel1985} \\
TiC                          & PM & \cite{Bernatowicz1991} \\
\hline
\multicolumn{3}{c}{S stars} \\
\hline
MgS                                                  & IR & \cite{Hony2009} \\
Amorphous silicate                           & IR $?$ & \cite{Hony2009} \\
Alumina (Al$_2$O$_3$)                   & IR $?$ & \cite{Smolders2012} \\Diopside (MgCaSi$_2$O$_6$)        & IR $?$ & \cite{Hony2009} \\
Gehlenite (Ca$_2$Al$_2$SiO$_7$) & IR $?$ & \cite{Smolders2012} \\
\hline
\end{tabular}
\tablenotea{\\
Notes: IR indicates identification through infrared observations, while PM stands for identification in presolar material from meteorites. A question mark indicates that the identification is not completely secure.
}
\end{table}

The dust in oxygen-rich AGB envelopes is mainly composed of silicates and oxides. Amorphous silicate is widely observed through the 9.7 $\mu$m band, and crystalline silicates of the families of olivine (Mg$_{(2-2x)}$Fe$_{2x}$SiO$_4$) and pyroxene (Mg$_{(1-x)}$Fe$_{x}$SiO$_3$) have also been identified through narrow bands at mid- and far-infrared wavelengths \citep{Waters1996,Blommaert2014}. Alumina (Al$_2$O$_3$), a highly refractory condensate, is also observed at infrared wavelengths \citep{Onaka1989} and has also been identified in presolar grains \citep{Nittler1994,Hutcheon1994}. There is also evidence of Mg-Fe oxides of the type Mg$_{1-x}$Fe$_x$O, with a high content of Fe \citep{Posch2002}. Another condensate that is highly refractory is hibonite (CaAl$_{12}$O$_{19}$), which has been identified in presolar grains \citep{Choi1999}. Spinel (MgAl$_2$O$_4$) has been proposed as a constituent of dust in O-rich envelopes \citep{Posch1999}, although the identification has been questioned \citep{DePew2006,Zeidler2013}. Further evidence for the presence of spinel comes from the analysis of presolar grains in meteorites \citep{Nittler1997}. \cite{Heras2005} found evidence of gehlenite (Ca$_2$Al$_2$SiO$_7$) based on modeling the spectral energy distribution of 28 O-rich AGB stars in the 2.4-45.2 $\mu$m range. The presence of metallic iron grains has been also inferred in the oxygen-rich envelope OH\,127.8+0.0 \citep{Kemper2002}. However, the identification is particularly uncertain because Fe lacks spectral features and it is only recognized by an excess of opacity in 3-8 $\mu$m range.

In carbon-rich AGB envelopes, dust is mostly composed of carbon, either amorphous or in the form of graphite. These two materials do not have spectral features but provide a smooth continuum at infrared wavelengths. \cite{Martin1987} modeled the spectral energy distribution of the prototypical carbon star IRC\,+10216 and found that amorphous carbon, rather than graphite, is the main form of carbonaceous dust in C-rich envelopes. Graphite must also be present to some extent because it has been identified in presolar meteoritic material, with isotopic ratios pointing to formation in the outflows of C-rich AGB stars \citep{Amari1990}. Silicon carbide (SiC) dust is widely identified toward C stars through a band centered at 11.3 $\mu$m \citep{Treffers1974}, and presolar SiC grains have also been identified in carbonaceous meteorites \citep{Bernatowicz1987}. \cite{Goebel1985} proposed MgS as the carrier of a band observed at 30 $\mu$m. The assignment to MgS has been disputed by \cite{Zhang2009}, who argued that the amount of MgS required to reproduce the observed band strength implies a sulfur abundance higher than solar. This problem vanishes if MgS is only present in the outer layers of the grains \citep{Lombaert2012}, as originally proposed by \cite{Zhukovska2008}. Currently, MgS remains the best candidate for the 30 $\mu$m feature \citep{Sloan2014}. Finally, there is strong evidence of the presence of TiC in grains formed in C-rich AGB ejecta from the analysis of presolar material in meteorites \citep{Bernatowicz1991}.

\begin{figure*}
\centering
\includegraphics[angle=0,width=\textwidth]{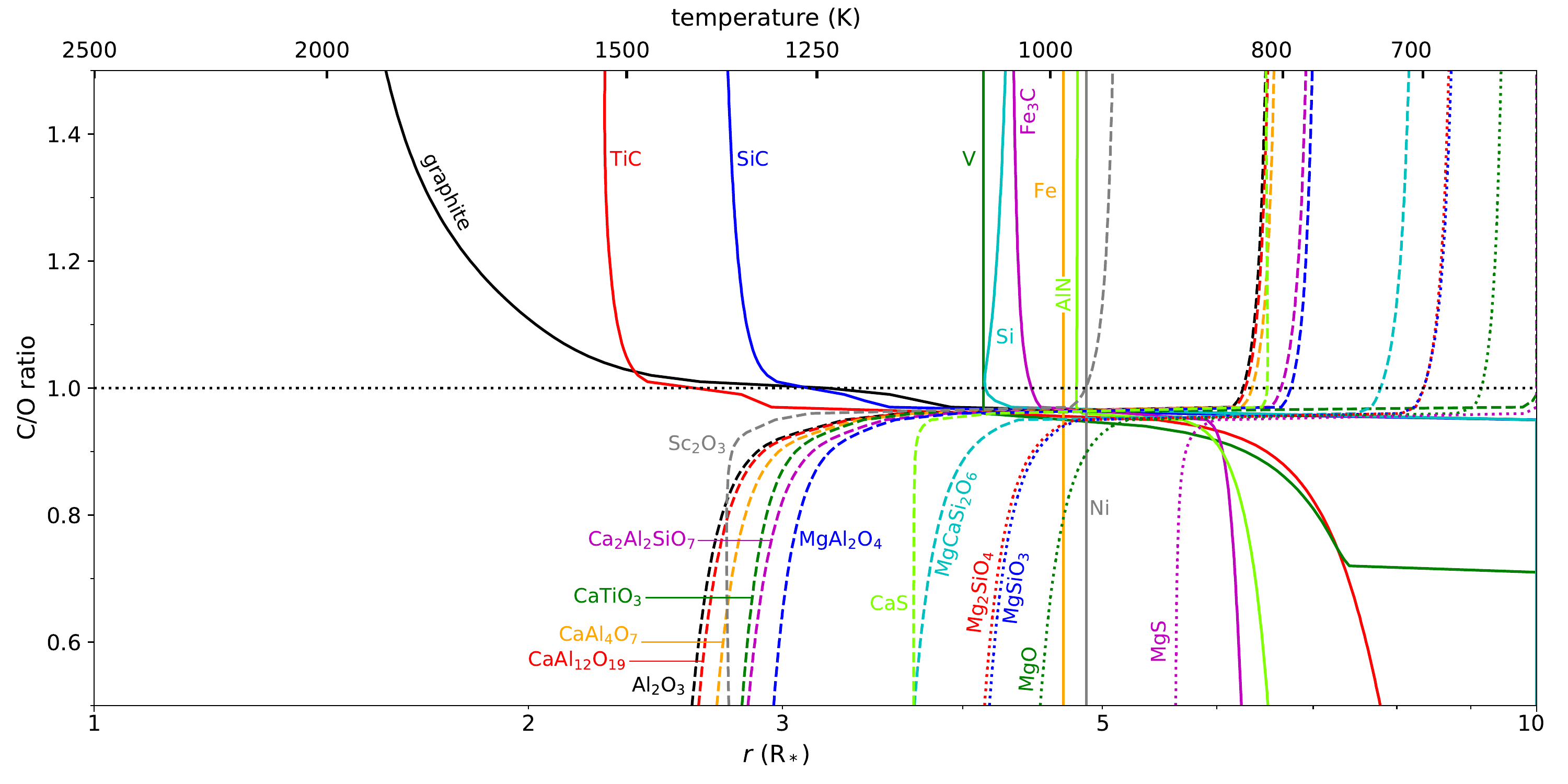}\caption{Radius (bottom $x$-axis) and temperature (top $x$-axis) at which some relevant condensates are predicted to appear in AGB atmospheres are shown as a function of the C/O ($y$-axis).} \label{fig:condensates_c_o_ratio}
\end{figure*}

The chemical composition of dust around S stars appears to contain features of both O-rich and C-rich stars. The study of \cite{Hony2009} revealed the 30 $\mu$m band, which is attributable to MgS, the amorphous silicate band, which appears shifted from 9.7 $\mu$m to redder wavelengths, however, and is proposed to be due to non-stoichiometric silicates and a series of emission bands in the 20-40 $\mu$m that were tentatively assigned to diopside (MgCaSi$_2$O$_6$). The presence of alumina (Al$_2$O$_3$) and gehlenite (Ca$_2$Al$_2$SiO$_7$) was also inferred by \cite{Smolders2012} by modeling the spectral energy distribution of a large sample of S stars.

\subsection{Expectations from chemical equilibrium}

Chemical equilibrium calculations can be very informative on the types of condensates that are expected to form in AGB atmospheres and on the sequence in which they are expected to appear \citep{Sharp1995,Lodders1997,Lodders1999,Gail2013}. Our main motivation to revisit the subject here is twofold. First, we aim to cross-check our calculations against previously published results. Second, we seek to establish a condensation sequence in M-, S-, and C-type atmospheres using a realistic pressure-temperature profile that serves us as starting point to discuss the most likely gas-phase precursors of selected condensates.

Here we present results from chemical equilibrium calculations in which condensates are considered. We collected thermochemical data for 185 condensed species. If all condensates are included simultaneously in the calculations, when multiple condensates having elements in common are thermodynamically favorable, only the most stable ones form at the expense of others that may never become abundant because their constituent elements are trapped by other more stable compounds. To circumvent this problem of competition between condensates with elements in common, calculations were run including only one condensed species each time. This allowed us to have a complete condensation sequence, without missing condensates, and to compare with previously published condensation sequences. For these calculations, we adopted the elemental composition given in Table~\ref{table:elements} and the pressure-temperature profile discussed in Sec.~\ref{sec:pt-profile}.

Carbonaceous dust in C-rich envelopes is expected to be mostly in the form of amorphous carbon \citep{Martin1987}. However, thermochemical data are not available for amorphous carbon, and we therefore used graphite as a proxy. Similarly, no thermochemical data for amorphous silicates or crystalline silicates with varying Mg/Fe, which are observed in O-rich envelopes \citep{Woolf1969,Waters1996}, are available. Therefore, forsterite (Mg$_2$SiO$_4$) and enstatite (MgSiO$_3$) were used in the chemical equilibrium calculations as proxies of olivine, pyroxene, and amorphous silicate. Similarly, we lack thermochemical data for the Mg-Fe oxide Mg$_{0.1}$Fe$_{0.9}$O identified in M stars. Nevertheless, the oxides MgO and FeO were included, and the latter was used as a proxy of Mg$_{0.1}$Fe$_{0.9}$O. We note, however, that there might be significant differences between the thermochemical properties of amorphous carbon and graphite, amorphous silicates and crystalline forsterite and enstatite, and Mg$_{0.1}$Fe$_{0.9}$O and FeO, which might lead to some changes in the condensation sequence calculated here for M-, S-, and C-type atmospheres.

\begin{figure*}
\centering
\includegraphics[angle=0,width=\textwidth]{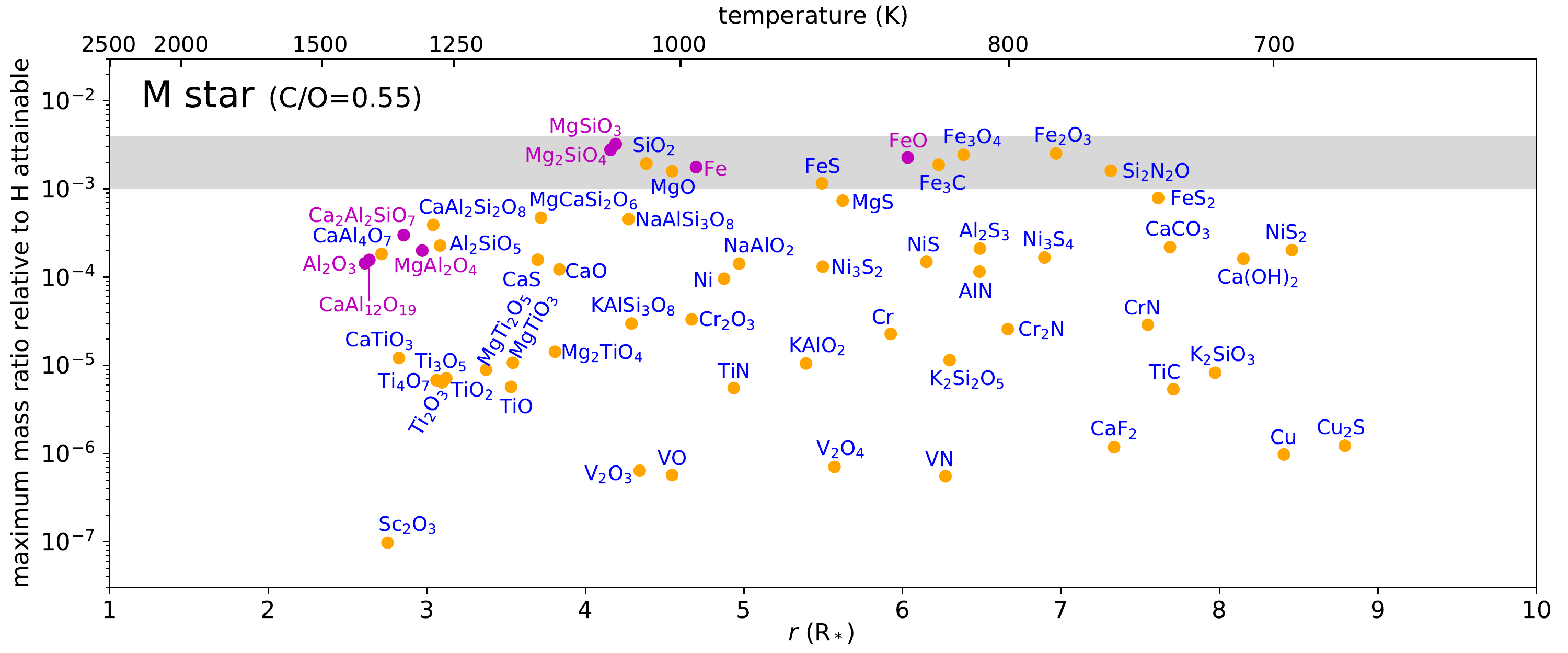} \includegraphics[angle=0,width=\textwidth]{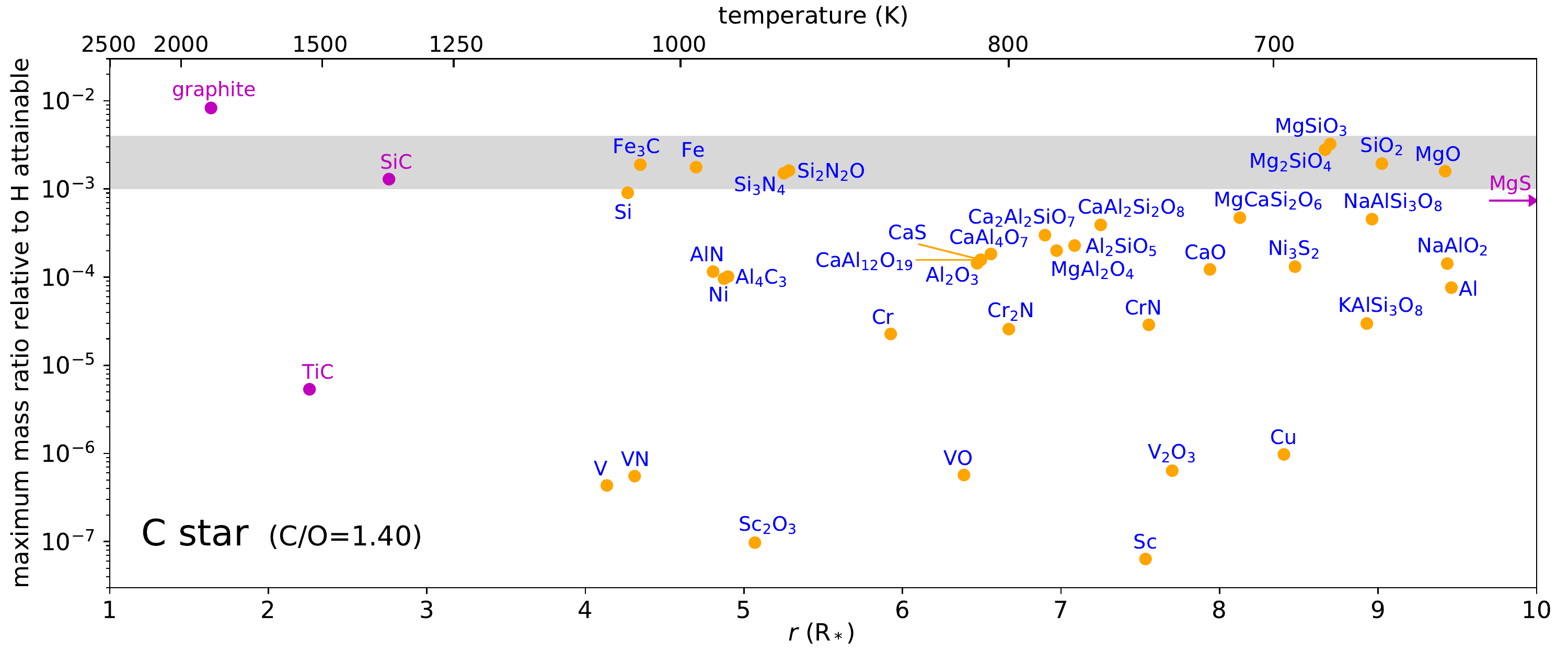}
\caption{All condensates predicted to appear in the 1-10 $R_*$ range in atmospheres of M- and C-type AGB stars (upper and lower panels, respectively). Condensates are located in the diagrams according to their condensation radius (bottom $x$-axis; the corresponding temperature is given in the top $x$-axis) and maximum mass ratio relative to H attainable. Condensates observed in AGB envelopes (see Table~\ref{table:condensates}) are indicated in magenta. The range of dust-to-gas mass ratios derived by \cite{Ramstedt2008} for envelopes of AGB stars is indicated by a gray horizontal band. In C-type AGB atmospheres, MgS is predicted to condense at temperatures below 628 K (see Table~\ref{table:condensates_element}), i.e., beyond 10 $R_*$.} \label{fig:abun_condensates}
\end{figure*}

Results from the chemical equilibrium calculations regarding condensates are shown in Fig.~\ref{fig:condensates_c_o_ratio} and Fig.~\ref{fig:abun_condensates}, where we show the radius (bottom $x$-axis) and temperature (top $x$-axis) at which each condensate appears. In Fig.~\ref{fig:condensates_c_o_ratio} the condensation radius (and temperature) of some relevant condensates is shown as a function of the C/O ($y$-axis). In Fig.~\ref{fig:abun_condensates} we show all the condensates that appear in the 1-10 $R_*$ range in M- and C-type atmospheres on an abundance scale ($y$-axis). The abundance scale is given by the maximum mass ratio relative to H that each condensate may attain, according to the abundances of its constituent elements (see Table~\ref{table:elements}). This is to be compared with the typical dust-to-gas mass ratios derived in AGB envelopes, in the range $(1-4)\times10^{-3}$ \citep{Ramstedt2008}, and is indicated by a gray horizontal band in Fig.~\ref{fig:abun_condensates}. This shows that condensates with maximum attainable mass ratios close to 10$^{-3}$ can be main constituents of dust, while those for which the maximum attainable mass ratio is substantially lower than 10$^{-3}$ can only be minor components of circumstellar dust. Still, minor condensates might be very important if they are predicted to be among the first condensates, in which case they can serve as condensation nuclei and accelerate the condensation of other compounds.

Chemical equilibrium predicts that the first condensates in carbon-rich atmospheres should be carbon, followed by TiC, and then SiC, while in oxygen-rich atmospheres Al$_2$O$_3$ should be the first condensate to appear, followed by minerals such as hibonite (CaAl$_{12}$O$_{19}$), grossite (CaAl$_4$O$_7$), scandia (Sc$_2$O$_3$), perovskite (CaTiO$_3$), gehlenite (Ca$_2$Al$_2$SiO$_7$), and spinel (MgAl$_2$O$_4$). However, these conclusions change for C-to-O ratios close to one (see Fig.~\ref{fig:condensates_c_o_ratio}). For example, the condensation sequence C-TiC-SiC changes to TiC-C-SiC for C/O below 1.02, and to TiC-SiC-C for even lower C/O, in the range 0.96-1.00. These values hold for the specific pressure-temperature profile adopted here, which yields pressures between a few 10$^{-10}$ and 10$^{-8}$ bar in the region where these compounds are expected to condense (see Fig.~\ref{fig:pt}). For higher pressures, the C-to-O ratios that separate the different condensation sequences shift to higher values. For example, at pressures of 10$^{-6}$-10$^{-5}$ bar, TiC condenses before carbon if C/O $<$ 1.1 \citep{Lodders1997}. Evidence of TiC serving as nucleation site for carbon dust has been found from the analysis of presolar grains \citep{Bernatowicz1991,Bernatowicz1996}, which implies formation at low C-to-O ratios and/or high pressures \citep{Lodders1997}. In oxygen-rich atmospheres with C/O in the range 0.82-0.96, Sc$_2$O$_3$, rather than Al$_2$O$_3$, is predicted to be the first condensate. Although Sc$_2$O$_3$ can only be a minor condensate because the elemental abundance of Sc is low (see the upper panel in Fig.~\ref{fig:abun_condensates}), it may provide the condensation nuclei for oxides and silicates. Therefore, depending on the C/O, either carbon, TiC, Sc$_2$O$_3$, or Al$_2$O$_3$ would be the first condensate according to chemical equilibrium. \cite{Lodders1999} reported that highly refractory condensates involving trace elements such as Hf and Zr are also likely to serve as condensation nuclei. Concretely, these authors mentioned HfO$_2$ and ZrO$_2$ in O-rich atmospheres and ZrC in C-rich stars. These condensates are not included in our calculations. Despite the low elemental abundance of Zr, there is evidence for ZrC in presolar grains \citep{Bernatowicz1996}. Therefore it would not be surprising if other highly refractory condensates involving trace elements such as Sc$_2$O$_3$ might also be identified in presolar grains.

It is worth noting that for slightly oxygen-rich conditions (C/O = 0.96-1.00), our calculations predict that condensates typical of carbon-rich conditions, such as TiC, SiC, and carbon, form well before minerals typical of oxygen-rich conditions, such as oxides and silicates (see Fig.~\ref{fig:condensates_c_o_ratio}). This behavior recalls that found previously for gas-phase molecules (see Sec.~\ref{sec:s_stars}). That is, for slightly oxygen-rich conditions, the chemical composition shares more features with a carbon-rich mixture than with an oxygen-rich one, and this applies to both gaseous species and condensates. This conclusion has direct consequences for S-type stars. For example, the predicted condensation sequence for an S-type atmosphere with C/O = 1 resembles that of a carbon-rich star far more than that of an M-type atmosphere (see Fig.~\ref{fig:condensates_c_o_ratio}).

Our calculations indicate that dust is expected to form very close to the star, at 1-3 $R_*$. Near-infrared polarimetric interferometric observations of R\,Dor and W\,Hya found that dust is already present as close to the star as 1.3-1.5 $R_*$ \citep{Khouri2016,Ohnaka2017}. The calculations also predict that carbon dust in C-rich atmospheres forms closer to the star than alumina dust in O-rich stars. Similarly, in S-type stars, condensation is shifted to slightly larger radii than in O- and C-rich stars.

\begin{table}
\caption{Condensates of each element in M and C stars.} \label{table:condensates_element}
\centering
\small
\begin{tabular}{ll@{\hspace{0.3cm}}rc@{\hspace{0.3cm}}l@{\hspace{0.3cm}}r}
\hline \hline
& \multicolumn{2}{c}{M stars} & & \multicolumn{2}{c}{C stars} \\
\cline{2-3}
\cline{5-6}
\multicolumn{1}{l}{Element$^a$} & \multicolumn{1}{l}{Condensate \hfill $f_{\rm el}$$^b$} & \multicolumn{1}{r}{$T_{\rm c}$$^c$} & & \multicolumn{1}{l}{Condensate \hfill $f_{\rm el}$$^b$} & \multicolumn{1}{r}{$T_{\rm c}$$^c$} \\
\multicolumn{1}{l}{} & \multicolumn{1}{l}{\hfill (\%)} & \multicolumn{1}{r}{(K)} & & \multicolumn{1}{l}{\hfill (\%)} & \multicolumn{1}{r}{(K)} \\
\hline
Mg \hfill 7.60 & \textcolor{magenta}{MgAl$_2$O$_4$} \hfill 4 & 1301 & & MgAl$_2$O$_4$ \hfill 4 & 780 \\
                     & MgCaSi$_2$O$_6$ \hfill 5 & 1137 & & MgCaSi$_2$O$_6$ \hfill 5 & 711 \\
                     & \textcolor{magenta}{Mg$_2$SiO$_4$} \hfill 100 & 1063 & & Mg$_2$SiO$_4$ \hfill 100 & 684 \\
                    & \textcolor{magenta}{MgSiO$_3$} \hfill 81 & 1058 & & MgSiO$_3$ \hfill 81 & 683 \\
                    & MgO \hfill 100 & 1008 & & MgO \hfill 100 & 651 \\
                    & MgS \hfill 33 & 887 & & \textcolor{magenta}{MgS} \hfill 33 & $<$ 628 \\
\hline
Si \hfill 7.51 & \textcolor{magenta}{Ca$_2$Al$_2$SiO$_7$} \hfill 3 & 1332 & & \textcolor{magenta}{SiC} \hfill 100 & 1359 \\
                   & CaAl$_2$Si$_2$O$_8$ \hfill 14 & 1283 & & Si \hfill 100 & 1047 \\
                   & Al$_2$SiO$_5$ \hfill 4 & 1272 & & Si$_3$N$_4$ \hfill 100 & 924 \\
                   & MgCaSi$_2$O$_6$ \hfill 14 & 1137 & & Si$_2$N$_2$O \hfill 100 & 921 \\
                  & \textcolor{magenta}{Mg$_2$SiO$_4$} \hfill 62 & 1063 & & Ca$_2$Al$_2$SiO$_7$ \hfill 3 & 785 \\
                  & \textcolor{magenta}{MgSiO$_3$} \hfill 100 & 1058 & & Al$_2$SiO$_5$ \hfill 4 & 772 \\
                  & NaAlSi$_3$O$_8$ \hfill 16 & 1046 & & CaAl$_2$Si$_2$O$_8$ \hfill 14 & 762 \\
                  & SiO$_2$ \hfill 100 & 1030 & & MgCaSi$_2$O$_6$ \hfill 14 & 711 \\
                  & Si$_2$N$_2$O \hfill 100 & 758 & & Mg$_2$SiO$_4$ \hfill 62 & 684 \\
                  & & & & MgSiO$_3$ \hfill 100 & 683 \\
                  & & & & NaAlSi$_3$O$_8$ \hfill 16 & 671 \\
\hline
Fe \hfill 7.50 & \textcolor{magenta}{Fe} \hfill 100 & 988 & & Fe$_3$C \hfill 100 & 1035 \\
                     & FeS \hfill 42 & 900 & & Fe \hfill 100 & 988 \\
                     & \textcolor{magenta}{FeO} \hfill 100 & 850 & & & \\                     & Fe$_3$C \hfill 100 & 834 & & & \\
                     & Fe$_3$O$_4$ \hfill 100 & 822 & & & \\
                     & Fe$_2$O$_3$ \hfill 100 & 780 & & & \\
                     & FeS$_2$ \hfill 21 & 740 & & & \\
\hline
S \hfill 7.12  & CaS \hfill 17 & 1140 & & CaS \hfill 17 & 814 \\
                   & FeS \hfill 100 & 900 & & Ni$_3$S$_2$ \hfill 8 & 693 \\
                   & Ni$_3$S$_2$ \hfill 8 & 899 & & \textcolor{magenta}{MgS} \hfill 100 & $<$ 628 \\
                   & MgS \hfill 100 & 887 & & &  \\
                  & NiS \hfill 13 & 841 & & & \\
                  & Al$_2$S$_3$ \hfill 32 & 814 & & & \\
                  & Ni$_3$S$_4$ \hfill 17 & 785 & & & \\
                  & FeS$_2$ \hfill 100 & 740 & & & \\
                  & NiS$_2$ \hfill 25 & 694 & & & \\
\hline
Al \hfill 6.45  & \textcolor{magenta}{Al$_2$O$_3$} \hfill 100 & 1406 & & AlN \hfill 100 & 975 \\
                    & \textcolor{magenta}{CaAl$_{12}$O$_{19}$} \hfill 100 & 1397 & & Al$_4$C$_3$ \hfill 100 & 964 \\
                    & CaAl$_4$O$_7$ \hfill 100 & 1373 & & Al$_2$O$_3$ \hfill 100 & 815 \\
                    & \textcolor{magenta}{Ca$_2$Al$_2$SiO$_7$} \hfill 78 & 1332 & & CaAl$_{12}$O$_{19}$ \hfill 100 & 814 \\
                    & \textcolor{magenta}{MgAl$_2$O$_4$} \hfill 100 & 1301 & & CaAl$_4$O$_7$ \hfill 100 & 809 \\
                   & CaAl$_2$Si$_2$O$_8$ \hfill 100 & 1283 & & Ca$_2$Al$_2$SiO$_7$ \hfill 78 & 785 \\
                   & Al$_2$SiO$_5$ \hfill 100 & 1272 & & MgAl$_2$O$_4$ \hfill 100 & 780 \\
                   & NaAlSi$_3$O$_8$ \hfill 62 & 1046 & & Al$_2$SiO$_5$ \hfill 100 & 772 \\
                   & KAlSi$_3$O$_8$ \hfill 4 & 1043 & & CaAl$_2$Si$_2$O$_8$ \hfill 100 & 762 \\
                   & NaAlO$_2$ \hfill 62 & 955 & & KAlSi$_3$O$_8$ \hfill 4 & 672 \\
                   & KAlO$_2$ \hfill 4 & 910 & & NaAlSi$_3$O$_8$ \hfill 62 & 671 \\
                   & AlN \hfill 100 & 814 & & NaAlO$_2$ \hfill 62 & 650 \\
                   & Al$_2$S$_3$ \hfill 100 & 814 & & Al \hfill 100 & 649 \\
\hline
Ca \hfill 6.34 & \textcolor{magenta}{CaAl$_{12}$O$_{19}$} \hfill 11 & 1397 & & CaS \hfill 100 & 814 \\
                     & CaAl$_4$O$_7$ \hfill 32 & 1373 & & CaAl$_{12}$O$_{19}$ \hfill 11 & 814 \\
                     & CaTiO$_3$ \hfill 4 & 1341 & & CaAl$_4$O$_7$ \hfill 32 & 809 \\
                     & \textcolor{magenta}{Ca$_2$Al$_2$SiO$_7$} \hfill 100 & 1332 & & Ca$_2$Al$_2$SiO$_7$ \hfill 100 & 785 \\
                     & CaAl$_2$Si$_2$O$_8$ \hfill 64 & 1283 & & CaAl$_2$Si$_2$O$_8$ \hfill 64 & 762 \\
                     & CaS \hfill 100 & 1140 & & CaO \hfill 100 & 721 \\
                     & MgCaSi$_2$O$_6$ \hfill 100 & 1137 & & MgCaSi$_2$O$_6$ \hfill 100 & 711 \\
                     & CaO \hfill 100 & 1116 & & & \\
                     & CaCO$_3$ \hfill 100 & 735 & & & \\
                     & Ca(OH)$_2$\hfill 100 & 710 & & & \\
\hline
\end{tabular}
\end{table}

\setcounter{table}{4}
\begin{table}
\caption{Continued.}
\centering
\small
\begin{tabular}{ll@{\hspace{0.3cm}}rc@{\hspace{0.3cm}}l@{\hspace{0.3cm}}r}
\hline \hline
& \multicolumn{2}{c}{M stars} & & \multicolumn{2}{c}{C stars} \\
\cline{2-3}
\cline{5-6}
\multicolumn{1}{l}{Element$^a$} & \multicolumn{1}{l}{Condensate \hfill $f_{\rm el}$$^b$} & \multicolumn{1}{r}{$T_{\rm c}$$^c$} & & \multicolumn{1}{l}{Condensate \hfill $f_{\rm el}$$^b$} & \multicolumn{1}{r}{$T_{\rm c}$$^c$} \\
\multicolumn{1}{l}{} & \multicolumn{1}{l}{\hfill (\%)} & \multicolumn{1}{r}{(K)} & & \multicolumn{1}{l}{\hfill (\%)} & \multicolumn{1}{r}{(K)} \\
\hline
Na \hfill 6.24 & NaAlSi$_3$O$_8$ \hfill 100 & 1046 & & NaAlSi$_3$O$_8$ \hfill 100 & 671 \\
                     & NaAlO$_2$ \hfill 100 & 955 & & NaAlO$_2$ \hfill 100 & 650 \\
\hline
Ni \hfill 6.22 & Ni \hfill 100 & 966 & & Ni \hfill 100 & 966 \\
                     & Ni$_3$S$_2$ \hfill 100 & 899 & & Ni$_3$S$_2$ \hfill 100 & 693 \\
                     & NiS$_2$ \hfill 100 & 694 & & & \\
\hline
Cr \hfill 5.64 & Cr$_2$O$_3$ \hfill 100 & 992 & & Cr \hfill 100 & 860 \\
                     & Cr \hfill 100 & 860 & & Cr$_2$N \hfill 100 & 801 \\
                     & Cr$_2$N \hfill 100 & 801 & & CrN \hfill 100 & 743 \\
                     & CrN \hfill 100 & 743 & & & \\
\hline
K \hfill 5.03   & KAlSi$_3$O$_8$ \hfill 100 & 1043 & & KAlSi$_3$O$_8$ \hfill 100 & 672 \\
                     & KAlO$_2$ \hfill 100 & 910 & & & \\
                     & K$_2$Si$_2$O$_5$ \hfill 100 & 829 & & & \\
                     & K$_2$SiO$_3$ \hfill 100 & 719 & & & \\
\hline
Ti \hfill 4.95   & CaTiO$_3$ \hfill 100 & 1341 & & \textcolor{magenta}{TiC} \hfill 100 & 1539 \\
                     & Ti$_4$O$_7$ \hfill 100 & 1278 & & & \\
                     & Ti$_3$O$_5$ \hfill 100 & 1273 & & & \\
                     & Ti$_2$O$_3$ \hfill 100 & 1269 & & & \\
                     & TiO$_2$ \hfill 100 & 1263 & & & \\
                     & MgTi$_2$O$_5$ \hfill 100 & 1205 & & & \\
                     & TiO \hfill 100 & 1173 & & & \\
                     & MgTiO$_3$ \hfill 100 & 1170 & & & \\
                     & Mg$_2$TiO$_4$ \hfill 100 & 1121 & & & \\
                     & TiN \hfill 100 & 959 & & & \\
                     & TiC \hfill 100 & 734 & & & \\
\hline
Cu \hfill 4.19 & Cu \hfill 100 & 697 & & Cu \hfill 100 & 697 \\
                     & Cu$_2$S \hfill 100 & 679 & & & \\
\hline
V \hfill 3.93   & V$_2$O$_3$ \hfill 100 & 1036 & & V \hfill 100 & 1067 \\                     & VO \hfill 100 & 1008 & & VN \hfill 100 & 1041 \\
                     & V$_2$O$_4$ \hfill 100 & 892 & & VO \hfill 100 & 822 \\
                     & VN \hfill 100 & 831 & & V$_2$O$_3$ \hfill 100 & 735 \\
\hline
Sc \hfill 3.15  & Sc$_2$O$_3$ \hfill 100 & 1362 & & Sc$_2$O$_3$ \hfill 100 & 944 \\
                     & & & & Sc \hfill 100 & 744 \\
\hline
\end{tabular}
\tablenotea{\\
Notes:\\
Condensates observed in AGB stars (see Table~\ref{table:condensates}) are highlighted in magenta. \\
$^a$ Element and abundance $\log \epsilon$, defined as $\log \epsilon$(X) = 12 + $\log$(X/H). \\
$^b$ Maximum fraction of elements that can be trapped by condensate. \\
$^c$ Condensation temperature. \\
}
\end{table}

The limited observational constraints on the composition of dust in AGB envelopes is roughly consistent with the expectations from chemical equilibrium. In the case of M-type stars (see Table~\ref{table:condensates} and the upper panel in Fig.~\ref{fig:abun_condensates}), observations indicate that the bulk of grains is composed of Mg-Fe silicates, which is consistent with the fact that MgSiO$_3$ and Mg$_2$SiO$_4$ are the first condensates of those that can attain dust-to-gas mass ratios higher than 10$^{-3}$. The detection of Al$_2$O$_3$, MgAl$_2$O$_4$, and CaAl$_{12}$O$_{19}$ and the possible presence of Ca$_2$Al$_2$SiO$_7$ are also in line with expectations from chemical equilibrium. These are among the first condensates predicted to appear, and all of them may attain moderately high dust-to-gas mass ratios higher than 10$^{-4}$. In C-type stars (see Table~\ref{table:condensates} and the lower panel in Fig.~\ref{fig:abun_condensates}), amorphous carbon and SiC are the main constituents of grains according to observations, and this is also in line with the predictions from chemical equilibrium, which clearly indicate that these two compounds are the first main condensates. In the case of S-type stars, observations seem to indicate a chemical composition of dust that is more similar to that of O-rich stars, with the exception of MgS (see Table~\ref{table:condensates}), while chemical equilibrium favors a dust composition that is more carbon-rich-like (see Fig.~\ref{fig:condensates_c_o_ratio}). We note, however, that the comparison between observations and chemical equilibrium for S stars is difficult because on the one hand, observational constraints are more uncertain, and on the other, the predicted condensation sequence is extremely sensitive to the exact C/O.

Table~\ref{table:condensates_element} presents similar information to that shown in Fig.~\ref{fig:abun_condensates}, but in a different manner. We list in order of appearance the condensates that are expected to trap each refractory element in M- and C-type atmospheres. The table also lists the maximum fraction of the element that each condensate can trap and the corresponding condensation temperature. In the main, the solid reservoirs of each element identified in M- and C-type atmospheres are similar to those presented by \cite{Lodders1999} in their Table~1, although there are some differences, which probably arise from differences in the thermochemical database of condensates and in the pressures and C-to-O ratios involved in the calculations. We now discuss each element individually, guided by Table~\ref{table:condensates_element}.

\emph{Magnesium.} The first Mg condensate expected in M-type atmospheres is spinel (MgAl$_2$O$_4$), for which there is evidence from infrared observations \citep{Posch1999} and analysis of presolar grains \citep{Nittler1997}. Spinel cannot be a main reservoir of Mg, however, because  it can only trap a small fraction of Mg (up to 4 \%) because the abundance of Al is lower. The next Mg condensate predicted is diopside (MgCaSi$_2$O$_6$), which has tentatively been identified in S-type stars \citep{Hony2009}, but not in O-rich envelopes. The presence of diopside in M-type envelopes, which would only trap as much as 5 \% of Mg, depends on whether some Ca is left after condensation of more refractory Ca compounds, such as CaTiO$_3$, Ca$_2$Al$_2$SiO$_7$, and CaAl$_2$Si$_2$O$_8$. Next in the condensation sequence of Mg, we have forsterite (Mg$_2$SiO$_4$) and enstatite (MgSiO$_3$), the Mg-rich end members of the olivine and pyroxene families, Mg$_{(2-2x)}$Fe$_{2x}$SiO$_4$ and Mg$_{(1-x)}$Fe$_{x}$SiO$_3$, respectively, which are expected to be main reservoirs of Mg, in agreement with observations \citep{Molster2002}. There is evidence of Mg in the form of Mg$_{0.1}$Fe$_{0.9}$O \citep{Posch2002}, which according to the condensation temperatures of MgO and FeO, should form from the Mg that is left after condensation of Mg-rich silicates. MgS is predicted to condense at even farther distances than MgO, and therefore little Mg is expected to be available to form it. In C-rich atmospheres, MgS is predicted to form at even larger distances from the star, beyond 10 $R_*$ (at temperatures below 628 K) for our radial pressure and temperature profiles. The lower condensation temperature of MgS in C-rich atmospheres, compared to O-rich ones, is related to the different main gaseous reservoirs of sulfur in these two types of sources (SiS in C-rich atmospheres and H$_2$S in O-rich ones), which compete differently with solid MgS for the sulfur. In spite of the large condensation radius of MgS in C-rich sources, this is the only Mg condensate identified so far in C-rich ejecta \citep{Goebel1985}. Several O-containing Mg condensates are predicted to appear earlier than MgS in C-rich atmospheres (see Table~\ref{table:condensates_element}). Their formation must therefore be inhibited either by the difficulty of competing for the oxygen, or because more refractory compounds would have trapped most of the Si, Al, and Ca. The formation of MgS at large distances from the AGB star, although somewhat surprising, is consistent with its presence in the outer layers of preexisting grains \citep{Zhukovska2008,Lombaert2012}. There is evidence that CS act as gas-phase precursor of MgS dust in high mass-loss rate C-rich envelopes, as indicated by the decrease in its fractional abundance with increasing envelope density and with increasing flux of the 30 $\mu$m feature attributed to MgS dust \citep{Massalkhi2019}. However, in some C-rich envelopes, gaseous CS and SiS trap most of the sulfur \citep{Danilovich2018,Massalkhi2019}, so that little would be left to form MgS dust.

\emph{Silicon.} In O-rich ejecta, silicon is predicted to condense first in the form of various Ca- and Al-containing silicates: gehlenite (Ca$_2$Al$_2$SiO$_7$), anorthite (CaAl$_2$Si$_2$O$_8$), andalusite (Al$_2$SiO$_5$), and diopside (MgCaSi$_2$O$_6$), although these can only take a small fraction of the silicon. Among them, there is only a tentative identification of the first expected condensate, gehlenite \citep{Heras2005}, but no observational indication of the others. Clearly, the first main Si condensates are forsterite (Mg$_2$SiO$_4$) and enstatite (MgSiO$_3$). Silica (SiO$_2$) might also be an important reservoir of Si because it is predicted to condense at only slightly lower temperatures than forsterite and enstatite. In C-rich atmospheres, the first and main Si condensate is clearly SiC, for which there is evidence from both infrared observations and analysis of presolar grains \citep{Treffers1974,Bernatowicz1987}. Other main Si condensates are pure Si, Si$_3$N$_4$, and Si$_2$N$_2$O, although they have much lower condensation temperatures.

\emph{Iron .} Metallic iron is predicted to condense at a temperature of 988 K independently of the C/O. In M-type atmospheres, this would be the first and main solid reservoir of the element, which is consistent with the inference of Fe grains from the modeling of the spectral energy distribution of the O-rich star OH\,127.8+0.0 \citep{Kemper2002}. Metallic iron is expected to condense after silicates. Other Fe condensates such as sulfides, oxides, and carbides (see Table~\ref{table:condensates_element}) can form later from the Fe that is left after condensation of pure iron. The detection of Mg$_{0.1}$Fe$_{0.9}$O \citep{Posch2002} is an indication of this. A potentially important reservoir of iron is troilite (FeS), which is predicted to condense at slightly higher temperatures than FeO. In C-rich envelopes, most iron is expected to be in the form of Fe$_3$C because this carbide is predicted to condense earlier than pure Fe. \cite{Lodders1999} found that (Fe,Ni)$_3$P and FeSi might also be important Fe condensates in O-rich and C-rich, respectively, atmospheres.

\emph{Sulfur.} The first S-containing condensate expected in both M- and C-type atmospheres is CaS. This compound, which would condense at a significantly higher temperature in O-rich atmospheres than in C-rich conditions, can take up to 17\% of the sulfur. Depending on the degree of depletion of this element from the gas phase, other S-containing condensates may form, such as FeS and Ni$_3$S$_2$ in M-type atmospheres and Ni$_3$S$_2$ and MgS in C-rich atmospheres. The presence of FeS in O-rich ejecta depends on whether some Fe is left after the condensation of pure iron, while similarly, the formation of Ni$_3$S$_2$ is conditioned on the condensation of pure Ni. The observational evidence of MgS in C-rich envelopes, together with the fact that CaS is the first condensate involving either Ca or S in C-rich atmospheres, strongly indicates that CaS is a very likely constituent of dust.

\emph{Aluminium.} Alumina (Al$_2$O$_3$) is predicted to be the first and main Al condensate in O-rich atmospheres, which is in line with observational evidence from infrared observations \citep{Onaka1989} and from the analysis of presolar meteoritic material \citep{Nittler1994,Hutcheon1994}. Other main condensates that appear later in the condensation sequence of Al and that can trap part of the Al that is not used by alumina are hibonite (CaAl$_{12}$O$_{19}$), grossite (CaAl$_4$O$_7$), gehlenite (Ca$_2$Al$_2$SiO$_7$), spinel (MgAl$_2$O$_4$), anorthite (CaAl$_2$Si$_2$O$_8$), and andalusite (Al$_2$SiO$_5$). Observational evidence for the presence of some of these condensates has been reported, concretely, hibonite \citep{Choi1999}, gehlenite \citep{Heras2005}, and spinel \citep{Posch1999,Nittler1997}. In C-rich atmospheres, aluminum is predicted to condense at temperatures below 1000 K. Main condensates, in order of appearance, are AlN, Al$_4$C$_3$, and Al$_2$O$_3$. None of them have been observed so far.

\emph{Calcium.} The first Ca condensates expected in O-rich atmospheres are the calcium aluminum oxides hibonite and grossite (already mentioned when discussing aluminum), and perovskite (CaTiO$_3$). These condensates can only trap a fraction of the calcium, however. In particular, the low abundance of CaTiO$_3$ may be behind the lack of observational evidence of this mineral in O-rich dust. Main Ca condensates that appear later in the condensation sequence of Ca are the calcium aluminum silicates gehlenite and anorthite. In C-rich atmospheres, calcium is expected to condense at relatively large distances from the AGB star ($>6.5$ $R_*$, corresponding to temperatures around or below 800 K). The first main Ca condensate is CaS, followed by the same calcium aluminum oxides and silicates expected in O-rich ejecta.

The condensates involving other refractory elements that are less abundant than calcium are given in Table~\ref{table:condensates_element}. For several elements, the first and main expected condensate is the same, regardless of the C/O. This is the case of Na and K, in which case albite (NaAlSi$_3$O$_8$) and orthoclase (KAlSi$_3$O$_8$) are the main expected condensates, of Ni and Cu, which are predicted to condense in pure metallic form, and of Sc, whose main condensate is Sc$_2$O$_3$. In the case of Cr and V, the oxides Cr$_2$O$_3$ and V$_2$O$_3$, respectively, are the main expected condensate in O-rich atmospheres, while in carbon-rich ejecta, these two elements are expected to condense in pure metallic form. Finally, for titanium, the main condensate in O-rich conditions is CaTiO$_3$, followed by several titanium oxides, while in C-rich atmospheres, TiC is clearly the main expected condensate, and this is supported by analysis of presolar grains \citep{Bernatowicz1991}.

\section{Gas-phase precursors of dust}

\begin{figure*}
\centering
\includegraphics[angle=0,width=\textwidth]{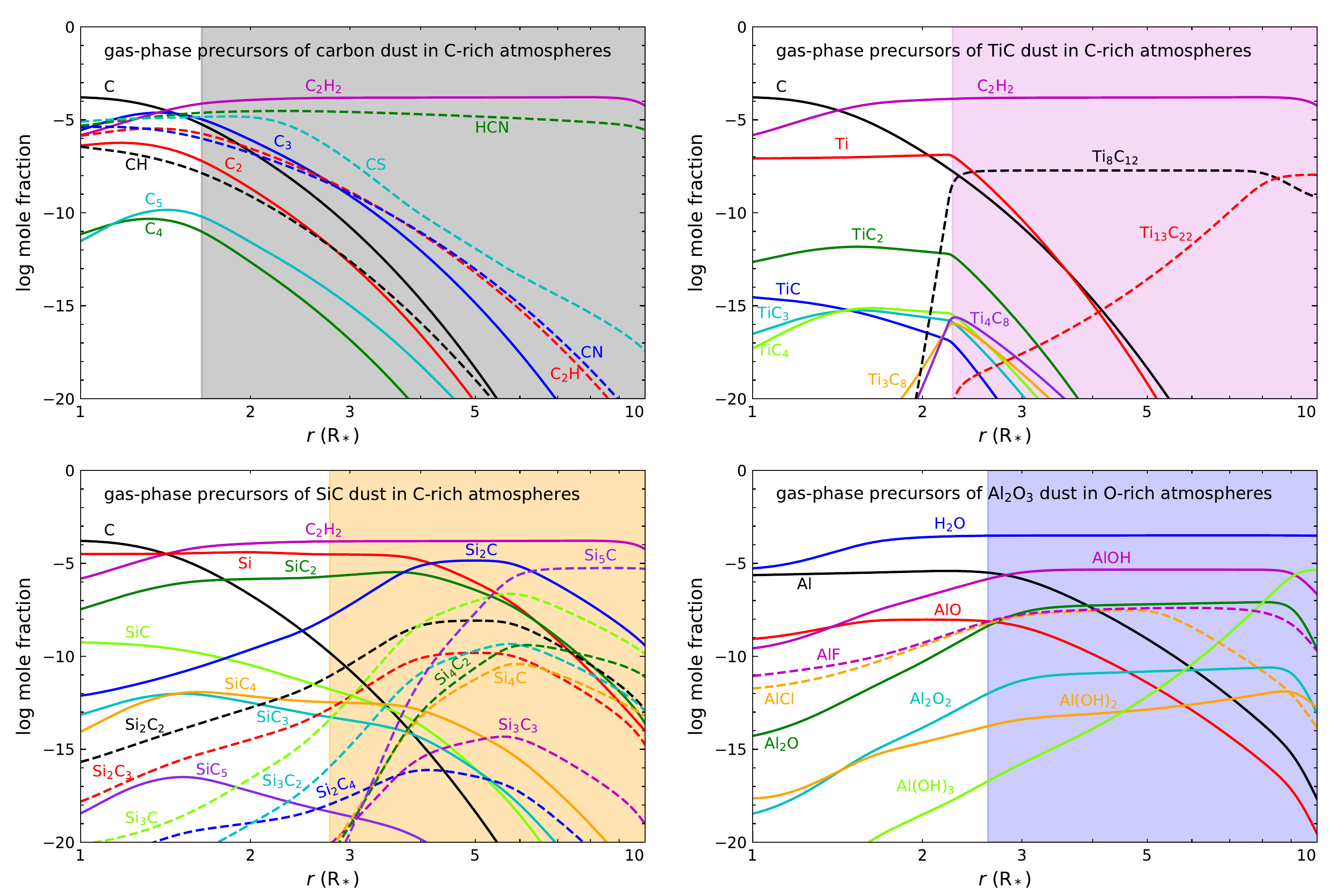}
\caption{Calculated chemical equilibrium abundances as a function of radius for gaseous reservoirs that can act as precursors of some of the first condensates expected in C- and O-rich AGB atmospheres: carbon, TiC, and SiC dust in C-rich stars (top left, top right, and bottom left, respectively), and Al$_2$O$_3$ dust in O-rich stars (bottom right). Abundances correspond to chemical equilibrium calculations in which only gaseous species are included. The shadowed areas indicate the regions where the condensation of the different types of dust is thermodynamically favorable: from 1.6 $R_*$ for graphite, 2.3 $R_*$ for TiC, 2.8 $R_*$ for SiC, and 2.6 $R_*$ for Al$_2$O$_3$.} \label{fig:precursors_dust}
\end{figure*}

Condensates can only form in regions in which the gas temperature and pressure make the appearance of solid compounds thermodynamically favorable. That is, condensates cannot appear earlier than predicted by chemical equilibrium, although they can appear later, depending on the kinetics of the process of condensation. The first condensates are predicted to form at a given distance from the star, and this process must necessarily occur at the expense of gas-phase atoms and small molecules. For our adopted radial profiles of pressure and temperature, the first condensates are expected to appear when temperatures are below 2000 K in the C-rich case and 1500 K in the O-rich case (see Fig.~\ref{fig:condensates_c_o_ratio}). Although condensation in the expanding and cooling outflow from AGB stars occurs in nonequilibrium conditions \citep{Gail2013}, chemical equilibrium can provide a useful starting point to examine the most likely gas-phase precursors of the first condensates.

Here we focus on the possible gas-phase precursors of the three condensates that are predicted to appear well before any other in C-rich outflows (carbon, TiC, and SiC) and of the first solid expected to condense in O-rich atmospheres: Al$_2$O$_3$. For each condensate, we examine the main gaseous reservoirs of the constituent elements, discuss the plausibility of the different reservoirs to act as precursors, and comment on the role that clusters of medium size may play in the formation of condensation nuclei.

\subsection{Carbon dust in C-rich atmospheres} \label{sec:carbon_dust}

Carbon dust is expected to be the first condensate in C-rich atmospheres, except for very low C/O. In the region where graphite is expected to form according to chemical equilibrium (1.6 $R_*$; see top left panel in Fig.~\ref{fig:precursors_dust}), the main reservoir of carbon is acetylene. Other abundant C-bearing molecules are HCN, CS, C$_3$, and atomic carbon. Because HCN and CS contain nitrogen and sulfur, respectively, they are less likely candidates to act as precursors of dust made up purely of carbon. Thus, chemical equilibrium indicates that C$_2$H$_2$, C$_3$, and C are the most likely precursors of carbon dust in C-type AGB atmospheres.

\cite{Gail1984} considered that C$_2$H$_2$ is the main source of carbon atoms in the synthesis of graphite in C-rich outflows, which would imply a depletion in the gas-phase abundance of acetylene in the region where carbon dust forms. This is in contrast with the study of \cite{Fonfria2008}, who modeled rovibrational lines of C$_2$H$_2$ in the carbon star IRC\,+10216 and found that acetylene maintains a constant abundance out to $\sim20$ $R_*$, a distance at which most carbon dust should have already formed. Atomic carbon is the main predicted reservoir of the element at the photosphere, although its abundance declines steeply with increasing radius. If, as predicted by chemical equilibrium, carbon dust forms inside 2 $R_*$, then atomic carbon may provide the necessary carbon to form dust. In a recent experiment designed to mimic the formation of carbon dust in evolved stars, atomic carbon was used as precursor, leading to the synthesis of amorphous carbon nanoparticles \citep{Martinez2019}.

If atomic carbon is the precursor of carbon dust, then the formation of the first condensation nuclei must proceed through clusters C$_n$ of increasing size. This is the theoretical scenario used by \cite{Gail1984} to describe the process of nucleation during the formation of graphite in C-rich atmospheres. Our calculations only include C$_n$ clusters up to C$_5$. The most abundant in the region around 1.6 $R_*$ of these is C$_3$, while C$_2$, C$_4$, and C$_5$ have lower abundances. It is unclear whether carbon clusters larger than C$_5$ might be stable enough to be present with a significant abundance. In any case, the large abundance of C$_3$ makes it a good candidate to act as precursor in the formation of carbon dust. Observational constraints of the radial variation of the C$_3$ abundance in the innermost regions of C-rich envelopes, currently restricted to the outer envelope \citep{Hinkle1988}, might shed light on this.

\subsection{Titanium carbide dust in C-rich atmospheres}

Titanium carbide is predicted to be the first condensate in the atmospheres of C-stars with very low C/O and in S-type atmospheres. The finding of TiC grains embedded in presolar graphitic material also indicates that TiC is a first condensate \citep{Bernatowicz1991}. Given the relative elemental abundances of titanium and carbon and the stoichiometry 1:1 of solid TiC, the formation of titanium carbide dust will be limited by the precursor providing Ti. Titanium carbide is expected to condense at a radius of 2.3 $R_*$ in our fiducial C-rich atmosphere. In this region, the main reservoir of titanium is atomic Ti (see the top right panel in Fig.~\ref{fig:precursors_dust}). \cite{Gail2013} suggested that atomic Ti reacting with C$_2$H$_2$ is the net reaction responsible for the formation of TiC dust. Acetylene is indeed the main reservoir of carbon at 2.3 $R_*$, although atomic carbon also has an abundance comparable to that of atomic Ti and might act as the carbon-supplier precursor.

Several gaseous molecules contain both titanium and carbon and might play a role in the formation of TiC dust. The most obvious is the diatomic molecule TiC, which is predicted to have a too low abundance, however. Therefore, a process of formation of TiC grains through a sequence of addition of TiC molecules to form clusters (TiC)$_n$ of increasing size $n$ is unlikely because of the low amount of Ti locked by TiC molecules. Of the small Ti-C molecules, TiC$_2$ is the most abundant, although it remains orders of magnitude below atomic Ti. Large Ti$_x$C$_y$ clusters become abundant at the expense of atomic Ti and small Ti-bearing molecules in the region where TiC dust condensation is expected. Although this coincidence may be accidental, it strongly suggests that the assembly of large Ti$_x$C$_y$ clusters might be related to the formation of TiC condensation nuclei. The most stable and abundant Ti$_x$C$_y$ clusters are Ti$_8$C$_{12}$ and Ti$_{13}$C$_{22}$, the former displacing atomic Ti as main reservoir of titanium at radii larger than 2.3 $R_*$. Therefore, Ti$_8$C$_{12}$ emerges as a very attractive candidate of a gas-phase precursor of TiC dust. It is clear that formation through addition of Ti$_8$C$_{12}$ monomers leading to (Ti$_8$C$_{12}$)$_n$ of increasing $n$ does not preserve the stoichiometry of solid TiC. If Ti$_8$C$_{12}$ acts as precursor, some rearrangement in which Ti atoms are incorporated or carbon atoms are lost is needed during the growth of TiC condensation nuclei.

The condensation of TiC in the outflows of carbon stars has been studied theoretically in conditions of nonequilibrium by \cite{Chigai1999}. These authors considered the growth of (TiC)$_n$ clusters of increasing size through chemical reactions involving atomic Ti and C$_2$H$_2$, following the formalism described by \cite{Gail1988} for the heteromolecular formation and growth of carbon grains. \cite{Chigai1999} found that formation of TiC cores covered by graphite mantles, in agreement with the constraints from presolar grains \citep{Bernatowicz1991}, is possible over certain ranges of mass-loss rate and gas outflow velocity. The study of \cite{Chigai1999}, however, did not investigate the detailed chemical pathways leading to the formation of TiC molecules and small (TiC)$_n$ clusters, mainly because the relevant reactions and rate constants are unknown.

\subsection{Silicon carbide dust in C-rich atmospheres}

Silicon carbide is expected to condense after carbon and TiC for C/O higher than 1.02 and after TiC for any C/O (see Fig.~\ref{fig:condensates_c_o_ratio}). This implies that SiC grains may nucleate heterogeneously, that is, on preexisting condensation nuclei of carbon and/or TiC. However, the analysis of a presolar SiC grain containing TiC crystals seems to indicate that SiC and TiC nucleated and grew independently \citep{Bernatowicz1992}, which implies that SiC can nucleate homogeneously. Silicon carbide is expected to condense at 2.8 $R_*$, and in this region, the main reservoir of silicon is atomic Si, while the main reservoir of carbon is C$_2$H$_2$ (see the bottom left panel in Fig.~\ref{fig:precursors_dust}). Therefore, these two species are candidates for gas-phase precursors of SiC dust, as suggested by \cite{Gail2013}. The role of acetylene in the formation of dust is in question, however, because of the lack of radial abundance decline inferred for IRC\,+10216 \citep{Fonfria2008}, as discussed in Sec.~\ref{sec:carbon_dust}.

Alternative candidates are molecules containing Si-C bonds, some of which are predicted to be abundant in C-rich atmospheres. Concretely, SiC$_2$, in the condensation region of SiC dust, Si$_2$C, slightly farther away, and Si$_5$C, at even larger radii, are predicted to be the most abundant carriers of Si-C bonds (see the bottom left panel in Fig.~\ref{fig:precursors_dust}). The molecules SiC$_2$ and Si$_2$C are indeed observed to be abundant in C-rich atmospheres \citep{Cernicharo2010,Fonfria2014,Cernicharo2015,Massalkhi2018}. Moreover, there is evidence that SiC$_2$ is a gas-phase precursor of SiC dust. On the one hand, \cite{Fonfria2014} inferred an abundance decline with increasing radius in the dust formation region of IRC\,+10216. On the other, \cite{Massalkhi2018} found that the abundance of SiC$_2$ in C-rich AGB envelopes decreases with increasing envelope density. Both observational facts indicate a depletion of gaseous SiC$_2$ to form SiC dust grains. The molecule Si$_2$C is as abundant as SiC$_2$ \citep{Cernicharo2015} and might also act as gas-phase precursor of SiC dust.

The formation of SiC dust in the outflows of C-rich AGB stars has been studied theoretically by \cite{Yasuda2012}. These authors presented chemical equilibrium abundances for Si$_x$C$_y$ species that agree with ours. This is expected because we used the same thermochemical properties for Si$_x$C$_y$ species as they did, that is, those of \cite{Deng2008}. \cite{Yasuda2012} investigated further the kinetics of formation of SiC dust in the framework of a dust-driven wind model considering a nucleation process consisting of addition of SiC and Si$_2$C$_2$ molecules to form (SiC)$_n$ clusters of increasing $n$. This clustering sequence  is also favored by the quantum chemical calculations of (SiC)$_n$ clusters by \cite{Gobrecht2017}. While small condensation nuclei may form at the expense of SiC and Si$_2$C$_2$, the abundances of these molecules is too low to provide the required amount of SiC dust. Observations derive dust-to-gas mass ratios of $(1-4)\times10^{-3}$ \citep{Ramstedt2008} and mass ratios between SiC and carbon dust of of 0.02-0.25 \citep{Groenewegen1998}, which results in a mass ratio between SiC dust and H$_2$ of $(0.2-10)\times10^{-4}$. The calculated mole fraction of Si$_2$C$_2$ is at most $10^{-8}$, which translates into a mass ratio relative to H$_2$ of $2\times10^{-7}$, at least two orders of magnitude below the observational value. The probability of the SiC molecule to act as main gas-phase precursor of SiC dust is even lower because the predicted abundance is low, which agrees with the abundance upper limit derived from observations \citep{Velilla-Prieto2015}. In summary, nucleation may occur by addition of SiC and Si$_2$C$_2$ molecules to (SiC)$_n$ clusters, although the mass of condensation nuclei grown by this process is limited by the low gas-phase abundance of SiC and Si$_2$C$_2$. The molecules SiC$_2$ and Si$_2$C thus emerge as the two most likely gas-phase precursors of SiC dust.

\subsection{Alumina dust in O-rich atmospheres}

Alumina is the first main condensate predicted to appear in O-rich atmospheres. Our chemical equilibrium calculations places its condensation radius at 2.6 $R_*$. In this region, the main carriers of aluminum are atomic Al and AlOH (see the bottom right panel in Fig.~\ref{fig:precursors_dust}). Other carriers of Al in the condensation region of alumina are AlO, Al$_2$O, AlF, and AlCl. From these, atomic Al and the molecules containing an Al-O bond arise as the most likely gas-phase precursors of Al$_2$O$_3$.

\cite{Gail2013} suggested that atomic Al reacting with water, the main carrier of oxygen other than CO, drive the condensation of Al$_2$O$_3$. \cite{Gobrecht2016} modeled the kinetics of formation of alumina dust in M-type atmospheres. In their chemical scheme, clusters (Al$_2$O$_3$)$_2$, the seed of condensation nuclei, form by three-body recombination of Al$_2$O$_3$ molecules. The formation of Al$_2$O$_3$ molecules relies on the oxidation, by reaction with H$_2$O, of Al$_2$O$_2$, which is formed by three-body recombination of AlO, which in turn is formed in the reaction of atomic Al with OH. In the scenario depicted by these authors, the starting reservoir of aluminum therefore is atomic Al, with AlO, Al$_2$O$_2$, and Al$_2$O$_3$ acting as intermediate species. In the more recent model by \cite{Boulangier2019}, in which the chemical kinetics scheme was revised with respect to \cite{Gobrecht2016}, Al$_2$O$_3$ molecules are not efficiently formed, which prevents the growth of large (Al$_2$O$_3$)$_n$ clusters. \cite{Boulangier2019} indicated that limitations in the chemical network for Al-bearing species are the reason for the low abundance of Al$_2$O$_3$.

From an observational point of view, two potential gas-phase precursors of alumina dust have been observed around M-type stars: AlO and AlOH \citep{Kaminski2016,Decin2017}. Reliable radial abundance distributions have not been derived, however, which makes it difficult to evaluate whether these molecules act as gas-phase precursors of alumina dust.

\section{Summary}

We investigated theoretically the chemical composition of AGB atmospheres of M-, S-, and C-type by means of chemical equilibrium calculations using a recently developed code. We compiled a large dataset of thermochemical properties for 919 gaseous and 185 condensed species involving 34 elements. We considered for the first time a large number of titanium-carbon clusters. Concretely, we computed thermochemical data for all Ti$_x$C$_y$ clusters with $x$ = 1-4 and $y$ = 1-4 and for various stable large clusters such as Ti$_3$C$_8$, Ti$_4$C$_8$, Ti$_6$C$_{13}$, Ti$_7$C$_{13}$, Ti$_8$C$_{12}$, Ti$_9$C$_{15}$, and Ti$_{13}$C$_{22}$. We studied the chemical composition in the 1-10 $R_*$ region of a generic AGB atmosphere by adopting realistic radial profiles of temperature and pressure based on constraints from recent observations and results from hydrodynamic models.

We compared the predictions of chemical equilibrium with the latest observational constraints. Chemical equilibrium reproduces the observed abundances of most of the parent molecules detected in AGB envelopes reasonably well. However, there are serious discrepancies between chemical equilibrium and observations for some parent molecules, which are observed with abundances several orders of magnitude above the expectations from chemical equilibrium. The concerned species are HCN, CS, NH$_3$, and SO$_2$ in M-type stars, H$_2$O and NH$_3$ in S-type stars, and the hydrides H$_2$O, NH$_3$, SiH$_4$, and PH$_3$ in C-type stars.

We systematically surveyed the budget of each element, examining the main reservoirs (see Appendix~\ref{app:budget}) and identifying several molecules that have not yet been observed in AGB atmospheres, but are predicted with non-negligible abundances. The most promising detectable molecules are SiC$_5$, SiNH, SiCl, PS, HBO, and the metal-containing molecules MgS, CaS, CaOH, CaCl, CaF, ScO, ZrO, VO, FeS, CoH, and NiS. For most of them, sensitive high-angular resolution observations with telescopes such as ALMA offer the best probabilities of detection.

We also investigated which condensates are predicted to appear and at which radius they are expected according to chemical equilibrium. In agreement with previous studies, we found that carbon, TiC, and SiC are the first condensates predicted to appear in C-rich outflows, while in O-rich atmospheres, Al$_2$O$_3$ is the first main expected condensate. Chemical equilibrium indicates that the most probable gas-phase precursors of carbonaceous dust are acetylene, atomic carbon, and/or C$_3$, while silicon carbide dust is most probably formed at the expense of the molecules SiC$_2$ and Si$_2$C. As concerns TiC dust, most titanium is atomic in the inner regions of AGB atmospheres and thus atomic Ti is a likely supplier of titanium during the formation of TiC dust. Interestingly, we found that according to chemical equilibrium, large titanium-carbon clusters such as Ti$_8$C$_{12}$ and Ti$_{13}$C$_{22}$ become the main reservoirs of titanium at the expense of atomic Ti in the region where TiC condensation is expected to occur. This strongly indicates that large Ti$_x$C$_y$ clusters are important intermediate species during the formation of the first condensation nuclei of TiC. Finally, in the case of Al$_2$O$_3$ dust, chemical equilibrium indicates that the main gas-phase precursor must be atomic Al or the molecules AlOH, AlO, and Al$_2$O, which are the main carriers of Al-O bonds in O-rich atmospheres.

\begin{acknowledgements}

We acknowledge funding support from the European Research Council (ERC Grant 610256: NANOCOSMOS) and from Spanish MINECO through grants AYA2016-75066-C2-1-P and MAT2017-85089-C2-1-R. M.A. and J.I.M also acknowledge funding support from the Ram\'on y Cajal programme of Spanish MINECO (grants RyC-2014-16277 and RyC-2015-17730, respectively). We thank the anonymous referee for a detailed report which helped to improve this manuscript. M.A. thanks Carlos Abia for an interesting discussion on elemental abundances and C/O, and Sara Bladh and Bernd Freytag for kindly providing radial profiles from their hydrodynamic models of AGB atmospheres. P.L.A. thanks Allan East for useful discussions.

\end{acknowledgements}

\begin{appendix}

\section{Element-by-element gas budget} \label{app:budget}

Here we review the main gas-phase reservoirs of each element in AGB atmospheres according to chemical equilibrium. The calculations include only gaseous species and use as input the elemental composition given in Table~\ref{table:elements} for AGB stars of M-, S-, and C-type and the pressure-temperature profile discussed in Sec.~\ref{sec:pt-profile}, which is taken as representative of AGB atmospheres. We do not discuss the noble gases He, Ne, and Ar, which are essentially present as neutral atoms, nor hydrogen, which is mostly present as H$_2$ with the exception of the hot inner atmosphere ($>$ 1700 K for our adopted pressure-temperature profile) where H is more abundant. The remaining elements we include comprise the nonmetals B, C, N, O, F, Si, P, S, and Cl, metals such as Al, the alkali metals Li, Na, K, and Rb, the alkali-earth metals Be, Mg, Ca, Sr, and Ba, and the transition metals Sc, Ti, V, Cr, Mn, Fe, Co, Ni, Cu, Zn, and Zr. 

The importance of the various types of reservoirs, either atomic, molecular, or in the form of condensates, is different for each element. Here we focus solely on the gas-phase budget and do not consider condensates. However, as discussed in Sec.~\ref{sec:condensates}, most elements, especially metals, tend to form thermodynamically stable condensates below a given temperature. We therefore recall that in regions in which the temperature has dropped below the relevant condensation temperature for each element, the abundance of the element in the gas phase must decrease, and all the abundances calculated here for the gaseous species should therefore be scaled down in these regions.

Nonmetals in the gas-phase budget in general tend to be in molecular form, although neutral atoms can also be a major reservoir in the hot inner atmosphere for many of them, such as C, Si, and B in carbon-rich stars, O and S in oxygen-rich stars, and P and Cl regardless of the C/O. For metals, the atomic reservoir tends to be more important than for nonmetals. Ionized atoms can be abundant in the hottest regions, and neutral atoms frequently largely dominate any molecular form throughout the entire extended atmosphere. We caution, however, that for some metals the number of molecules for which thermochemical data are available is small and therefore important molecular reservoirs of metals might be missed.

\subsection{Carbon}

\begin{figure*}
\centering
\includegraphics[angle=0,width=\textwidth]{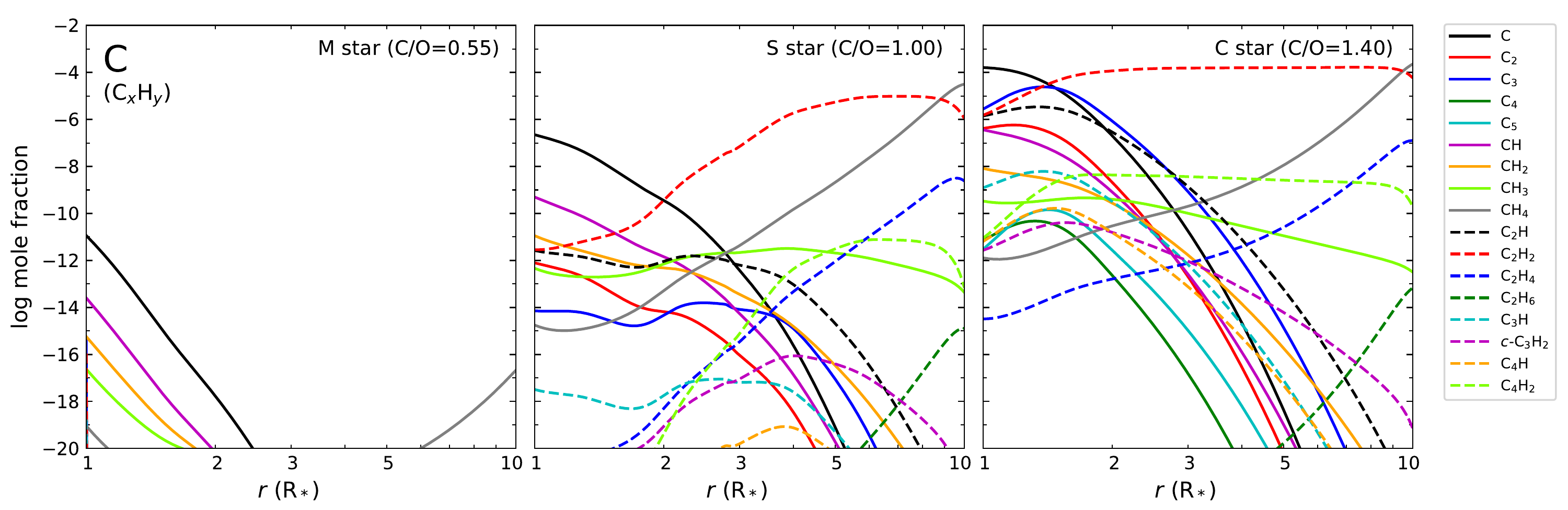} \includegraphics[angle=0,width=\textwidth]{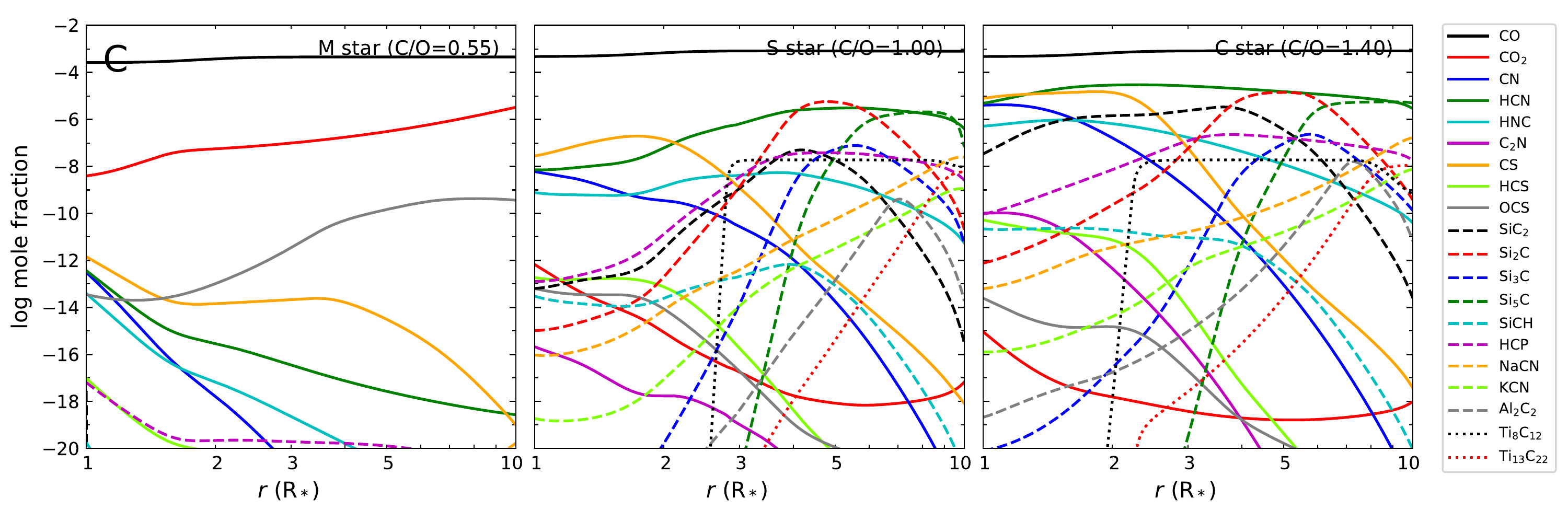} \includegraphics[angle=0,width=\textwidth]{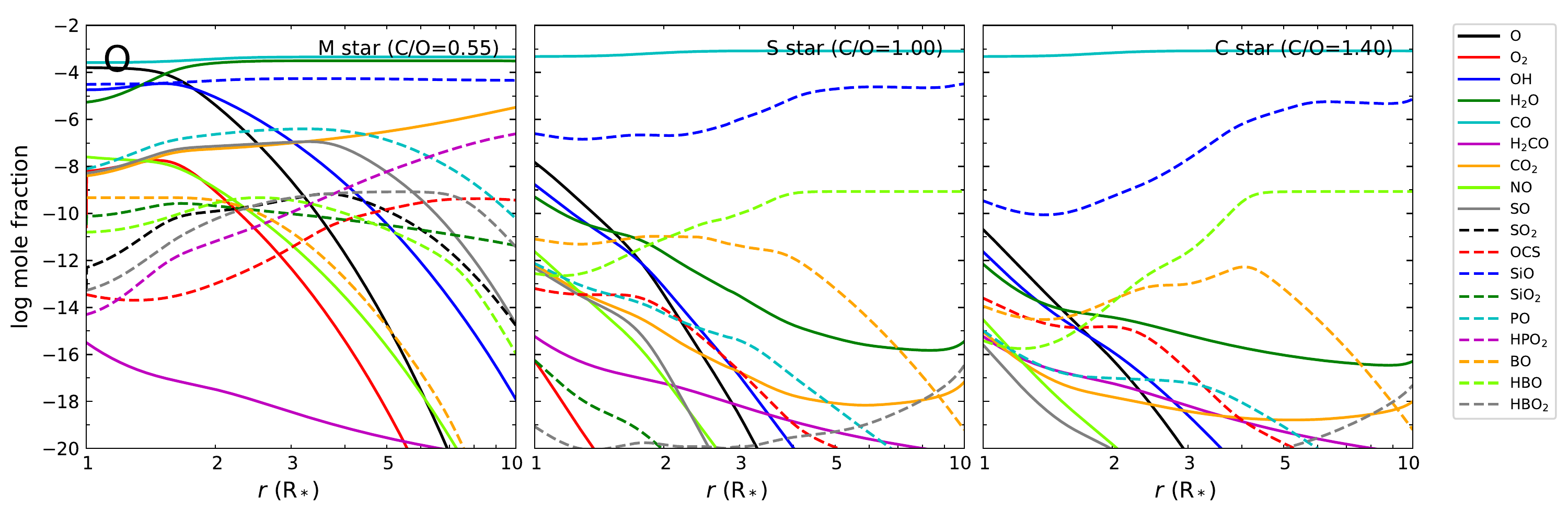} \includegraphics[angle=0,width=\textwidth]{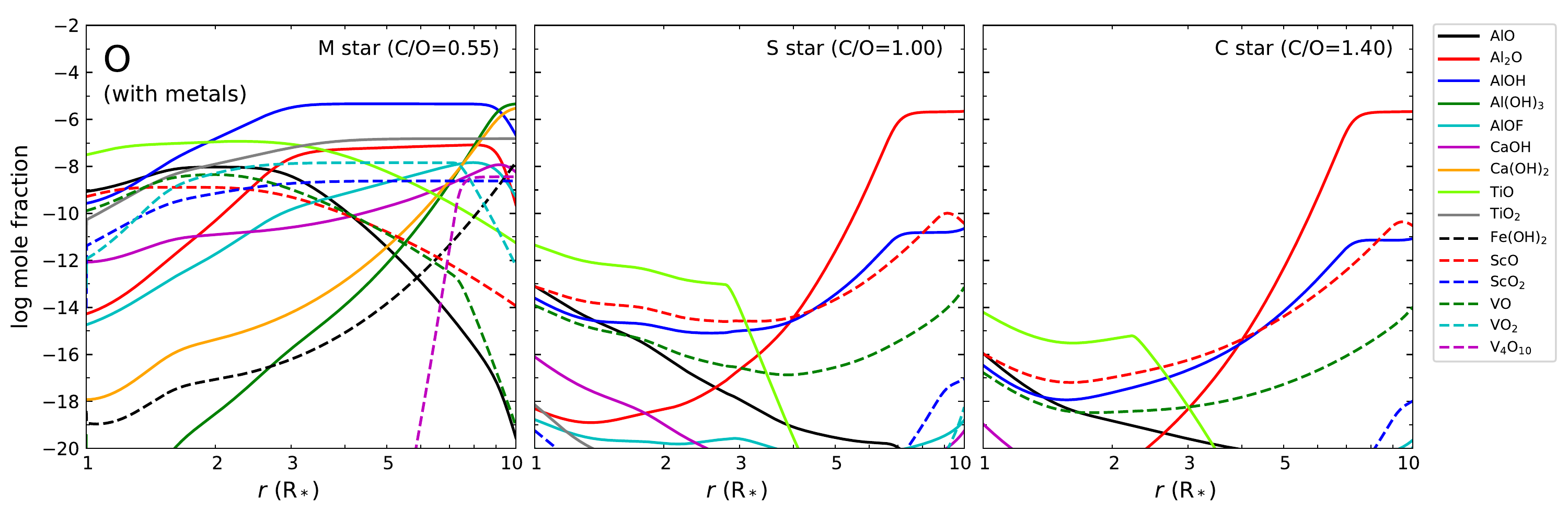}
\caption{Chemical equilibrium abundances of species containing C (upper panels) and O (lower panels) in M-, S-, and C-type AGB atmospheres.} \label{fig:c}
\end{figure*}

The main gas reservoirs of carbon are shown in the two upper panels of Fig.~\ref{fig:c}. Most carbon is in the form of CO. This implies that M-type stars, which have C/O $<$ 1, have no carbon-bearing molecule with a significant abundance, with the exception of CO$_2$. On the other hand, in C-type stars, where the C/O is higher than one, a great variety of carbon-bearing molecules are formed with large abundances. These include pure carbon clusters, hydrocarbons, and different stable molecules in which a carbon atom is bonded to a nonmetal, such as N, S, Si, or P, or to a metal, such as Na, K, Al, or Ti. The situation for S-type stars with C/O equalling one resembles that of C-type stars, but the abundances of carbon-bearing molecules are scaled down by some orders of magnitude.

In C-type stars, the main reservoir of carbon that is not locked into CO is atomic in the hot innermost atmosphere and C$_2$H$_2$ elsewhere. Only at large radii ($>$10 $R_*$) does CH$_4$ become the main reservoir, although it is unlikely that chemical equilibrium regulates the chemical composition at such large radii. At large radii, polycyclic aromatic hydrocarbons (PAHs) have also been predicted to become important carriers of carbon \citep{Tejero1991,Cherchneff1992}. However, PAHs are not observed in envelopes around AGB stars, and thus it is uncertain whether they effectively form in these environments. Other main reservoirs of carbon are HCN, CS, and C$_3$, while at a given distance from the star (3-5 $R_*$), silicon, titanium, and aluminum carbides of medium to large size (SiC$_2$, Si$_2$C, Si$_3$C, Si$_5$C, Ti$_8$C$_{12}$, and Al$_2$C$_2$) become increasingly abundant.

\subsection{Oxygen}

In the two lower panels of Fig.~\ref{fig:c} we show the calculated abundances of the most abundant oxygen-bearing molecules. Similarly to the case of carbon, the very high abundance of CO in this case causes a lack of O-bearing species in C-type stars, with the exception of SiO and Al$_2$O, which become abundant reservoirs of silicon and aluminum, respectively, at large radii. In contrast, in oxygen-rich atmospheres, many different O-bearing molecules are formed abundantly, mostly consisting of oxides, hydroxides, and inorganic acids. Excluding CO, most oxygen is atomic in the surroundings of the AGB star and in the form of H$_2$O elsewhere. Additional important reservoirs of oxygen are the radical OH, which is very abundant in the hot inner atmosphere, and SiO, which is a very stable molecule that locks most of the silicon. Other molecules present at a lower level of abundance are AlOH, CO$_2$, SO, and PO. At large radii ($>$5 $R_*$), polyatomic molecules such as the hydroxides Al(OH)$_3$ and Ca(OH)$_2$ , also become quite abundant. In S-type stars the situation resembles that of carbon stars, with a paucity of oxygen-bearing molecules, and only SiO and Al$_2$O reach relatively high abundances.

\subsection{Nitrogen}

\begin{figure*}
\centering
\includegraphics[angle=0,width=\textwidth]{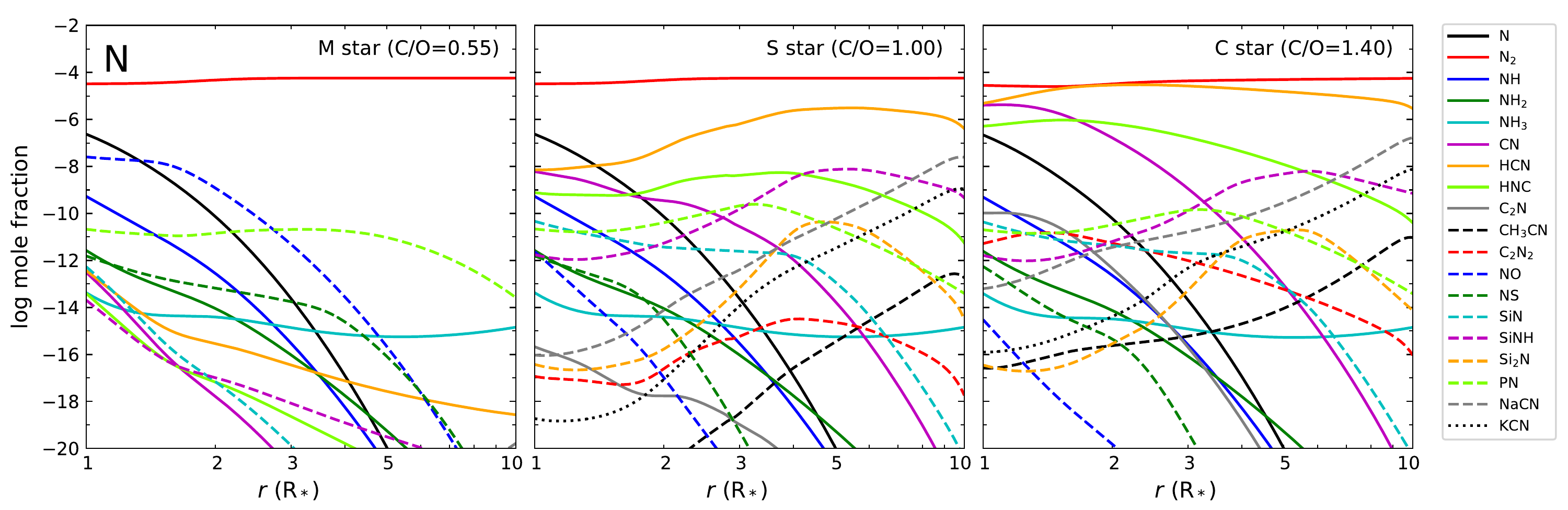} \includegraphics[angle=0,width=\textwidth]{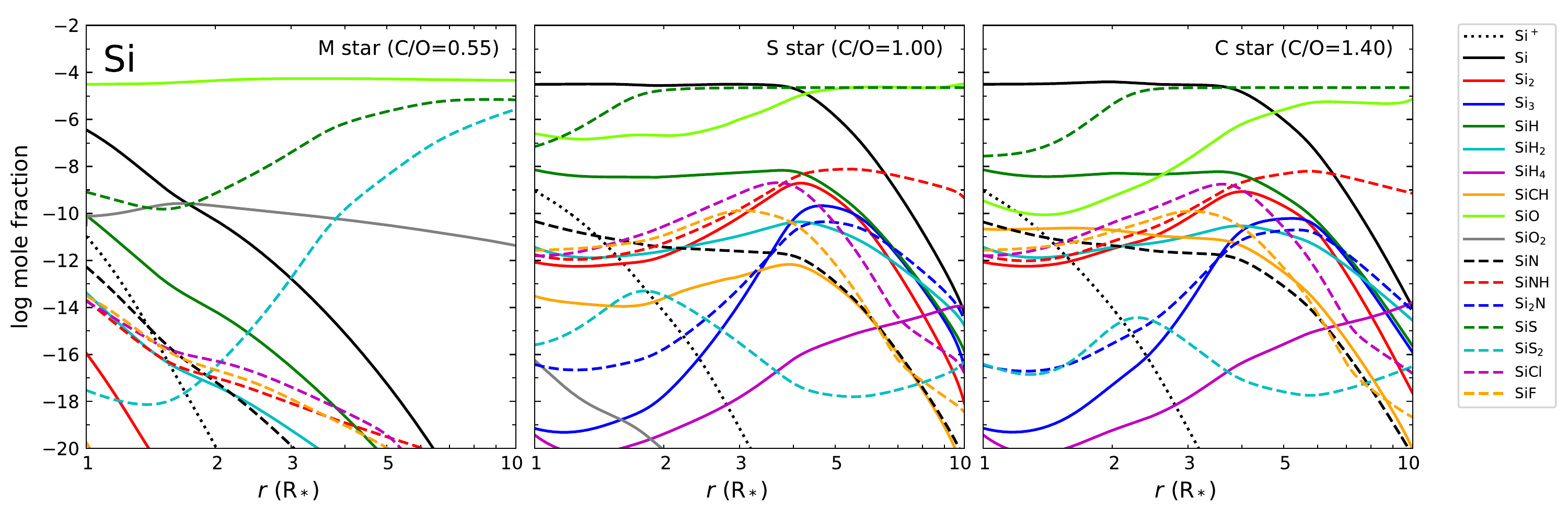} \includegraphics[angle=0,width=\textwidth]{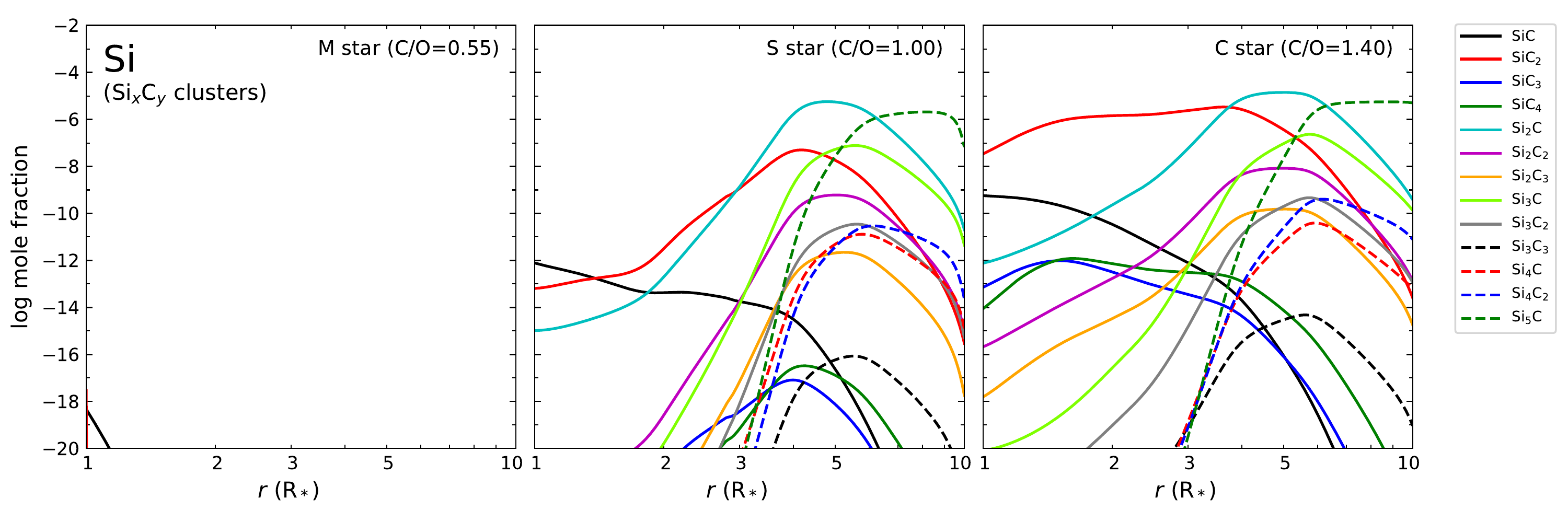}
\caption{Chemical equilibrium abundances of species containing N and Si in M-, S-, and C-type AGB atmospheres.} \label{fig:n}
\end{figure*}

The main reservoir of nitrogen in AGB atmospheres, regardless of the C/O, is clearly N$_2$ (see the top panel in Fig.~\ref{fig:n}). Unfortunately, this species is very difficult to detect and has never been observed in the atmosphere or envelope of an AGB star. Only HCN competes with N$_2$ in abundance in carbon-rich atmospheres, and to a lower extent, in S-type atmospheres. Neutral atoms are not an important reservoir of nitrogen because they would need temperatures in excess of 3000 K to compete in abundance with N$_2$. Other nitrogen-bearing molecules are present at a lower level. The metastable isomer HNC and the radical CN reach relatively high abundances, comparable to or somewhat lower than that of HCN, in the hottest inner atmosphere of S- and C-type stars, although the HNC-to-HCN and CN-to-HCN abundance ratios experience a steep decline with decreasing temperature, and thus with increasing radius. At large radii ($>$ 5 $R_*$), molecules such as SiNH and the metal cyanides NaCN and KCN reach non-negligible abundances in S- and C-type stars. In oxygen-rich atmospheres, the only N-bearing molecule that reaches a non-negligible abundance except for N$_2$ is NO, which is calculated with a mole fraction of $\sim10^{-8}$ in the inner atmosphere, although this rapidly decreases with increasing radius.
\subsection{Silicon} \label{app:si}

The calculated abundances of Si-bearing species are shown in the two lower panels of Fig.~\ref{fig:n}. The main carrier of silicon in M-type atmospheres is clearly SiO, while SiS and SiS$_2$ also become abundant at large radii ($>$ 5 $R_*$). In C-type atmospheres, atomic silicon in the inner atmosphere and SiS in the outer parts are the most abundant reservoirs. SiO is also an important Si-bearing species in carbon-rich atmospheres, but it only reaches a high abundance, slightly below that of SiS, beyond $\sim5$ $R_*$. In the atmospheres around S-type stars, the three species Si, SiS, and SiO are all main reservoirs of silicon, each in a different region.

In carbon-rich atmospheres, the availability of carbon that is not locked into CO brings a variety of silicon-carbon clusters of the type Si$_x$C$_y$, some of them with very high abundances (see bottom panel of Fig.~\ref{fig:n}). The most abundant are clearly SiC$_2$, Si$_2$C, and Si$_5$C, the latter being a main reservoir only at large radii ($>$ 6 $R_*$). Other clusters predicted at a lower level are Si$_3$C and Si$_2$C$_2$. The calculated abundances are on the same order as those reported by \cite{Yasuda2012}, who also used thermochemical data for Si$_x$C$_y$ clusters from \cite{Deng2008}. These silicon-carbon clusters are also present in S-type atmospheres with abundances comparable to or somewhat lower than in carbon-rich atmospheres, while they are completely absent in oxygen-rich atmospheres.

\subsection{Sulfur}

\begin{figure*}
\centering
\includegraphics[angle=0,width=\textwidth]{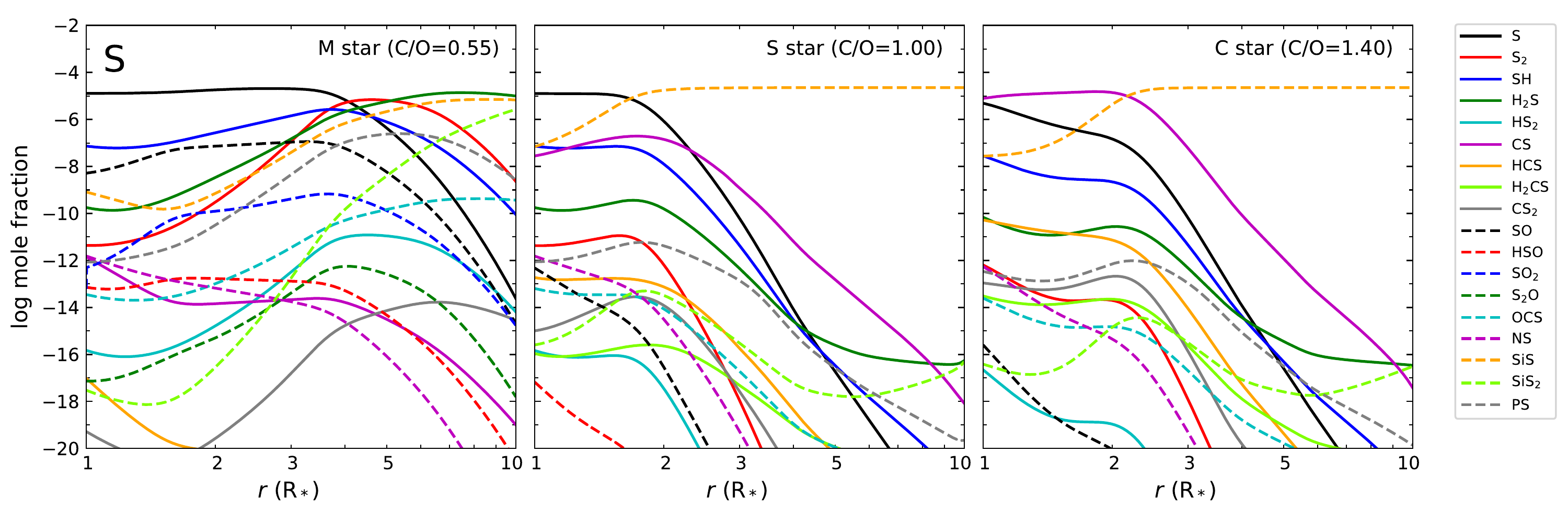} \includegraphics[angle=0,width=\textwidth]{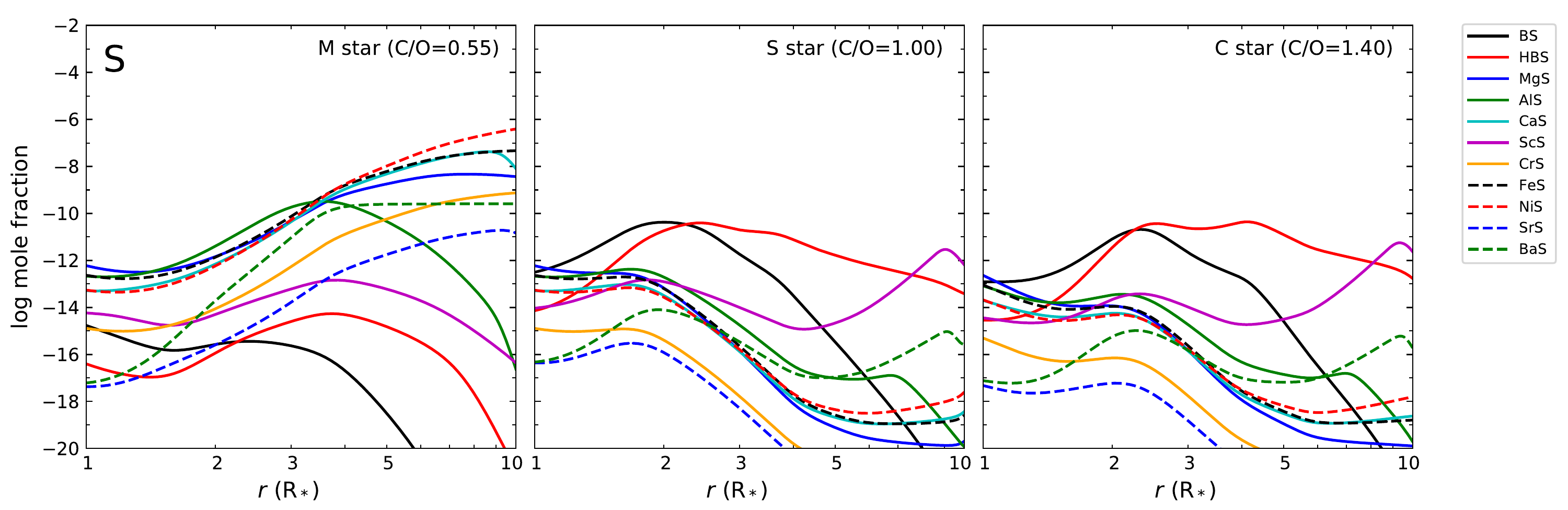} \includegraphics[angle=0,width=\textwidth]{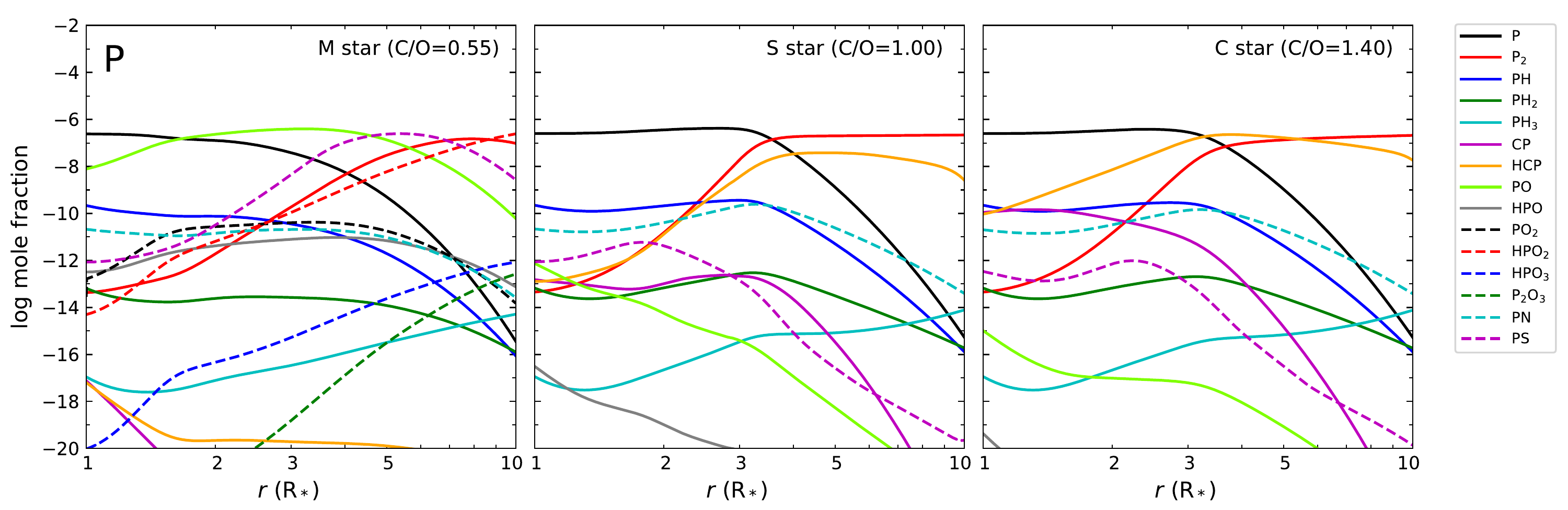}
\caption{Chemical equilibrium abundances of species containing S and P in M-, S-, and C-type AGB atmospheres.} \label{fig:s}
\end{figure*}

The calculated abundances of sulfur-bearing species are shown in the two upper panels of Fig.~\ref{fig:s}. Atomic sulfur is the main reservoir of this element in the inner atmosphere of M-type stars and also for S-type stars, although it is restricted to a smaller region around the star. In carbon-rich atmospheres, CS replaces atomic sulfur as the main reservoir in the inner atmosphere. The other main carrier of sulfur in S- and C-type stars is SiS, while in M-type stars molecules such as S$_2$, H$_2$S, and SiS lock most of the sulfur in the region in which atomic S drops in abundance. The radical SH is also an important S-bearing species, especially in M-type stars, where SO, PS, and SiS$_2$ also trap a non-negligible amount of sulfur.

\subsection{Phosphorus}

In the bottom panel of Fig.~\ref{fig:s} we show the calculated abundances of the most abundant P-bearing molecules. Most of the phosphorus is atomic in the inner atmosphere of S- and C-type stars, while at larger radii ($>$ 3 $R_*$), the molecules HCP and P$_2$ become the main carriers of this element. In M-type stars, atomic P is the main reservoir only very closely to the star, and molecules such as PO, PS, P$_2$, and HPO$_2$ (the latter only at large radii, $\sim10$ $R_*$) lock most of the phosphorus.

\begin{figure*}
\centering
\includegraphics[angle=0,width=\textwidth]{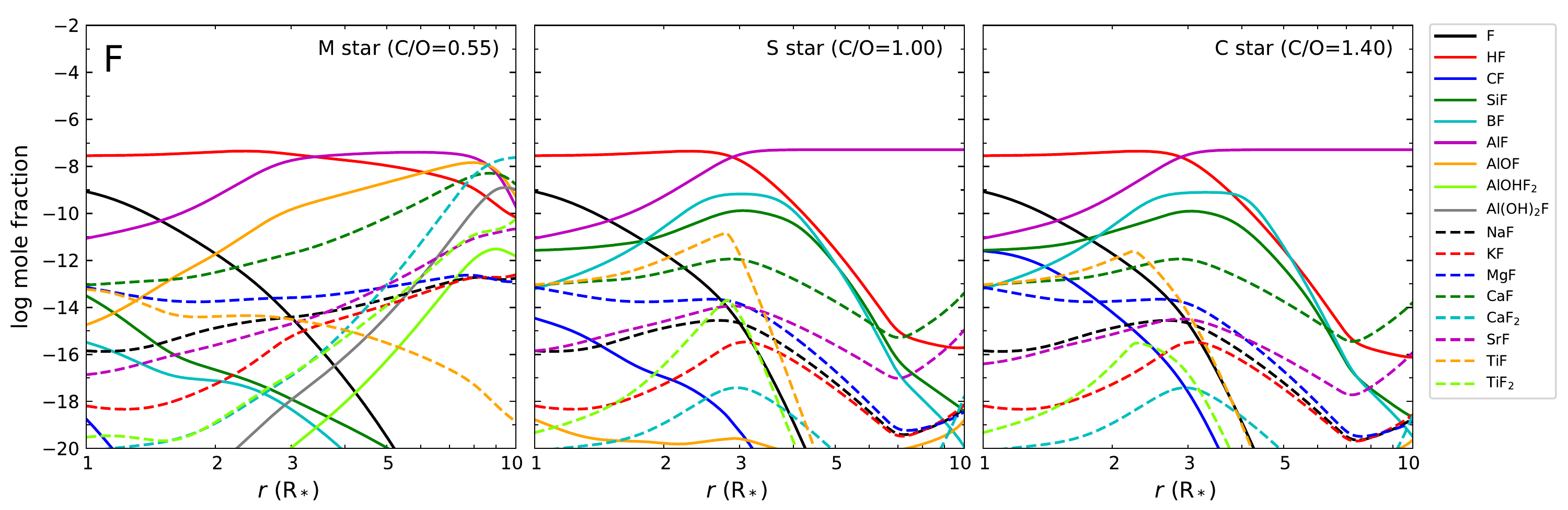} \includegraphics[angle=0,width=\textwidth]{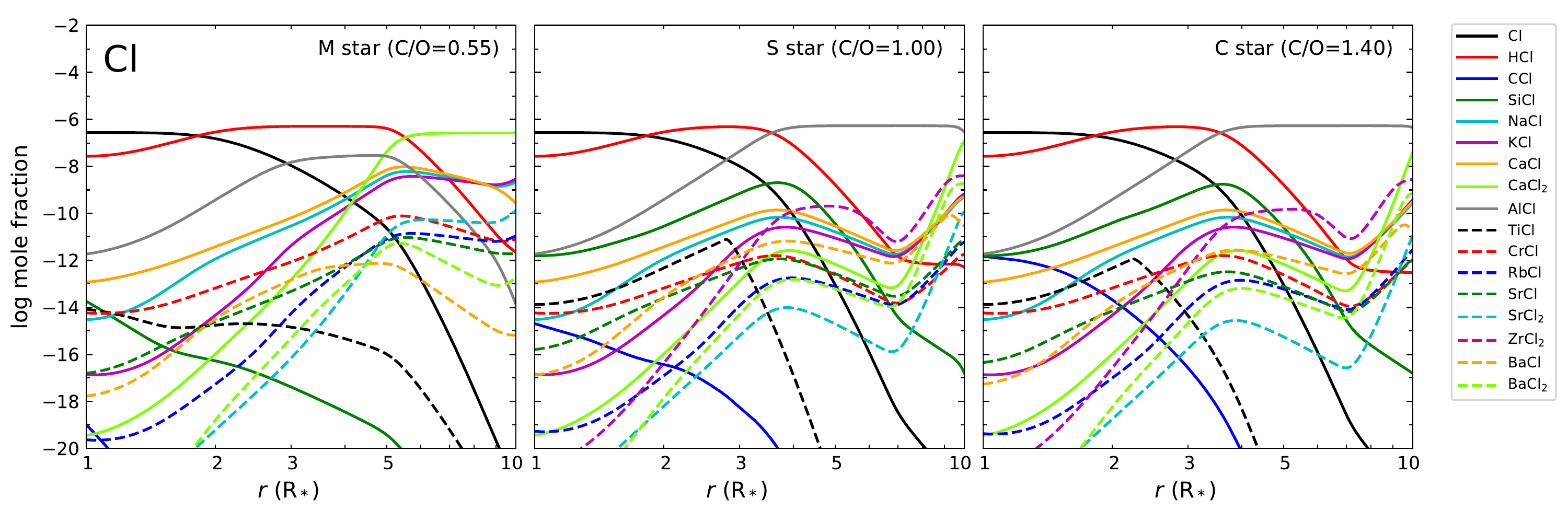}
\caption{Chemical equilibrium abundances of species containing F and Cl in M-, S-, and C-type AGB atmospheres.} \label{fig:f}
\end{figure*}

The calculations presented here differ somewhat from those presented in a previous study by \cite{Agundez2007}. For example, our calculations indicate that in carbon-rich atmospheres, P, P$_2$, and HCP are all main carriers of phosphorus, while the calculations carried out by \cite{Agundez2007} indicated that HCP is the main and almost exclusive reservoir of this element. The reason for the difference most likely is the formation enthalpy of HCP. The value at 298.15 K given in the NIST-JANAF compilation \citep{Chase1998} is 149.9 $\pm$ 63 kJ mol$^{-1}$ (note the substantial uncertainty). The thermochemical data adopted here for HCP is taken from the Third Millenium Thermochemical Database \citep{Goos} and use a substantially higher formation enthalpy, 216 kJ mol$^{-1}$, based on a more recent revision (see NIST CCCBDB\footnote{\texttt{https://cccbdb.nist.gov/}}). Another important difference concerns P$_4$O$_6$, for which \cite{Agundez2007} found that it is the main reservoir of phosphorus beyond $\sim4$ $R_*$ in oxygen-rich atmospheres. Here we find a negligible abundance for this species. The reason is again related to the formation enthalpy. The NIST-JANAF compilation \citep{Chase1998} gives a value at 298.15 K of $-$2214.3 $\pm$ 33.5 kJ mol$^{-1}$. The Third Millenium Thermochemical Database \citep{Goos} states that the latter value is erroneous and uses a much higher value, $-$1606 kJ mol$^{-1}$, from the compilation of \cite{Gurvich1989}.

\subsection{Fluorine}

The main gas reservoirs of fluorine are shown in the top panel of Fig.~\ref{fig:f}. The budget of this element in AGB atmospheres is relatively simple and remarkably independent of the C/O. Atomic fluorine only becomes the main reservoir at temperatures above $\sim3250$ K and thus is not an important reservoir in cool AGB stars. Most of the fluorine is locked by HF in the inner atmosphere, while AlF becomes the main carrier of this element at radii larger than $\sim3$ $R_*$. This applies to M-, S-, and C-type stars. In M-type stars, at radii as large as 10 $R_*$, metal-containing molecules such as AlOF, CaF, and CaF$_2$ can trap important amounts of fluorine. 

\subsection{Chlorine}

The calculated abundances of Cl-bearing species are shown in the lower panel of Fig.~\ref{fig:f}. The situation for chlorine, with HCl and AlCl as major reservoirs, somewhat resembles that of fluorine, although there are some important differences. In this case, atomic chlorine is an important reservoir of the element in the inner atmosphere. For our adopted pressure-temperature profile, atomic chlorine is the main carrier for temperatures above $\sim1750$ K and HCl dominates at lower temperatures, regardless of the C/O. At radii larger than 3-4 $R_*$, the chlorine budget is different depending on the C/O. In this region, AlCl becomes the most abundant Cl-bearing species in S- and C-type atmospheres, while for M-type stars, CaCl$_2$ replaces AlCl as the most important reservoir of this element. Chlorine tends to form relatively stable metal chlorides. In addition to the aforementioned AlCl and CaCl$_2$, metal chlorides such as CaCl, NaCl, KCl, SiCl, and ZrCl$_2$ are formed to different extents depending on the C/O.

\subsection{Boron}

\begin{figure*}
\centering
\includegraphics[angle=0,width=\textwidth]{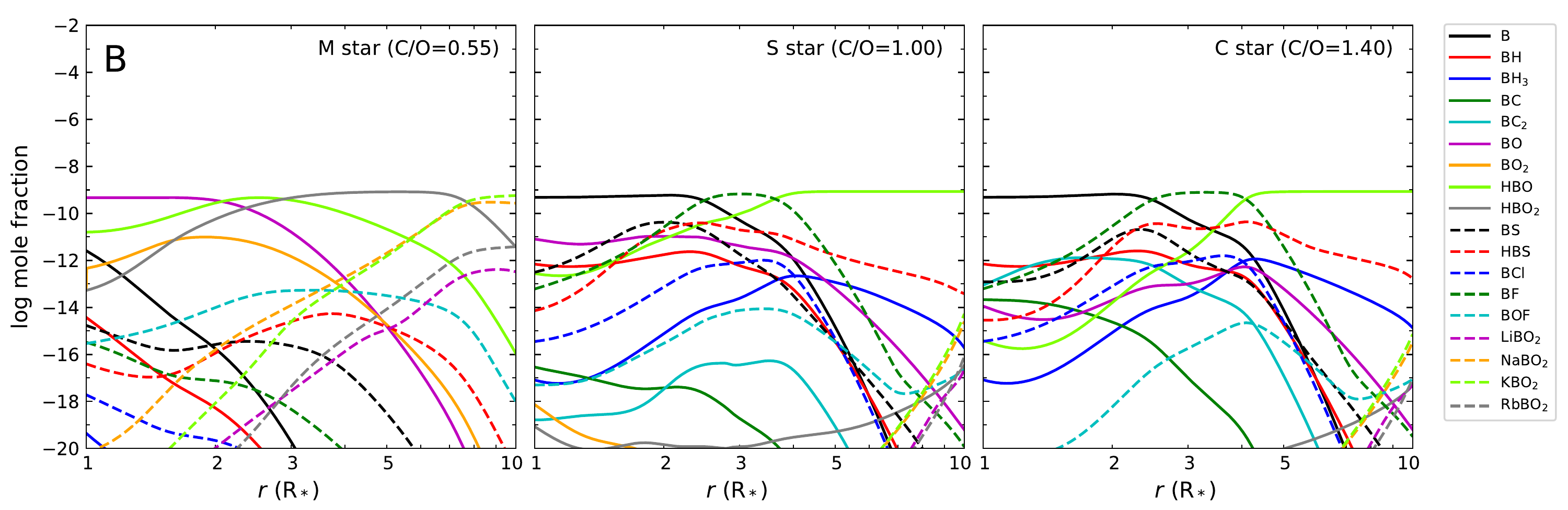} \includegraphics[angle=0,width=\textwidth]{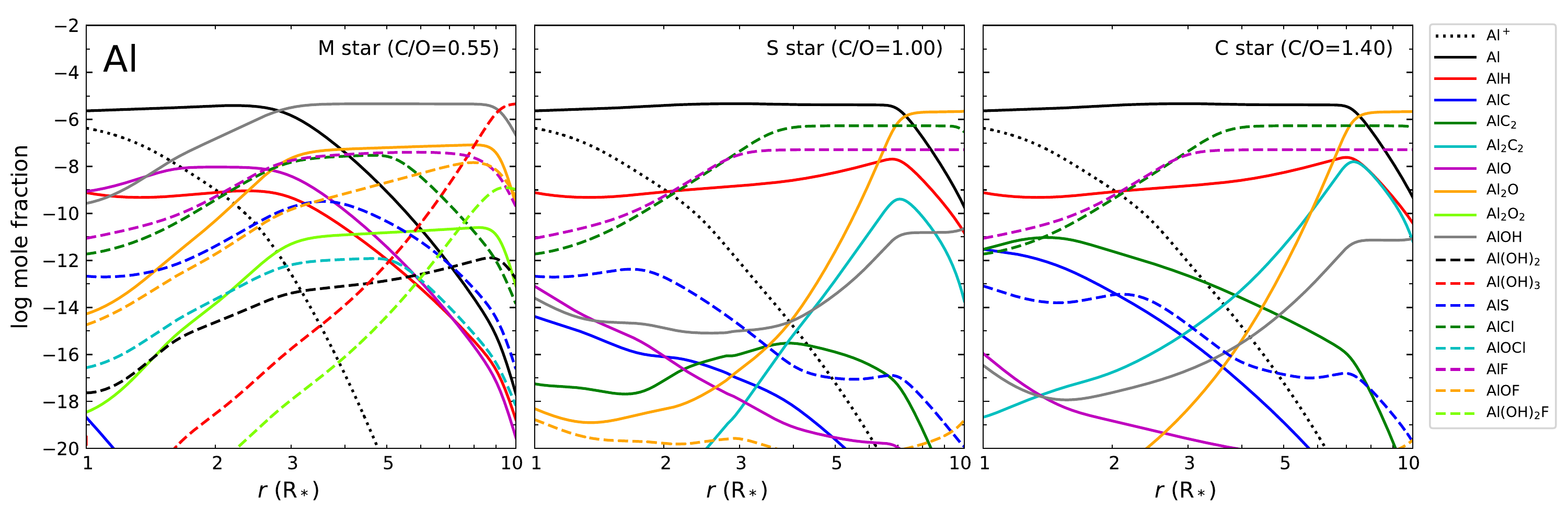}
\caption{Chemical equilibrium abundances of species containing B and Al in M-, S-, and C-type AGB atmospheres.} \label{fig:b}
\end{figure*}

In the upper panel of Fig.~\ref{fig:b} we show the most abundant species containing boron. The low abundance of this element, $5\times10^{-10}$ relative to H makes the observation of B-containing species difficult. No such species has been observed to date in the atmosphere or envelope of an AGB star.  Boron is essentially in atomic form in the inner atmosphere of S- and C-type atmospheres. However, at a relatively short distance from the star ($\sim2$ $R_*$), the molecules BF and HBO become the main carriers of the element. In the atmospheres of M-type stars the situation is different. Atomic boron is not a main carrier, and the molecules BO, HBO, and HBO$_2$ take most of the element. At radii larger than 7-8 $R_*$, where chemical equilibrium is less likely to hold, the alkali metaborate species NaBO$_2$ and KBO$_2$ are predicted to be the main reservoirs of boron in oxygen-rich atmospheres.

\subsection{Aluminum}

The aluminum budget is shown in the lower panel of Fig.~\ref{fig:b}. This element is mostly in the form of neutral atoms at the photosphere of AGB stars, regardless of the C/O, while Al$^+$ is the second most abundant carrier and becomes the main reservoir for temperatures above $\sim3000$ K. At distances larger than a few stellar radii, atomic aluminum begins to become less abundant in favor of molecules. In M-type atmospheres, hydroxides such as AlOH and Al(OH)$_3$, the latter only at large radii ($>$ 9 $R_*$), are main carriers of aluminum, while molecules such as AlO, Al$_2$O, AlCl, and AlF are also predicted to trap a significant fraction of the element. In the atmospheres of S- and C-type stars, the main molecular reservoirs of aluminum are the halides AlCl and AlF, together with Al$_2$O at radii larger than $\sim7$ $R_*$.

\subsection{Alkali metals: Li, Na, K, and Rb}

\begin{figure*}
\centering
\includegraphics[angle=0,width=\textwidth]{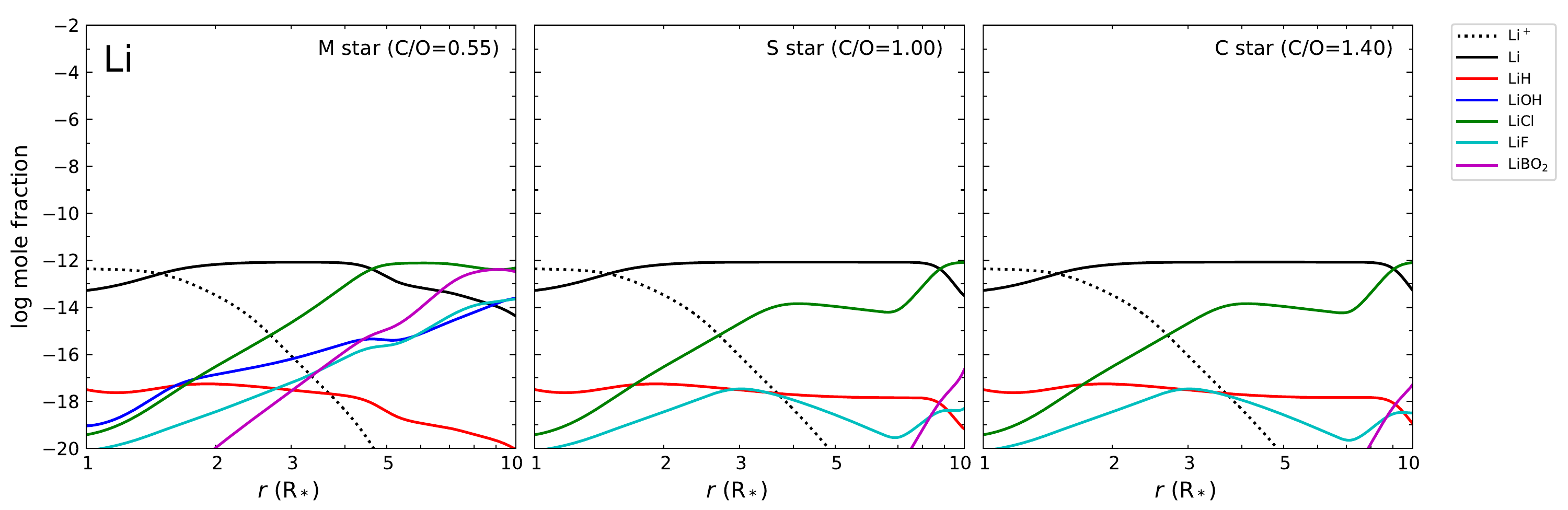} \includegraphics[angle=0,width=\textwidth]{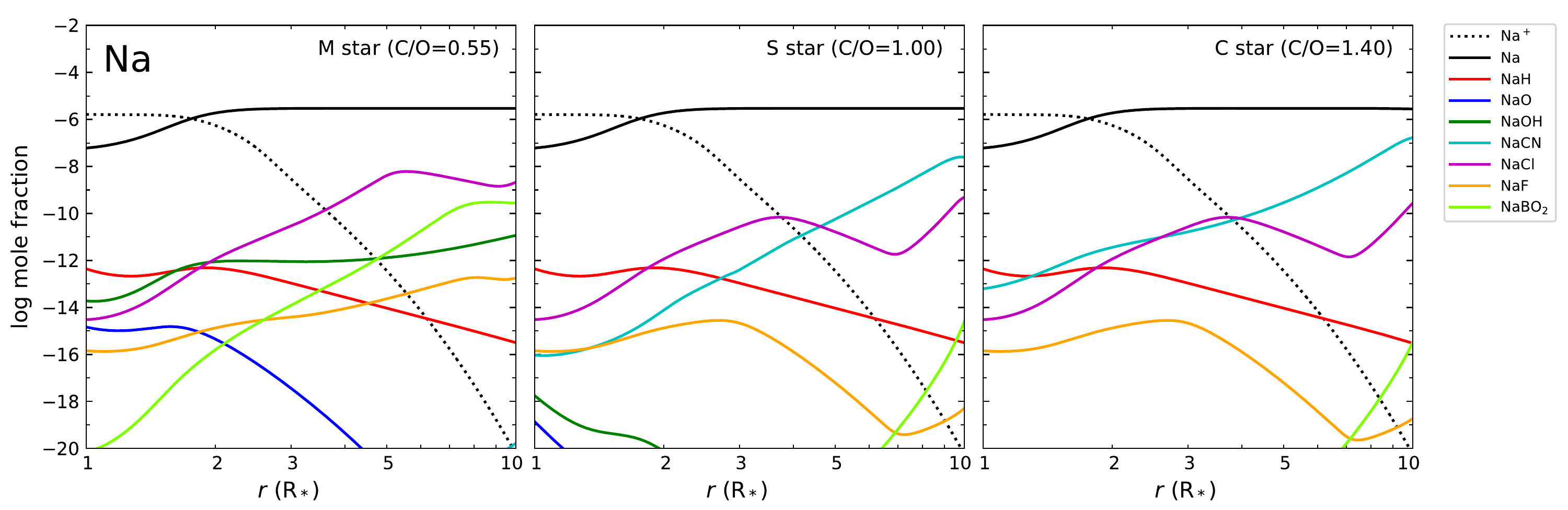} \includegraphics[angle=0,width=\textwidth]{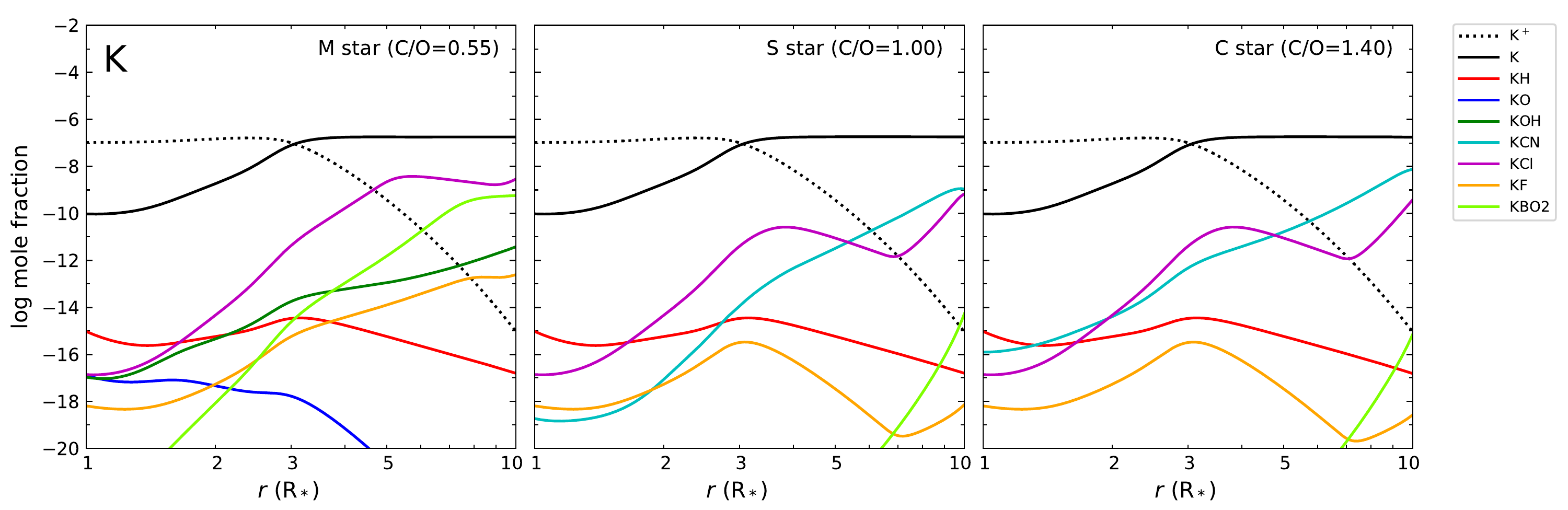} \includegraphics[angle=0,width=\textwidth]{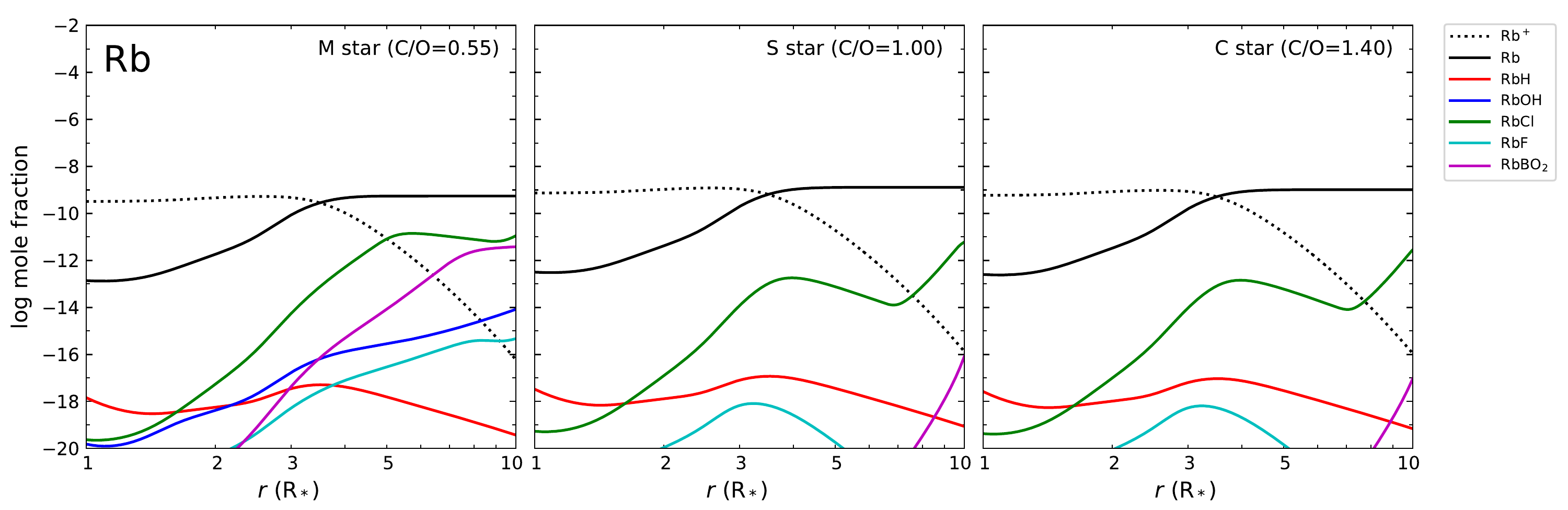}
\caption{Chemical equilibrium abundances of species containing alkali metals in M-, S-, and C-type AGB atmospheres.} \label{fig:li}
\end{figure*}

The calculated abundances of species containing the alkali metals Li, Na, K, and Rb are shown in Fig.~\ref{fig:li}. The four elements show a similar behavior. That is, most of the element is in atomic rather than in molecular form. Ionized atoms are the main reservoir at the photosphere, while neutral atoms dominate from radial distances not too far from the AGB star. The importance of ionized atoms increases with atomic number, following the decrease in the ionization energy. The most important molecular reservoir for all the alkali metals are chlorides. In the case of lithium, LiCl even becomes the main carrier of Li at large radii, especially in M-type atmospheres. For Na, K, and Rb, the corresponding chloride (NaCl, KCl, and RbCl, respectively) trap a significant fraction of the alkali metal regardless of the C/O. In S- and C-type atmospheres, the cyanides NaCN and KCN also become abundant at large radii, while in M-type atmospheres, the alkali metaborate species LiBO$_2$, NaBO$_2$, KBO$_2$, and RbBO$_2$ are also important carriers of alkali metals.

In general, molecules are not the main reservoirs of alkali metals. However, the high elemental abundance of Na and K ($\sim10^{-6}$ and $\sim10^{-7}$ relative to H, respectively) makes it possible to observe molecules such as NaCl, KCl, NaCN, and KCN in envelopes of AGB stars. In the case of Li and Rb, they can experience significant abundance enhancements compared to the Sun in some AGB stars, but their mean abundances remain low ($\sim10^{-12}$ for Li and $\sim10^{-10}$ for Rb, relative to H), which makes it difficult to detect molecules such as LiCl and RbCl.

\subsection{Alkali-earth metals: Be, Mg, Ca, Sr, and Ba}

\begin{figure*}
\centering
\includegraphics[angle=0,width=\textwidth]{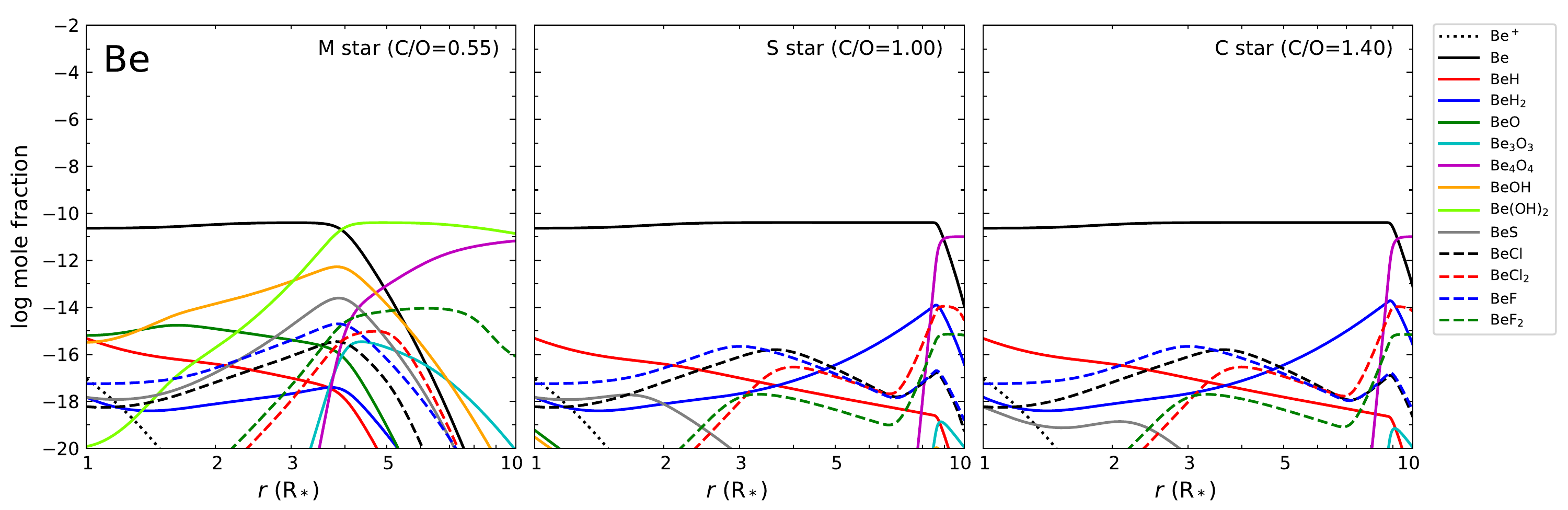} \includegraphics[angle=0,width=\textwidth]{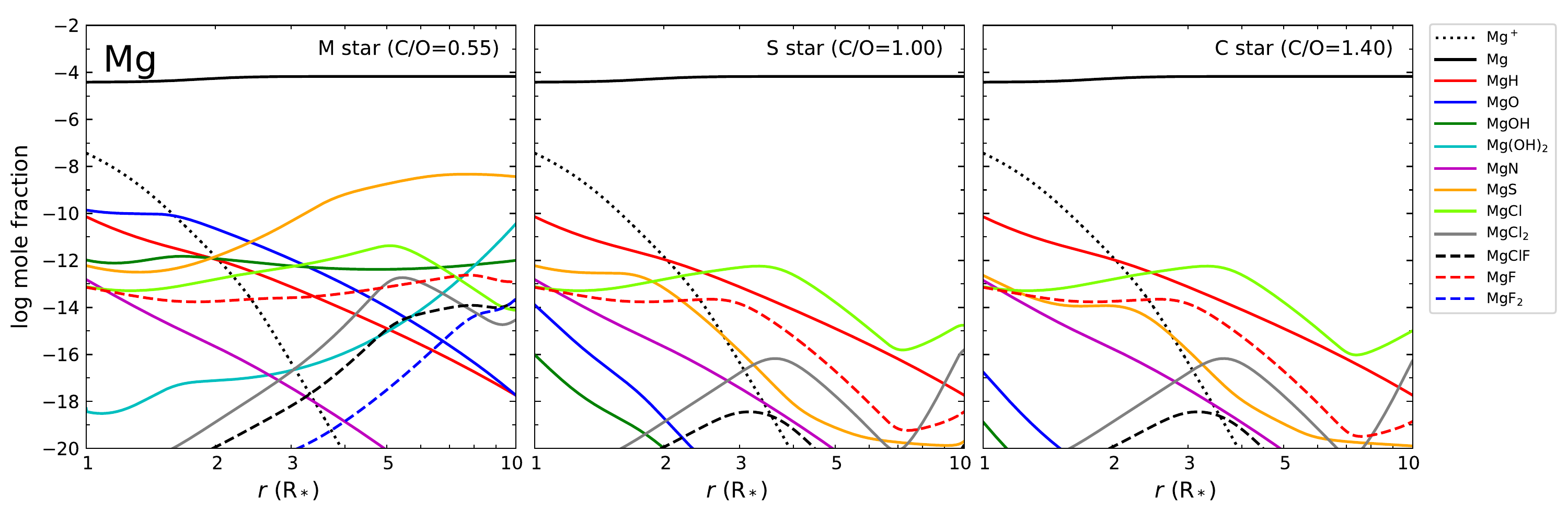} \includegraphics[angle=0,width=\textwidth]{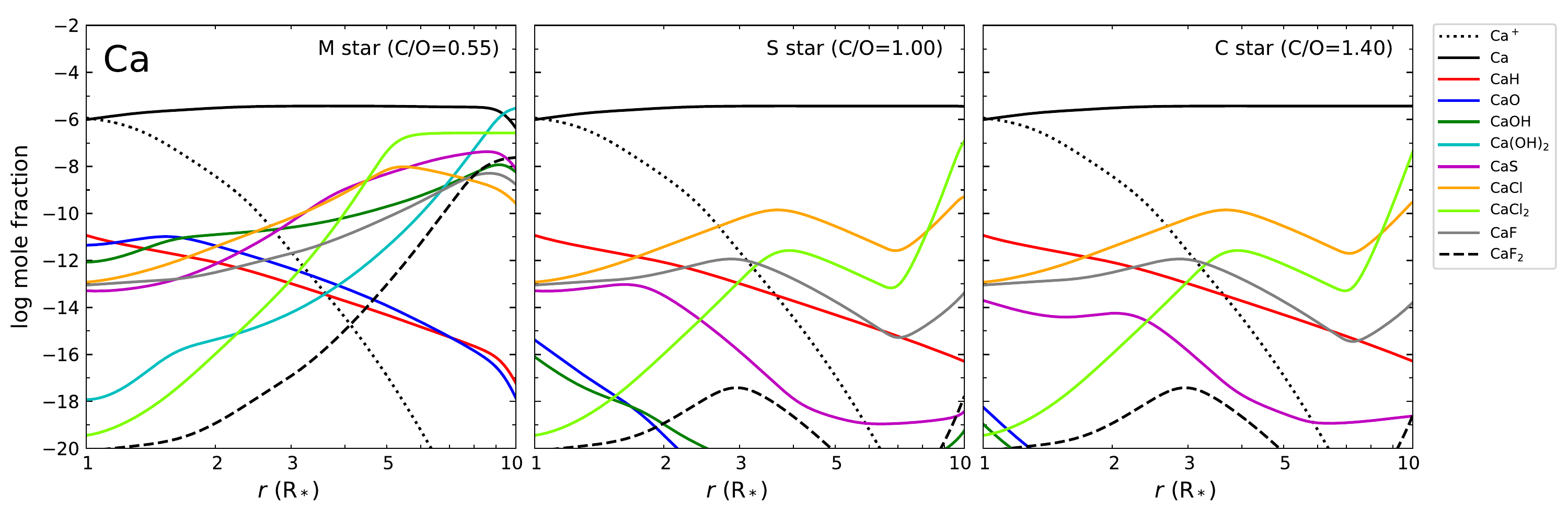} \includegraphics[angle=0,width=\textwidth]{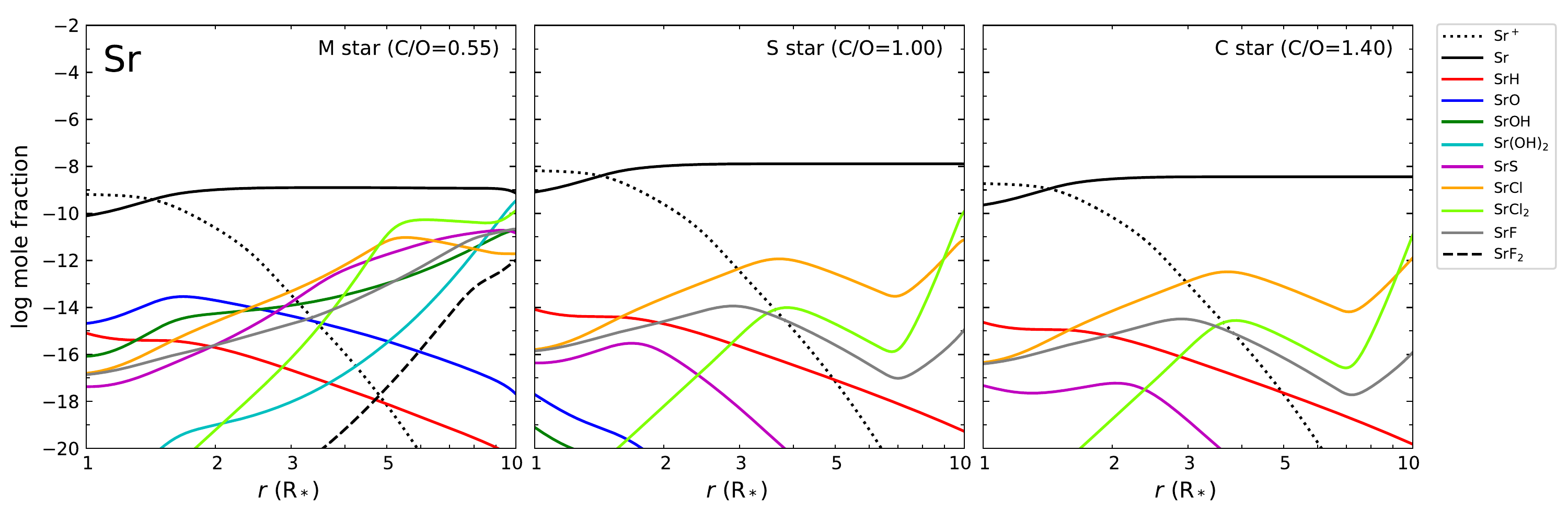}\caption{Chemical equilibrium abundances of species containing alkali-earth metals in M-, S-, and C-type AGB atmospheres.} \label{fig:be}
\end{figure*}

\setcounter{figure}{6}
\begin{figure*}
\centering
\includegraphics[angle=0,width=\textwidth]{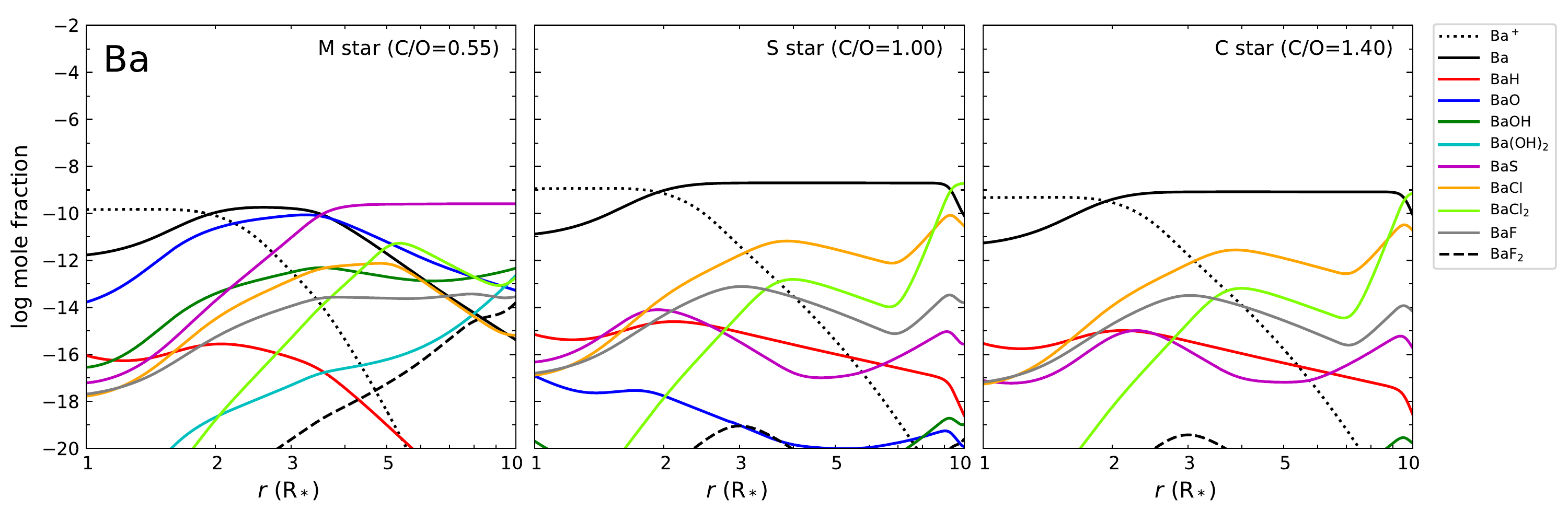}
\caption{Continued.} \label{fig:be}
\end{figure*}

In Fig.~\ref{fig:be} we show the calculated abundances of species containing the alkali-earth metals Be, Mg, Ca, Sr, and Ba. As with alkali metals, alkali-earth metals are mostly in atomic form in the photosphere of AGB stars and also to a large extent in the remaining extended atmosphere. Moreover, as occurs for alkali metals, ionized atoms become increasingly important with increasing atomic number, and thus decreasing ionization energy. They are of little importance for Be, but are a main reservoir in the case of Ba. Molecules can also trap a larger or smaller fraction of the alkali-earth metal depending on each element and on the radial distance from the AGB star.

Beryllium is essentially in the form of neutral atoms in S- and C-type atmospheres, while in M-type stars, molecules such as Be(OH)$_2$ and Be$_4$O$_4$ become main reservoirs of this element beyond $\sim4$ $R_*$. The abundances reached by these molecules are low, however, because the intrinsic abundance of Be is low ($\sim10^{-11}$ relative to H).

In the case of magnesium, neutral atoms are clearly the main reservoir throughout the entire extended atmosphere for any C/O. The only Mg-bearing molecules that are present with non-negligible abundances are MgS and MgO, which reach mole fractions between $\sim10^{-10}$ and a few times 10$^{-9}$ in oxygen-rich atmospheres. In S- and C-type atmospheres, Mg-bearing molecules are largely absent.

Calcium is also mostly atomic in AGB atmospheres regardless of the C/O. However, some molecules such as CaOH, Ca(OH)$_2$, CaS, CaCl, CaCl$_2$, CaF, and CaF$_2$ form with relatively high abundances, especially in oxygen-rich atmospheres. In S- and C-type atmospheres, no Ca-bearing molecule is predicted with a significant abundance, except for CaCl and CaCl$_2$ at large radii ($\sim10$ $R_*$).

The situation of Sr resembles that of Ca; atoms are the main reservoir, and some hydroxides, halides, and monosulfide trap a fraction of Sr in oxygen-rich atmospheres. As an $s$-process element, the abundance of Sr is higher in AGB atmospheres than in the Sun, but it is still substantially lower than that of Ca, resulting in very low mole fractions for Sr-bearing molecules ($<$ 10$^{-10}$) and low probabilities for detecting any of them.

The last alkali-earth metal included is Ba. Similarly to Ca and Sr, most barium is atomic in AGB atmospheres, although in M-type atmospheres BaO and BaS emerge as two important reservoirs of this element, with mole fractions of about 10$^{-10}$. There is evidence of BaO in M-type atmospheres from near-infrared spectra \citep{Dubois1977}. Dubois (1977) also identified BaF and BaCl in M- and S-type atmospheres, respectively, although our calculations show low abundances for these halides. Their presence in such atmospheres might be a consequence of an enhancement in the abundance of the $s$-process element Ba over the values adopted by us.

\subsection{Transition metals: Sc, Ti, Zr, V, Cr, Mn, Fe, Co, Ni, Cu, and Zn} \label{app:transition-metals}

\begin{figure*}
\centering
\includegraphics[angle=0,width=\textwidth]{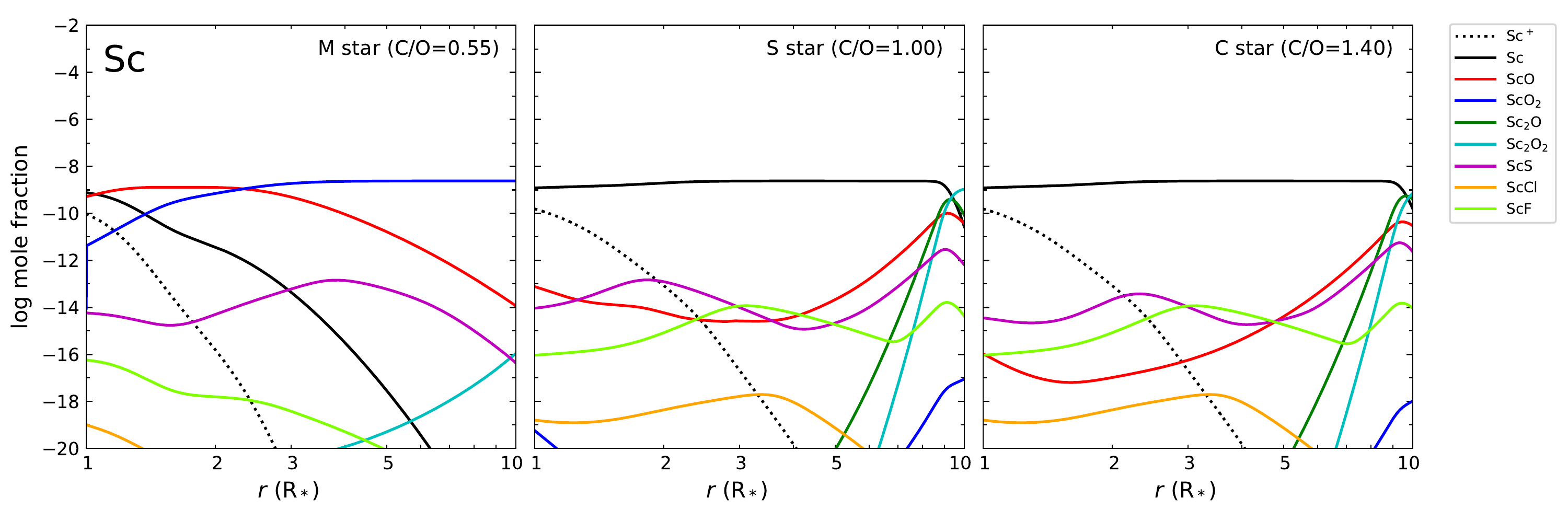} \includegraphics[angle=0,width=\textwidth]{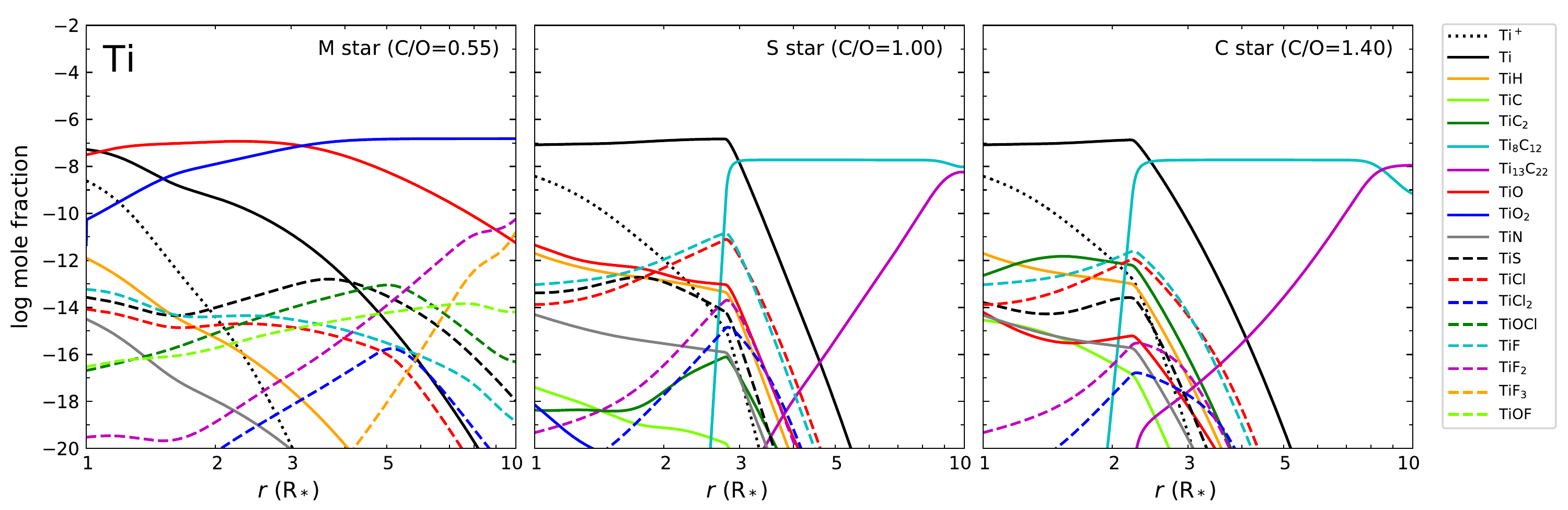} \includegraphics[angle=0,width=\textwidth]{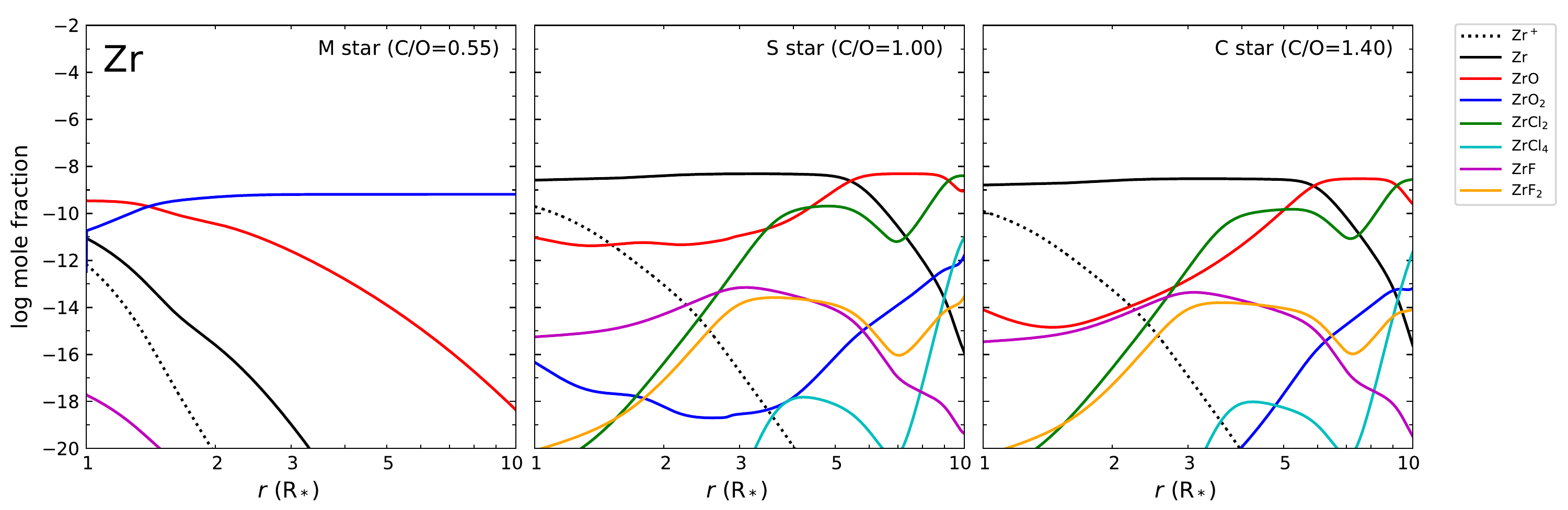} \includegraphics[angle=0,width=\textwidth]{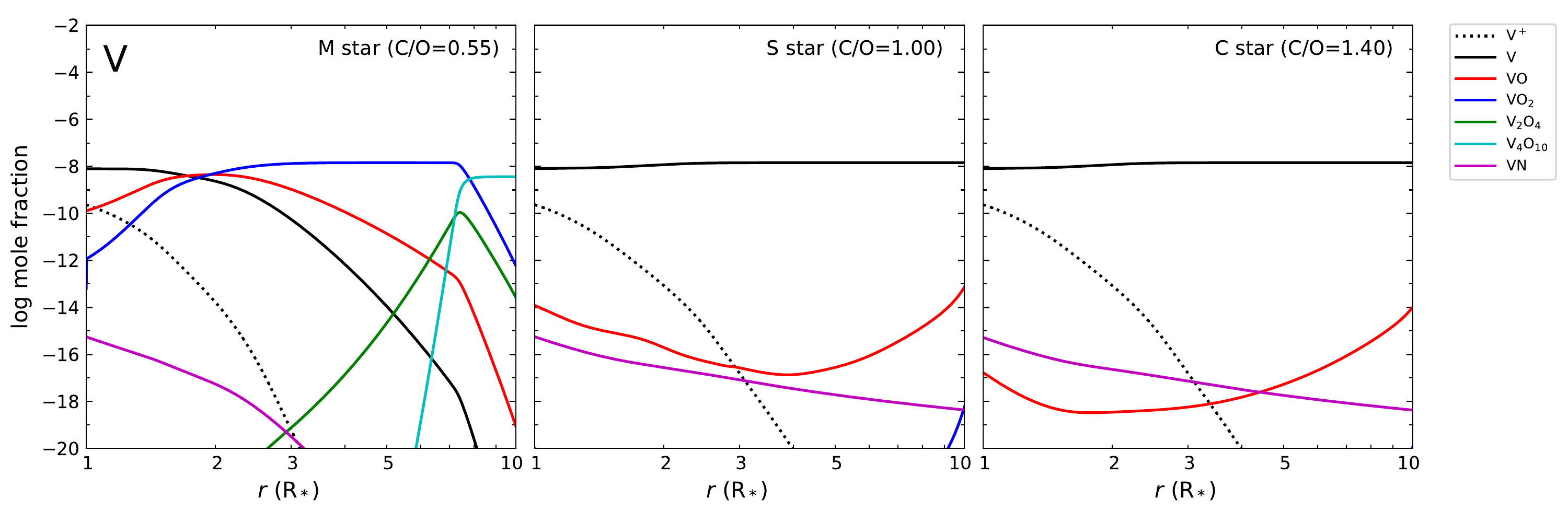}
\caption{Chemical equilibrium abundances of species containing Sc, Ti, Zr, and V in M-, S-, and C-type AGB atmospheres.} \label{fig:sc}
\end{figure*}

\begin{figure*}
\centering
\includegraphics[angle=0,width=\textwidth]{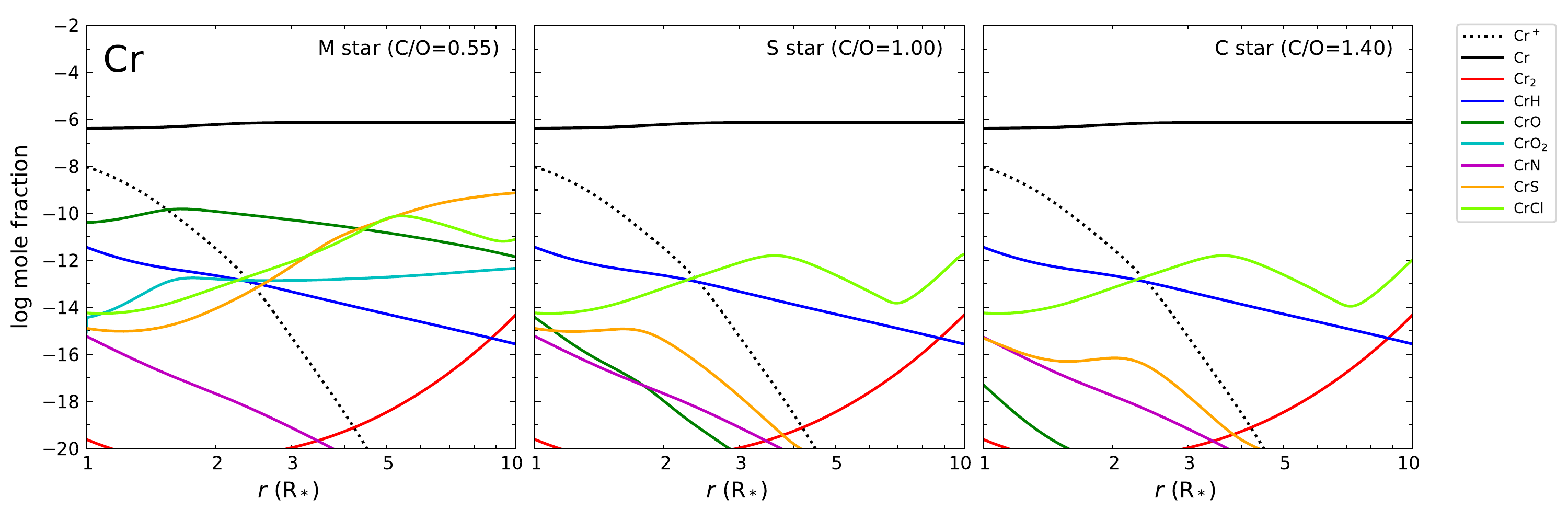} \includegraphics[angle=0,width=\textwidth]{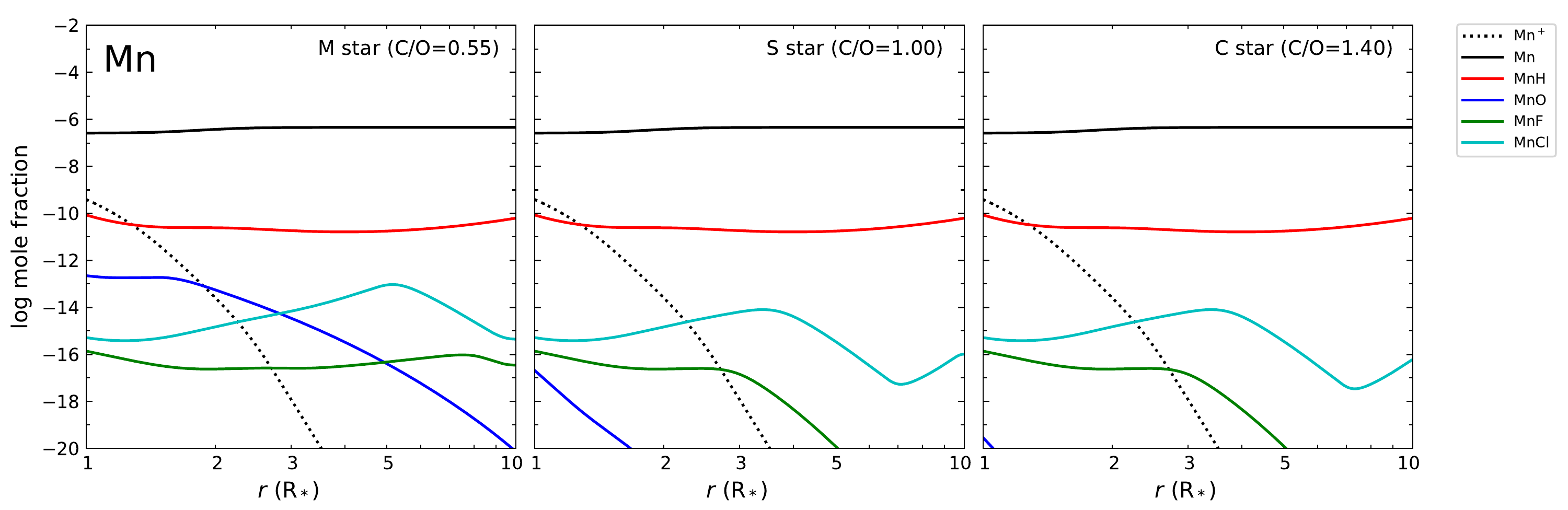} \includegraphics[angle=0,width=\textwidth]{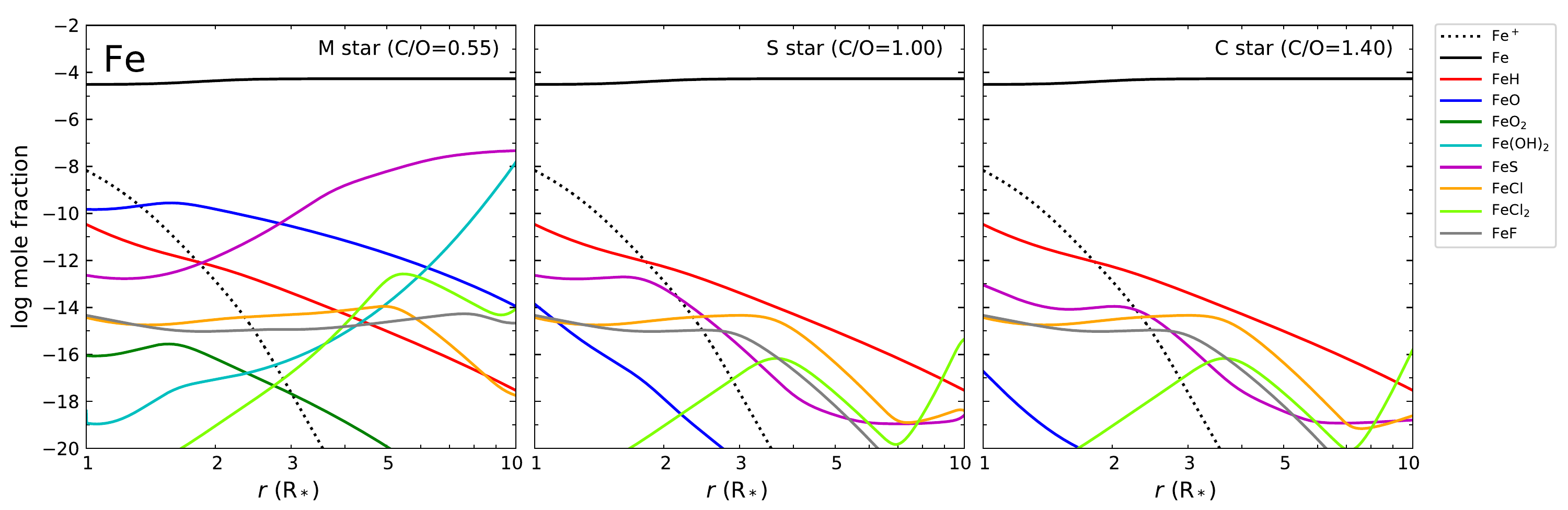} \includegraphics[angle=0,width=\textwidth]{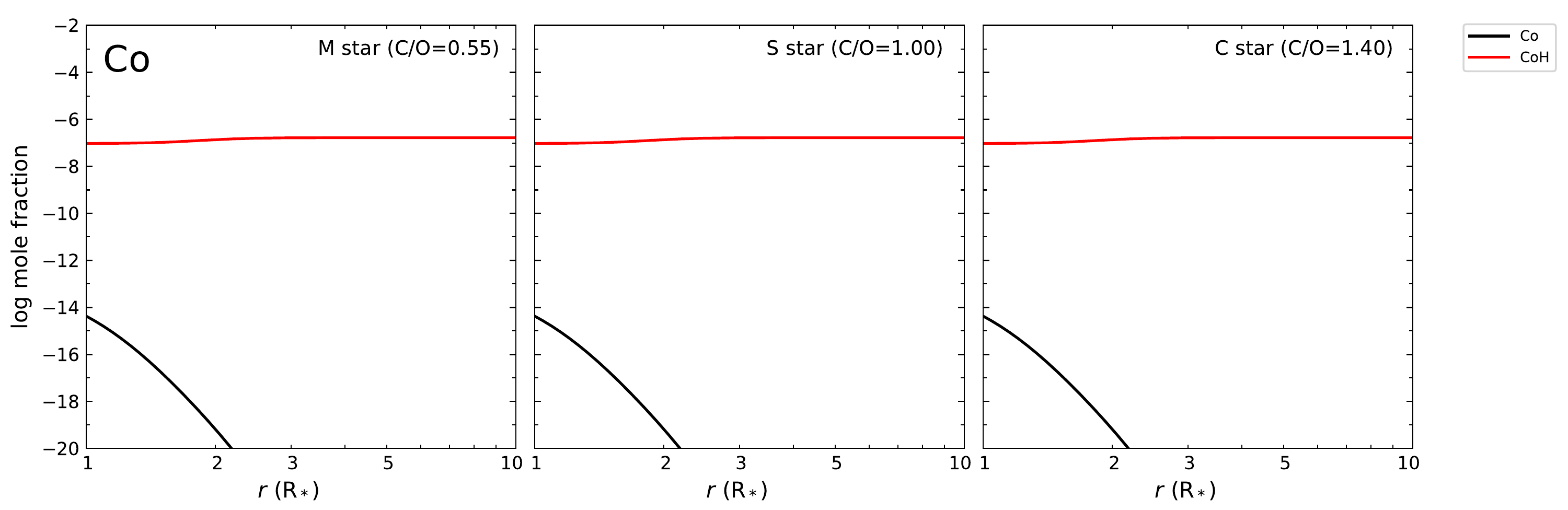}\caption{Chemical equilibrium abundances of species containing Cr, Mn, Fe, and Co in M-, S-, and C-type AGB atmospheres.} \label{fig:cr}
\end{figure*}

\begin{figure*}
\centering
\includegraphics[angle=0,width=\textwidth]{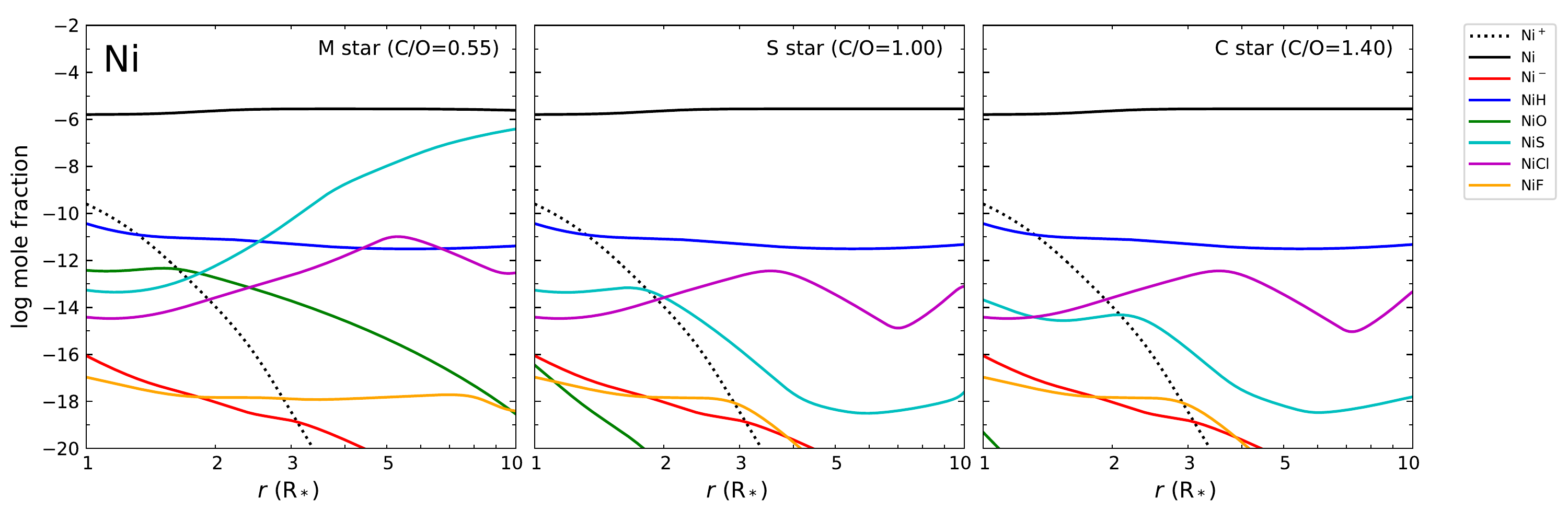} \includegraphics[angle=0,width=\textwidth]{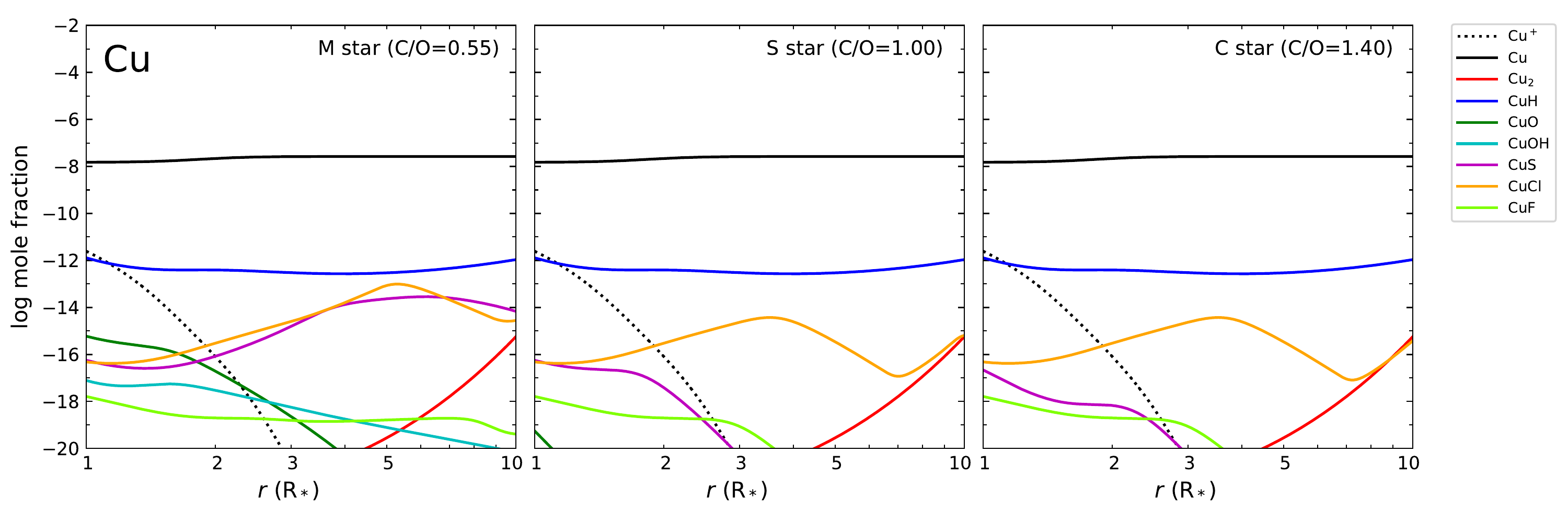} \includegraphics[angle=0,width=\textwidth]{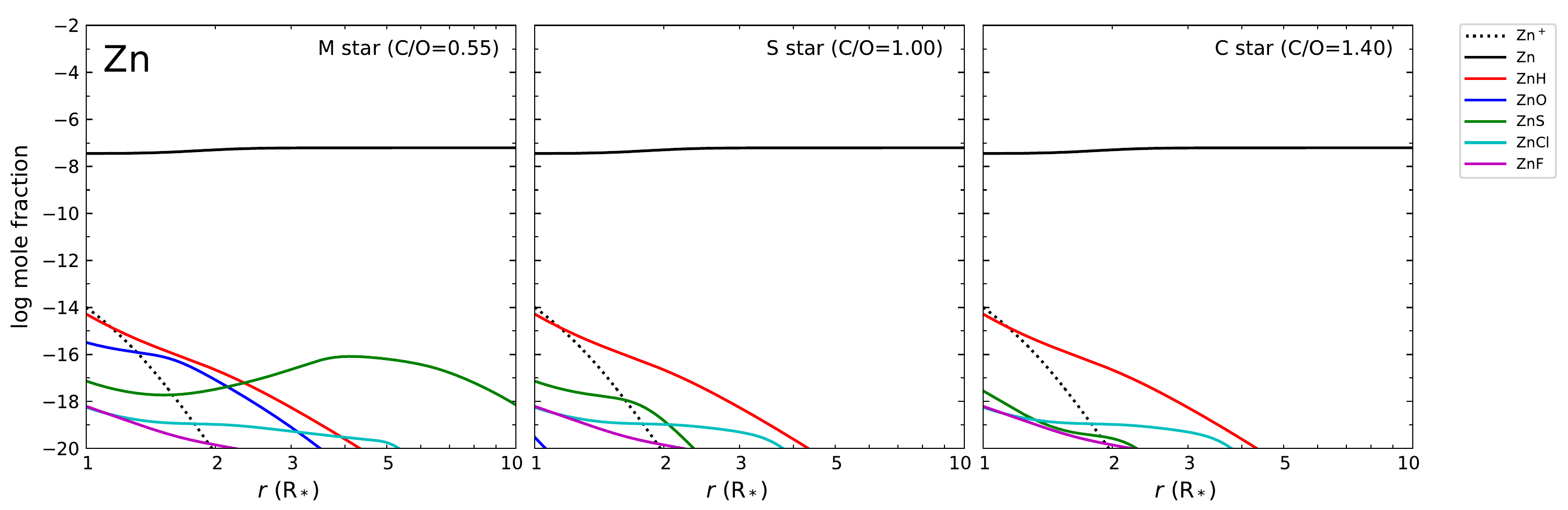}
\caption{Chemical equilibrium abundances of species containing Ni, Cu, and Zn in M-, S-, and C-type AGB atmospheres.} \label{fig:ni}
\end{figure*}

The calculated abundances of species containing transition metals are shown in Fig.~\ref{fig:sc} (Sc, Ti, Zr, and V), Fig.~\ref{fig:cr} (Cr, Mn, Fe, and Co), and Fig.~\ref{fig:ni} (Ni, Cu, and Zn). In general, transition metals tend to be mostly as neutral atoms in S- and C-type atmospheres, while in oxygen-rich atmospheres, neutral atoms and oxides are important reservoirs. There seems to be a trend: from left to right in the periodic table, the abundance of oxides in M-type atmospheres decreases in favor of atoms or other molecules such as sulfides and hydrides. For example, oxides are a main reservoir for Sc, Ti, Zr, and V, but not for Cr, Mn, Fe, Ni, Cu, or Zn. These generic conclusions are to be taken with caution, however, because there might be an important problem of completeness regarding the metal-bearing molecules for which thermochemical data is available. An example of this is illustrated by the case of titanium, in which the availability of thermochemical data for the many titanium-carbon clusters computed in this work reveals that some large Ti$_x$C$_y$ clusters become the main reservoir of titanium in S- and C-type stars over a large part of the atmosphere. The same might also hold true for other metals for which thermochemical data of metal-carbon clusters are currently lacking. Another example is provided by cobalt, for which thermochemical data are only available for a few Co-bearing molecules such as CoH and some halides. Unlike the remaining transition metals, cobalt is predicted to be essentially present as CoH. However, the situation may change when  Co-bearing molecules such as oxides or sulfides, for which thermochemical data are currently lacking, are included in the calculation.

The Sc, Ti, Zr, and V budgets (shown in Fig.~\ref{fig:sc}) share some similarities. In S- and C-type atmospheres, the metal is mostly in the form of neutral atoms. This is clearly the case of Sc and V, although for Ti and Zr some molecules become main carriers of the metal over a certain region of the atmosphere. In the case of Ti, large titanium-carbon clusters (mostly Ti$_8$C$_{12}$) trap most of the Ti at radii larger than $\sim2$ $R_*$, while for Zr, the molecules ZrO and ZrCl$_2$ are also main carriers of Zr at radial distances $>$ 6 $R_*$. We note, however, that if thermochemical data were available for metal-carbon clusters M$_x$C$_y$ (where M stands for the metal) involving Sc, Zr, and V, the budget of these elements in S- and C-type atmospheres might be different. In oxygen-rich atmospheres, the oxides MO and MO$_2$ are the main reservoirs of the transition metal. The oxides of the most abundant of these four metals, TiO and TiO$_2$, have been observed in the atmospheres of M stars \citep{Kaminski2017}. The oxides ZrO and VO have long been detected in the optical and near-infrared toward M- and S-type AGB stars \citep{Keenan1952,Keenan1954,Joyce1998}, and ScO has also recently been detected at optical and near-infrared wavelengths toward the remnant of a stellar merger V1309\,Sco \citep{Kaminski2015}. Thus, detecting ZrO, VO, and ScO in M-type atmospheres through their rotational spectrum may be simply a matter of sensitivity. The sulfides TiS and ZrS have been observed through near-infrared observations toward S-type stars \citep{Hinkle1989,Jonsson1992,Joyce1998}. Our calculations result in a low mole fraction ($<10^{-12}$) for TiS in S-type atmospheres, however, while in the case of ZrS, we lack thermochemical data. We note that similarly to the case of titanium-carbon clusters in carbon-rich atmospheres, large metal-oxygen clusters with specific stoichiometries might also be fairly abundant in oxygen-rich atmospheres, as is illustrated by the case of V, where V$_4$O$_{10}$ is a main reservoir at large radii. Calculations of thermochemical data for such clusters (M$_x$O$_y$) are expected to allow us to shed light on this.

For Cr, Mn, Fe, Ni, Cu, and Zn, neutral atoms are clearly the main reservoir of the metal throughout the extended atmosphere and for any C/O. For the most abundant of these elements, some molecules can reach non-negligible abundances. In the case of Cr (top panel in Fig.~\ref{fig:cr}), the molecules CrO, CrS, and CrCl reach mole fractions of about 10$^{-10}$ in oxygen-rich atmospheres. The only Mn-bearing molecule that is predicted with a non-negligible abundance is MnH, which has a calculated mole fraction of about 10$^{-10}$ in atmospheres of any chemical type (see the second panel from the top in Fig.~\ref{fig:cr}). For iron (third panel from the top in Fig.~\ref{fig:cr}), the molecules FeS and FeO are present with mole fractions up to $\sim4\times10^{-8}$ and $\sim3\times10^{-10}$ in M-type atmospheres, while Fe(OH)$_2$ also becomes abundant at large radii. The hydride FeH has been observed at near-infrared wavelengths toward S-type stars, although its abundance has not been constrained \citep{Clegg1978}. According to our calculations, the abundance of FeH is insensitive to the C/O and thus is expected with the same abundance in M-, S-, and C-type atmospheres. The maximum predicted abundance, which is reached at the stellar photosphere, is somewhat low (slightly below 10$^{-10}$), however. The only Ni-bearing molecule with a non-negligible abundance is NiS, which becomes increasingly abundant with increasing radius in oxygen-rich atmospheres (see the top panel in Fig.~\ref{fig:ni}). Finally, for Cu and Zn, no molecule is predicted with a significant abundance.

Cobalt (see the bottom panel in Fig.~\ref{fig:cr})  is the only transition metal for which a hydride such as CoH is found to be the main carrier of the element by far. The Co budget is completely different to that of any other transition metal discussed here. Chemical equilibrium predicts that CoH is more abundant than atomic Co by orders of magnitude, and this applies to atmospheres with any C/O. For the other transition metals discussed here, neutral atoms are the main reservoir of the metal, or at least an important reservoir in the hottest regions of the atmosphere. This implies that CoH is a rather stable species. The thermochemical data for this molecule are taken from \cite{Barklem2016}. The other Co-containing molecules included in the calculations are the halides CoCl, CoCl$_2$, CoCl$_3$, Co$_2$Cl$_4$, and CoF$_2$, which are all predicted to have negligible abundances compared to CoH. These halides are thus much less stable than CoH. Cobalt clearly has an incompleteness problem in the set of molecules for which thermochemical data are available. It would be desirable to have such data for potentially abundant molecules such as oxides, hydroxides, sulfides, and carbides. If some of them were found to be especially stable, then it might become a main carrier of Co at the expense of CoH.

\clearpage

\section{Thermochemical data of Ti$_x$C$_y$ clusters} \label{app:tixcy}

\begin{table*}
\caption{Properties of Ti$_x$C$_y$ clusters.} \label{table:tixcy_properties}
\centering
\small
\begin{tabular}{lrlcrrrr}
\hline \hline
& \multicolumn{1}{r}{$\sigma$ $^a$} & \multicolumn{1}{l}{G $^b$} & \multicolumn{1}{c}{$m$ $^c$} & \multicolumn{2}{c}{$\Delta_{\rm at}H$ $^d$} & \multicolumn{1}{c}{$\Delta_{\rm at}H^o$ (298.15 K) $^e$} & \multicolumn{1}{c}{$\Delta_{f}H^o$ (298.15 K) $^f$} \\
\cline{5-8}
& & & \multicolumn{5}{c}{kJ mol$^{-1}$} \\
\hline
TiC                         &   1 & C$_v$      & 1 &    366 &   372 $^g$ & 363 & 826.67 \\
TiC$_2$                 &   2 & C$_{2v}$  & 3 &   1146 & 1157 $^h$ & 1143 & 763.35 \\
TiC$_3$                 &   2 & C$_{2v}$  & 1 &   1740 & & 1738 & 885.03 \\
TiC$_4$                 &   2 & C$_{2v}$  & 3 &   2405 & 2398 $^h$ & 2403 & 936.70 \\
Ti$_2$C                 &   2 & C$_{2v}$  & 3 &     726 & & 722 & 940.67 \\
Ti$_2$C$_2$         &   4 & D$_{2h}$  & 1 &   1379 & & 11376 & 1003.35 \\Ti$_2$C$_3$         &   1 & C$_{2v}$  & 1 &   2024 & & 2022 & 1074.02 \\
Ti$_2$C$_4$         &   1 & C$_s$      & 1 &   2708 & & 2704 & 1108.70 \\Ti$_3$C                 &   1 & C$_s$      & 1 &     909 & & 902 & 1233.67 \\
Ti$_3$C$_2$         &   2 & C$_{2v}$ & 1 &   1714 & & 1708 & 1144.34 \\
Ti$_3$C$_3$         &   1 & C$_s$      & 1 &   2403 & & 2399 & 1170.02 \\Ti$_3$C$_4$         &   2 & C$_{2v}$ & 1 &   3070 & & 3068 & 1217.70 \\
Ti$_4$C                 &   1 & C$_s$      & 1 &   1185 & & 1178 & 1430.67 \\
Ti$_4$C$_2$         &   2 & C$_{2v}$ & 1 &   1990 & & 1985 & 1340.34 \\
Ti$_4$C$_3$         &   1 & C$_s$     & 1 &   2741 & & 2737 & 1305.02 \\
Ti$_4$C$_4$         & 12 & T$_d$     & 1 &   3589 & & 3587 & 1171.69 \\
\hline
Ti$_3$C$_8$         &   1 & C$_s$      & 1 &  6348 & 6436 $^i$ & 6345 & 807.40 \\
Ti$_4$C$_8$         &   2 & C$_{2v}$ & 1 &   7077 & 7063 $^i$ & 7075 & 550.40 \\
Ti$_6$C$_{13}$     &  1 & C$_1$      & 1 & 11682 & 11694 $^i$ & 11682 & 472.77 \\
Ti$_7$C$_{13}$     &  1 & C$_1$      & 1 &  11975 & 12533 $^i$ & 11973 & 654.77 \\
Ti$_8$C$_{12}$     &  1 & C$_1$      & 1 & 12723 & 12698 $^i$ & 12727 & $-$342.91 \\
Ti$_9$C$_{15}$     &  1 & C$_1$      & 1 & 14961 & & 14958 & 49.11 \\
Ti$_{13}$C$_{22}$ & 1 & C$_1$      & 1 & 22513 & & 22520 & $-$604.17 \\
\hline
\end{tabular}
\tablenoteb{\\
$^a$ Order of the rotational subgroup. $^b$ Symmetry group. $^c$ Spin multiplicity. $^d$ Enthalpy of atomization (at 0 K and 0 bar). Ground state reference energies for the atoms are $-$37.8585747 Ha for C and $-$849.3765897 Ha for Ti using gaussians (upper block) and $-$5.6859 Ha for C and $-$59.65385 Ha for Ti using plane-waves (lower block). $^e$ Standard enthalpy of atomization (at 298.15 K and 1 bar). $^f$ Standard enthalpy of formation (at 298.15 K and 1 bar). Calculated from $\Delta_{\rm at}H^o$ (298.15 K) adopting standard enthalpies of formation of Ti and C atoms of 472.9973 kJ mol$^{-1}$ and 716.6759 kJ mol$^{-1}$, respectively \citep{Goos}. $^g$ Experimental value from \cite{Sevy2018}. $^h$ Experimental value from \cite{Stearns1974}. $^i$ DFT calculation by \cite{Munoz1999}. \\
}
\end{table*}

For each cluster Ti$_x$C$_y$, we compute a partition function $Z$ \citep{Kardar2007} which takes into account electronic ($e$), translational ($t$), rotational ($r$), and vibrational ($v$) contributions \citep{Ochterski2000},
\begin{equation}
Z (N,V,T) = \sum \exp \Bigg(- \frac{U_e + U_t + U_r + U_v}{k T} \Bigg) = Z_e Z_t Z_r Z_v,
\end{equation}
where $N$ is the number of particles, $V$ the volume, $T$ the absolute temperature (i.e., we work in the microcanonical ensemble), the different terms $U$ are the internal energy contributions as labelled above, and $k$ is the Boltzmann constant.

The main thermodynamical magnitudes per unit mol, enthalpy ($H$), entropy ($S$), and heat capacity at constant pressure ($C_P$), are derived as
\begin{equation}
H =  \frac{R T^2}{Z} \frac{\partial Z}{\partial V} \bigg|_{V}
\end{equation}
\begin{equation}
S =  R \ln{Z} + \frac{R T}{Z} \frac{\partial Z}{\partial V} \bigg|_{V}
\end{equation}
\begin{equation}
C_P = T \frac{\partial S}{\partial T} \bigg|_{P}
\end{equation}
where $R$ is the ideal gas constant and $P$ is the pressure.

Here, we are only interested in values for not too low temperatures, $T \ge 50$ K. Accordingly, we write:
\begin{equation}
Z_e = m
\end{equation}
\begin{equation}
Z_t = \Bigg(\frac{2 \pi M k T}{h^2}\Bigg)^{3/2} \Bigg( \frac{k T}{P} \Bigg) \\
\end{equation}
\begin{equation}
Z_r = 
\begin{cases}
\frac{1}{\sigma} \frac{T}{\Theta} ~,~\text{for linear}\\
\sqrt{ \frac{\pi}{\sigma^2} \frac{T^3}{\Theta} } ~,~\text{for non-linear}\end{cases}
\end{equation}
\begin{equation}
Z_v = \Pi_i \Bigg[ \frac{\exp \Big( - \frac{\hbar w_i}{2 k T} \Big)} {1-\exp \Big(-\frac{\hbar w_i}{k T} \Big)} \Bigg]
\end{equation}
where $h$ is the Planck constant ($\hbar$ = $\frac{h}{2\pi}$), $M$ the total mass, $m$ the spin multiplicity of the electronic state (see Table~\ref{table:tixcy_properties}), and $w_i$ is the vibrational frequency of the mode $i$ (the product is taken over all modes $i$ with positive frequencies). For linear clusters, $\Theta=\frac{h^2}{8 \pi^2 k I}$ ($I$ is the moment of inertia) and $\sigma=1$ or $2$ depending on whether it is heteronuclear or homonuclear. The only linear molecule is TiC, in which case $\sigma=1$ since it is heteronuclear. For non-linear clusters (all Ti$_x$C$_y$ clusters except TiC), $\Theta=\Theta_x \Theta_y \Theta_z$ ($x$, $y$, and $z$ are the principal axes of the moment of inertia tensor) and $\sigma$ is the order of the rotational subgroup in the point group associated to the cluster (see Table~\ref{table:tixcy_properties}).

For each cluster, self-consistent many-body wavefunctions $\Psi$ and their associated ground-state variational total energies $U_0$ for optimized geometrical configurations have been obtained from {\it ab initio} Density Functional Theory calculations \citep{Hohenberg1964,Kohn1965}. The optimized geometries of the clusters are given in Table~\ref{table:tixcy_geometries}. For the sake of efficiency and accuracy we apply two different strategies. For small Ti$_x$C$_y$ clusters, with $x+y \le 10$, we use an all-electron localized basis formed with linear combinations of gaussians (cc-pVTZ; \citealt{Frisch2009,Duning1989}) and a chemistry model based on the hybrid exchange and correlation functional B3LYP \citep{Becke1988}. For large Ti$_x$C$_y$ clusters, with $x+y > 10$, the system is big enough to require the use of pseudo-potentials for Ti and we favor the use of an extended basis formed with linear combinations of plane-waves \citep{Giannozzi2009} and a chemistry model based on a generalized-gradients approximation for exchange and correlation \citep{Perdew1996}, with $E_c=490$ eV and $\Gamma$ point. Both approaches provide a reasonable representation of equilibrium geometries, but their combined use yields more accurate values of the enthalpy of atomization
\begin{equation}
{\rm Ti}_x{\rm C}_y {\rm (g)} \rightarrow x~{\rm Ti} {\rm (g)} + y~{\rm C} {\rm (g)}
\end{equation}
for all the clusters studied here. In order to work with a minimal set of geometrical parameters, symmetrized models have been preferred whenever it has been possible.

Finally, the electric dipole moment $\vec p$ has been obtained from $\vec p = <\Psi \mid \vec r \mid \Psi >$ \citep{Snyder1974}. Since this is the expectation value of a one-electron operator, its value should not critically depend on the choice for the exchange-correlation functional, although it should be noted that in practice one cannot expect in computed values a precision better than $\approx 10$\%, because the contribution of the tails of wave functions require the use of large basis sets with diffuse functions.

The calculated enthalpies of atomization agree well with literature values, either experimental or theoretical, when these are available (see Table~\ref{table:tixcy_properties}). The thermochemical properties of all Ti$_x$C$_y$ clusters calculated at 1 bar and as a function of temperature are given in Tables~\ref{table:tic}-\ref{table:ti13c22}.

\begin{table}
\caption{Equilibrium geometries of Ti$_x$C$_y$ clusters.} \label{table:tixcy_geometries}
\centering
\small
\begin{tabular}{lrrr}
\hline \hline
 & \multicolumn{1}{c}{$x$ ({\AA})} & \multicolumn{1}{c}{$y$ ({\AA})} & \multicolumn{1}{c}{$z$ ({\AA})} \\
\hline
 & \multicolumn{3}{c}{TiC} \\
\hline
C &  0.000000 &  0.000000 & $-$1.252485 \\
Ti &  0.000000 &  0.000000 &  0.341587 \\
\hline
 & \multicolumn{3}{c}{TiC$_2$} \\
\hline
 C&  0.000000&  0.643420& $-$1.208026\\
 C&  0.000000& $-$0.643420& $-$1.208026\\
Ti&  0.000000&  0.000000&  0.658923\\
\hline
\end{tabular}
\end{table}

\setcounter{table}{1}
\begin{table}
\caption{Continued}
\centering
\small
\begin{tabular}{lrrr}
\hline \hline
 & \multicolumn{1}{c}{$x$ ({\AA})} & \multicolumn{1}{c}{$y$ ({\AA})} & \multicolumn{1}{c}{$z$ ({\AA})} \\
\hline
 & \multicolumn{3}{c}{Ti$_4$C} \\
\hline
 C&  1.684138& $-$0.077493&  0.000000\\
Ti&  0.395001& $-$0.915142&  1.180567\\
Ti&  0.395001& $-$0.915142& $-$1.180567\\
Ti&  0.395001&  1.435395&  0.000000\\
Ti& $-$1.644313&  0.416023&  0.000000\\
\hline
 & \multicolumn{3}{c}{Ti$_4$C$_2$} \\
\hline
\end{tabular}
\end{table}

\setcounter{table}{1}
\begin{table}
\caption{Continued}
\centering
\small
\begin{tabular}{lrrr}
\hline \hline
 & \multicolumn{1}{c}{$x$ ({\AA})} & \multicolumn{1}{c}{$y$ ({\AA})} & \multicolumn{1}{c}{$z$ ({\AA})} \\
\hline
 & \multicolumn{3}{c}{Ti$_6$C$_{13}$} \\
\hline
Ti& $-$2.065543& $-$1.427553& $-$1.139204\\
Ti&  0.233151&  0.044000& $-$1.635614\\
 C& $-$1.079408&  1.644746& $-$1.163851\\
 C&  1.148345& $-$1.740708& $-$0.954883\\
 C& $-$1.834590&  0.550348& $-$1.402548\\
 C& $-$0.145894& $-$2.025830& $-$1.219164\\
\hline
\end{tabular}
\end{table}

\setcounter{table}{1}
\begin{table}
\caption{Continued}
\centering
\small
\begin{tabular}{lrrr}
\hline \hline
 & \multicolumn{1}{c}{$x$ ({\AA})} & \multicolumn{1}{c}{$y$ ({\AA})} & \multicolumn{1}{c}{$z$ ({\AA})} \\
\hline
 & \multicolumn{3}{c}{Ti$_9$C$_{15}$} \\
\hline
Ti&  0.000543&  0.000289& $-$1.225500\\
Ti& $-$2.586230&  1.811839& $-$1.014003\\
 C& $-$2.397501& $-$0.077244& $-$1.045225\\
 C& $-$0.580749& $-$2.224080& $-$1.445109\\
 C& $-$0.730783&  2.211305& $-$1.375106\\
 C&  0.731832& $-$2.210641& $-$1.376045\\
\hline
\end{tabular}
\end{table}

\clearpage

\begin{table*}
\caption{Thermochemical properties of TiC at 1 bar.} \label{table:tic}
\centering
\small
\begin{tabular}{rrrrrrrrrr}
\hline \hline
\multicolumn{1}{c}{$T$} & \multicolumn{1}{c}{$C_P^o$} & \multicolumn{1}{c}{$S^o$} & \multicolumn{1}{c}{$-$[$G^o-H^o$(298.15 K)]/$T$} & & \multicolumn{1}{c}{$H^o-H^o$(298.15 K)} & \multicolumn{1}{c}{$\Delta_f H^o$} & \multicolumn{1}{c}{$\Delta_f G^o$} & \multicolumn{1}{c}{$\log K_f$} \\
\cline{2-4}
\cline{6-8}
\multicolumn{1}{c}{(K)} & \multicolumn{3}{c}{J K$^{-1}$ mol$^{-1}$} & & \multicolumn{3}{c}{kJ mol$^{-1}$} & \\
\hline
 100 &    29.102 &   183.974 &   242.724 & &    $-$5.875 &   826.055 &   808.579 & $-$Infinity \\
 200 &    29.461 &   204.216 &   218.981 & &    $-$2.953 &   826.734 &   790.753 &  $-$206.519 \\
 298.15 &    30.776 &   216.211 &   216.211 & &     0.000 &   826.670 &   773.089 &  $-$135.439 \\
 300 &    30.805 &   216.401 &   216.211 & &     0.057 &   826.664 &   772.756 &  $-$134.546 \\
 400 &    32.334 &   225.483 &   217.438 & &     3.218 &   826.189 &   754.850 &   $-$98.571 \\
 500 &    33.561 &   232.842 &   219.806 & &     6.518 &   825.422 &   737.099 &   $-$77.003 \\
\hline
\end{tabular}
\end{table*}

\begin{table*}
\caption{Thermochemical properties of TiC$_2$ at 1 bar.} \label{table:tic2}
\centering
\small
\begin{tabular}{rrrrrrrrrr}
\hline \hline
\multicolumn{1}{c}{$T$} & \multicolumn{1}{c}{$C_P^o$} & \multicolumn{1}{c}{$S^o$} & \multicolumn{1}{c}{$-$[$G^o-H^o$(298.15 K)]/$T$} & & \multicolumn{1}{c}{$H^o-H^o$(298.15 K)} & \multicolumn{1}{c}{$\Delta_f H^o$} & \multicolumn{1}{c}{$\Delta_f G^o$} & \multicolumn{1}{c}{$\log K_f$} \\
\cline{2-4}
\cline{6-8}
\multicolumn{1}{c}{(K)} & \multicolumn{3}{c}{J K$^{-1}$ mol$^{-1}$} & & \multicolumn{3}{c}{kJ mol$^{-1}$} & \\
\hline
 100 &    33.417 &   221.573 &   294.093 & &    $-$7.252 &   762.349 &   741.208 & $-$Infinity \\
 200 &    36.434 &   245.483 &   264.408 & &    $-$3.785 &   763.247 &   719.629 &  $-$187.944 \\
 298.15 &    40.569 &   260.825 &   260.825 & &     0.000 &   763.350 &   698.179 &  $-$122.316 \\
 300 &    40.641 &   261.076 &   260.826 & &     0.075 &   763.346 &   697.774 &  $-$121.491 \\
 400 &    44.102 &   273.270 &   262.465 & &     4.322 &   762.934 &   675.966 &   $-$88.270 \\
 500 &    46.818 &   283.422 &   265.668 & &     8.877 &   762.096 &   654.314 &   $-$68.354 \\
\hline
\end{tabular}
\end{table*}

\begin{table*}
\caption{Thermochemical properties of TiC$_3$ at 1 bar.} \label{table:tic3}
\centering
\small
\begin{tabular}{rrrrrrrrrr}
\hline \hline
\multicolumn{1}{c}{$T$} & \multicolumn{1}{c}{$C_P^o$} & \multicolumn{1}{c}{$S^o$} & \multicolumn{1}{c}{$-$[$G^o-H^o$(298.15 K)]/$T$} & & \multicolumn{1}{c}{$H^o-H^o$(298.15 K)} & \multicolumn{1}{c}{$\Delta_f H^o$} & \multicolumn{1}{c}{$\Delta_f G^o$} & \multicolumn{1}{c}{$\log K_f$} \\
\cline{2-4}
\cline{6-8}
\multicolumn{1}{c}{(K)} & \multicolumn{3}{c}{J K$^{-1}$ mol$^{-1}$} & & \multicolumn{3}{c}{kJ mol$^{-1}$} & \\
\hline
 100 &    33.997 &   221.112 &   303.592 & &    $-$8.248 &   884.024 &   863.024 & $-$Infinity \\
 200 &    41.497 &   246.665 &   269.255 & &    $-$4.518 &   884.859 &   841.621 &  $-$219.804 \\
 298.15 &    50.307 &   264.929 &   264.929 & &     0.000 &   885.030 &   820.346 &  $-$143.718 \\
 300 &    50.459 &   265.241 &   264.931 & &     0.093 &   885.028 &   819.944 &  $-$142.762 \\
 400 &    57.716 &   280.805 &   267.002 & &     5.521 &   884.774 &   798.277 &  $-$104.242 \\
 500 &    63.254 &   294.317 &   271.143 & &    11.587 &   884.121 &   776.722 &   $-$81.142 \\
\hline
\end{tabular}
\end{table*}

\begin{table*}
\caption{Thermochemical properties of TiC$_4$ at 1 bar.} \label{table:tic4}
\centering
\small
\begin{tabular}{rrrrrrrrrr}
\hline \hline
\multicolumn{1}{c}{$T$} & \multicolumn{1}{c}{$C_P^o$} & \multicolumn{1}{c}{$S^o$} & \multicolumn{1}{c}{$-$[$G^o-H^o$(298.15 K)]/$T$} & & \multicolumn{1}{c}{$H^o-H^o$(298.15 K)} & \multicolumn{1}{c}{$\Delta_f H^o$} & \multicolumn{1}{c}{$\Delta_f G^o$} & \multicolumn{1}{c}{$\log K_f$} \\
\cline{2-4}
\cline{6-8}
\multicolumn{1}{c}{(K)} & \multicolumn{3}{c}{J K$^{-1}$ mol$^{-1}$} & & \multicolumn{3}{c}{kJ mol$^{-1}$} & \\
\hline
 100 &    38.860 &   239.783 &   349.093 & &   $-$10.931 &   934.002 &   911.231 & $-$Infinity \\
 200 &    56.171 &   272.079 &   302.934 & &    $-$6.171 &   935.541 &   887.836 &  $-$231.874 \\
 298.15 &    68.560 &   297.017 &   297.017 & &     0.000 &   936.700 &   864.161 &  $-$151.394 \\
 300 &    68.745 &   297.442 &   297.019 & &     0.127 &   936.716 &   863.709 &  $-$150.382 \\
 400 &    77.014 &   318.437 &   299.825 & &     7.445 &   937.329 &   839.264 &  $-$109.595 \\
 500 &    83.017 &   336.311 &   305.379 & &    15.466 &   937.305 &   814.741 &   $-$85.114 \\
\hline
\end{tabular}
\end{table*}

\begin{table*}
\caption{Thermochemical properties of Ti$_2$C at 1 bar.} \label{table:ti2c}
\centering
\small
\begin{tabular}{rrrrrrrrrr}
\hline \hline
\multicolumn{1}{c}{$T$} & \multicolumn{1}{c}{$C_P^o$} & \multicolumn{1}{c}{$S^o$} & \multicolumn{1}{c}{$-$[$G^o-H^o$(298.15 K)]/$T$} & & \multicolumn{1}{c}{$H^o-H^o$(298.15 K)} & \multicolumn{1}{c}{$\Delta_f H^o$} & \multicolumn{1}{c}{$\Delta_f G^o$} & \multicolumn{1}{c}{$\log K_f$} \\
\cline{2-4}
\cline{6-8}
\multicolumn{1}{c}{(K)} & \multicolumn{3}{c}{J K$^{-1}$ mol$^{-1}$} & & \multicolumn{3}{c}{kJ mol$^{-1}$} & \\
\hline
 100 &    36.104 &   237.616 &   323.426 & &    $-$8.581 &   941.618 &   919.604 & $-$Infinity \\
 200 &    43.796 &   265.145 &   287.965 & &    $-$4.564 &   941.475 &   897.553 &  $-$234.412 \\
 298.15 &    48.753 &   283.647 &   283.647 & &     0.000 &   940.670 &   876.154 &  $-$153.495 \\
 300 &    48.825 &   283.949 &   283.649 & &     0.090 &   940.650 &   875.752 &  $-$152.479 \\
 400 &    51.876 &   298.459 &   285.604 & &     5.142 &   939.453 &   854.293 &  $-$111.557 \\
 500 &    53.740 &   310.260 &   289.394 & &    10.433 &   937.936 &   833.171 &   $-$87.039 \\
\hline
\end{tabular}
\end{table*}

\begin{table*}
\caption{Thermochemical properties of Ti$_2$C$_2$ at 1 bar.} \label{table:ti2c2}
\centering
\small
\begin{tabular}{rrrrrrrrrr}
\hline \hline
\multicolumn{1}{c}{$T$} & \multicolumn{1}{c}{$C_P^o$} & \multicolumn{1}{c}{$S^o$} & \multicolumn{1}{c}{$-$[$G^o-H^o$(298.15 K)]/$T$} & & \multicolumn{1}{c}{$H^o-H^o$(298.15 K)} & \multicolumn{1}{c}{$\Delta_f H^o$} & \multicolumn{1}{c}{$\Delta_f G^o$} & \multicolumn{1}{c}{$\log K_f$} \\
\cline{2-4}
\cline{6-8}
\multicolumn{1}{c}{(K)} & \multicolumn{3}{c}{J K$^{-1}$ mol$^{-1}$} & & \multicolumn{3}{c}{kJ mol$^{-1}$} & \\
\hline
 100 &    38.772 &   230.105 &   329.375 & &    $-$9.927 &  1003.943 &   982.775 & $-$Infinity \\
 200 &    50.435 &   260.531 &   287.881 & &    $-$5.470 &  1003.914 &   961.531 &  $-$251.121 \\
 298.15 &    60.453 &   282.649 &   282.649 & &     0.000 &  1003.350 &   940.843 &  $-$164.828 \\
 300 &    60.611 &   283.024 &   282.651 & &     0.112 &  1003.336 &   940.454 &  $-$163.744 \\
 400 &    67.562 &   301.497 &   285.122 & &     6.550 &  1002.502 &   919.612 &  $-$120.087 \\
 500 &    72.016 &   317.099 &   289.999 & &    13.550 &  1001.368 &   899.014 &   $-$93.918 \\
\hline
\end{tabular}
\end{table*}

\begin{table*}
\caption{Thermochemical properties of Ti$_2$C$_3$ at 1 bar.} \label{table:ti2c3}
\centering
\small
\begin{tabular}{rrrrrrrrrr}
\hline \hline
\multicolumn{1}{c}{$T$} & \multicolumn{1}{c}{$C_P^o$} & \multicolumn{1}{c}{$S^o$} & \multicolumn{1}{c}{$-$[$G^o-H^o$(298.15 K)]/$T$} & & \multicolumn{1}{c}{$H^o-H^o$(298.15 K)} & \multicolumn{1}{c}{$\Delta_f H^o$} & \multicolumn{1}{c}{$\Delta_f G^o$} & \multicolumn{1}{c}{$\log K_f$} \\
\cline{2-4}
\cline{6-8}
\multicolumn{1}{c}{(K)} & \multicolumn{3}{c}{J K$^{-1}$ mol$^{-1}$} & & \multicolumn{3}{c}{kJ mol$^{-1}$} & \\
\hline
 100 &    37.417 &   239.168 &   350.888 & &   $-$11.172 &  1074.359 &  1052.380 & $-$Infinity \\
 200 &    57.293 &   271.006 &   303.316 & &    $-$6.462 &  1074.257 &  1030.396 &  $-$269.107 \\
 298.15 &    73.197 &   297.084 &   297.084 & &     0.000 &  1074.020 &  1008.920 &  $-$176.755 \\
 300 &    73.436 &   297.538 &   297.085 & &     0.136 &  1074.014 &  1008.515 &  $-$175.595 \\
 400 &    83.703 &   320.197 &   300.105 & &     8.037 &  1073.620 &   986.735 &  $-$128.852 \\
 500 &    90.285 &   339.647 &   306.115 & &    16.766 &  1072.889 &   965.093 &  $-$100.821 \\
\hline
\end{tabular}
\end{table*}

\begin{table*}
\caption{Thermochemical properties of Ti$_2$C$_4$ at 1 bar.} \label{table:ti2c4}
\centering
\small
\begin{tabular}{rrrrrrrrrr}
\hline \hline
\multicolumn{1}{c}{$T$} & \multicolumn{1}{c}{$C_P^o$} & \multicolumn{1}{c}{$S^o$} & \multicolumn{1}{c}{$-$[$G^o-H^o$(298.15 K)]/$T$} & & \multicolumn{1}{c}{$H^o-H^o$(298.15 K)} & \multicolumn{1}{c}{$\Delta_f H^o$} & \multicolumn{1}{c}{$\Delta_f G^o$} & \multicolumn{1}{c}{$\log K_f$} \\
\cline{2-4}
\cline{6-8}
\multicolumn{1}{c}{(K)} & \multicolumn{3}{c}{J K$^{-1}$ mol$^{-1}$} & & \multicolumn{3}{c}{kJ mol$^{-1}$} & \\
\hline
 100 &    52.387 &   258.183 &   412.063 & &   $-$15.388 &  1105.814 &  1082.029 & $-$Infinity \\
 200 &    79.831 &   303.534 &   346.959 & &    $-$8.685 &  1107.379 &  1057.629 &  $-$276.219 \\
 298.15 &    95.395 &   338.659 &   338.659 & &     0.000 &  1108.700 &  1032.916 &  $-$180.959 \\
 300 &    95.606 &   339.250 &   338.660 & &     0.177 &  1108.719 &  1032.445 &  $-$179.761 \\
 400 &   104.414 &   368.084 &   342.531 & &    10.221 &  1109.445 &  1006.891 &  $-$131.484 \\
 500 &   110.223 &   392.060 &   350.108 & &    20.976 &  1109.414 &   981.242 &  $-$102.508 \\
\hline
\end{tabular}
\end{table*}

\begin{table*}
\caption{Thermochemical properties of Ti$_3$C at 1 bar.} \label{table:ti3c}
\centering
\small
\begin{tabular}{rrrrrrrrrr}
\hline \hline
\multicolumn{1}{c}{$T$} & \multicolumn{1}{c}{$C_P^o$} & \multicolumn{1}{c}{$S^o$} & \multicolumn{1}{c}{$-$[$G^o-H^o$(298.15 K)]/$T$} & & \multicolumn{1}{c}{$H^o-H^o$(298.15 K)} & \multicolumn{1}{c}{$\Delta_f H^o$} & \multicolumn{1}{c}{$\Delta_f G^o$} & \multicolumn{1}{c}{$\log K_f$} \\
\cline{2-4}
\cline{6-8}
\multicolumn{1}{c}{(K)} & \multicolumn{3}{c}{J K$^{-1}$ mol$^{-1}$} & & \multicolumn{3}{c}{kJ mol$^{-1}$} & \\
\hline
 100 &    49.719 &   260.044 &   383.204 & &   $-$12.316 &  1235.152 &  1211.721 & $-$Infinity \\
 200 &    63.126 &   299.107 &   332.122 & &    $-$6.603 &  1234.788 &  1188.319 &  $-$310.351 \\
 298.15 &    70.650 &   325.870 &   325.870 & &     0.000 &  1233.670 &  1165.736 &  $-$204.228 \\
 300 &    70.756 &   326.308 &   325.871 & &     0.131 &  1233.644 &  1165.313 &  $-$202.895 \\
 400 &    75.066 &   347.325 &   328.707 & &     7.447 &  1232.098 &  1142.761 &  $-$149.226 \\
 500 &    77.566 &   364.380 &   334.190 & &    15.095 &  1230.197 &  1120.639 &  $-$117.070 \\
\hline
\end{tabular}
\end{table*}

\begin{table*}
\caption{Thermochemical properties of Ti$_3$C$_2$ at 1 bar.} \label{table:ti3c2}
\centering
\small
\begin{tabular}{rrrrrrrrrr}
\hline \hline
\multicolumn{1}{c}{$T$} & \multicolumn{1}{c}{$C_P^o$} & \multicolumn{1}{c}{$S^o$} & \multicolumn{1}{c}{$-$[$G^o-H^o$(298.15 K)]/$T$} & & \multicolumn{1}{c}{$H^o-H^o$(298.15 K)} & \multicolumn{1}{c}{$\Delta_f H^o$} & \multicolumn{1}{c}{$\Delta_f G^o$} & \multicolumn{1}{c}{$\log K_f$} \\
\cline{2-4}
\cline{6-8}
\multicolumn{1}{c}{(K)} & \multicolumn{3}{c}{J K$^{-1}$ mol$^{-1}$} & & \multicolumn{3}{c}{kJ mol$^{-1}$} & \\
\hline
 100 &    47.514 &   255.633 &   389.673 & &   $-$13.404 &  1145.725 &  1122.830 & $-$Infinity \\
 200 &    69.004 &   295.408 &   333.113 & &    $-$7.541 &  1145.185 &  1100.072 &  $-$287.304 \\
 298.15 &    83.328 &   325.893 &   325.893 & &     0.000 &  1144.340 &  1078.110 &  $-$188.877 \\
 300 &    83.532 &   326.409 &   325.896 & &     0.154 &  1144.321 &  1077.698 &  $-$187.640 \\
 400 &    91.952 &   351.721 &   329.288 & &     8.973 &  1143.255 &  1055.645 &  $-$137.850 \\
 500 &    96.895 &   372.831 &   335.945 & &    18.443 &  1141.850 &  1033.897 &  $-$108.008 \\
\hline
\end{tabular}
\end{table*}

\begin{table*}
\caption{Thermochemical properties of Ti$_3$C$_3$ at 1 bar.} \label{table:ti3c3}
\centering
\small
\begin{tabular}{rrrrrrrrrr}
\hline \hline
\multicolumn{1}{c}{$T$} & \multicolumn{1}{c}{$C_P^o$} & \multicolumn{1}{c}{$S^o$} & \multicolumn{1}{c}{$-$[$G^o-H^o$(298.15 K)]/$T$} & & \multicolumn{1}{c}{$H^o-H^o$(298.15 K)} & \multicolumn{1}{c}{$\Delta_f H^o$} & \multicolumn{1}{c}{$\Delta_f G^o$} & \multicolumn{1}{c}{$\log K_f$} \\
\cline{2-4}
\cline{6-8}
\multicolumn{1}{c}{(K)} & \multicolumn{3}{c}{J K$^{-1}$ mol$^{-1}$} & & \multicolumn{3}{c}{kJ mol$^{-1}$} & \\
\hline
 100 &    47.450 &   261.284 &   408.364 & &   $-$14.708 &  1171.092 &  1147.727 & $-$Infinity \\
 200 &    76.035 &   303.138 &   345.648 & &    $-$8.502 &  1170.569 &  1124.527 &  $-$293.691 \\
 298.15 &    95.415 &   337.469 &   337.469 & &     0.000 &  1170.020 &  1102.050 &  $-$193.071 \\
 300 &    95.690 &   338.060 &   337.470 & &     0.177 &  1170.008 &  1101.627 &  $-$191.807 \\
 400 &   107.207 &   367.327 &   341.382 & &    10.378 &  1169.301 &  1078.933 &  $-$140.891 \\
 500 &   114.298 &   392.087 &   349.113 & &    21.487 &  1168.209 &  1056.459 &  $-$110.365 \\
\hline
\end{tabular}
\end{table*}

\begin{table*}
\caption{Thermochemical properties of Ti$_3$C$_4$ at 1 bar.} \label{table:ti3c4}
\centering
\small
\begin{tabular}{rrrrrrrrrr}
\hline \hline
\multicolumn{1}{c}{$T$} & \multicolumn{1}{c}{$C_P^o$} & \multicolumn{1}{c}{$S^o$} & \multicolumn{1}{c}{$-$[$G^o-H^o$(298.15 K)]/$T$} & & \multicolumn{1}{c}{$H^o-H^o$(298.15 K)} & \multicolumn{1}{c}{$\Delta_f H^o$} & \multicolumn{1}{c}{$\Delta_f G^o$} & \multicolumn{1}{c}{$\log K_f$} \\
\cline{2-4}
\cline{6-8}
\multicolumn{1}{c}{(K)} & \multicolumn{3}{c}{J K$^{-1}$ mol$^{-1}$} & & \multicolumn{3}{c}{kJ mol$^{-1}$} & \\
\hline
 100 &    49.256 &   261.892 &   421.362 & &   $-$15.947 &  1218.524 &  1195.194 & $-$Infinity \\
 200 &    82.370 &   306.123 &   352.983 & &    $-$9.372 &  1218.044 &  1172.021 &  $-$306.095 \\
 298.15 &   106.529 &   343.928 &   343.928 & &     0.000 &  1217.700 &  1149.516 &  $-$201.386 \\
 300 &   106.879 &   344.589 &   343.929 & &     0.198 &  1217.693 &  1149.091 &  $-$200.071 \\
 400 &   121.810 &   377.573 &   348.325 & &    11.699 &  1217.263 &  1126.282 &  $-$147.074 \\
 500 &   131.338 &   405.871 &   357.075 & &    24.398 &  1216.435 &  1103.624 &  $-$115.293 \\
\hline
\end{tabular}
\end{table*}

\begin{table*}
\caption{Thermochemical properties of Ti$_4$C at 1 bar.} \label{table:ti4c}
\centering
\small
\begin{tabular}{rrrrrrrrrr}
\hline \hline
\multicolumn{1}{c}{$T$} & \multicolumn{1}{c}{$C_P^o$} & \multicolumn{1}{c}{$S^o$} & \multicolumn{1}{c}{$-$[$G^o-H^o$(298.15 K)]/$T$} & & \multicolumn{1}{c}{$H^o-H^o$(298.15 K)} & \multicolumn{1}{c}{$\Delta_f H^o$} & \multicolumn{1}{c}{$\Delta_f G^o$} & \multicolumn{1}{c}{$\log K_f$} \\
\cline{2-4}
\cline{6-8}
\multicolumn{1}{c}{(K)} & \multicolumn{3}{c}{J K$^{-1}$ mol$^{-1}$} & & \multicolumn{3}{c}{kJ mol$^{-1}$} & \\
\hline
 100 &    56.425 &   272.031 &   426.841 & &   $-$15.481 &  1433.256 &  1409.453 & $-$Infinity \\
 200 &    80.219 &   319.392 &   361.942 & &    $-$8.510 &  1432.233 &  1385.953 & $-$Infinity \\
 298.15 &    91.748 &   353.859 &   353.859 & &     0.000 &  1430.670 &  1363.562 &  $-$238.886 \\
 300 &    91.898 &   354.427 &   353.860 & &     0.170 &  1430.636 &  1363.144 &  $-$237.340 \\
 400 &    97.833 &   381.785 &   357.550 & &     9.694 &  1428.685 &  1340.933 &  $-$175.104 \\
 500 &   101.110 &   404.016 &   364.690 & &    19.663 &  1426.364 &  1319.255 &  $-$137.819 \\
\hline
\end{tabular}
\end{table*}

\begin{table*}
\caption{Thermochemical properties of Ti$_4$C$_2$ at 1 bar.} \label{table:ti4c2}
\centering
\small
\begin{tabular}{rrrrrrrrrr}
\hline \hline
\multicolumn{1}{c}{$T$} & \multicolumn{1}{c}{$C_P^o$} & \multicolumn{1}{c}{$S^o$} & \multicolumn{1}{c}{$-$[$G^o-H^o$(298.15 K)]/$T$} & & \multicolumn{1}{c}{$H^o-H^o$(298.15 K)} & \multicolumn{1}{c}{$\Delta_f H^o$} & \multicolumn{1}{c}{$\Delta_f G^o$} & \multicolumn{1}{c}{$\log K_f$} \\
\cline{2-4}
\cline{6-8}
\multicolumn{1}{c}{(K)} & \multicolumn{3}{c}{J K$^{-1}$ mol$^{-1}$} & & \multicolumn{3}{c}{kJ mol$^{-1}$} & \\
\hline
 100 &    53.151 &   263.864 &   426.574 & &   $-$16.271 &  1343.127 &  1320.235 & $-$Infinity \\
 200 &    84.348 &   310.990 &   357.445 & &    $-$9.291 &  1341.787 &  1297.803 & $-$Infinity \\
 298.15 &   103.078 &   348.538 &   348.538 & &     0.000 &  1340.340 &  1276.529 &  $-$223.638 \\
 300 &   103.337 &   349.177 &   348.540 & &     0.191 &  1340.311 &  1276.132 &  $-$222.190 \\
 400 &   113.847 &   380.512 &   352.740 & &    11.109 &  1338.731 &  1254.973 &  $-$163.880 \\
 500 &   119.854 &   406.638 &   360.982 & &    22.828 &  1336.834 &  1234.245 &  $-$128.938 \\
\hline
\end{tabular}
\end{table*}

\begin{table*}
\caption{Thermochemical properties of Ti$_4$C$_3$ at 1 bar.} \label{table:ti4c3}
\centering
\small
\begin{tabular}{rrrrrrrrrr}
\hline \hline
\multicolumn{1}{c}{$T$} & \multicolumn{1}{c}{$C_P^o$} & \multicolumn{1}{c}{$S^o$} & \multicolumn{1}{c}{$-$[$G^o-H^o$(298.15 K)]/$T$} & & \multicolumn{1}{c}{$H^o-H^o$(298.15 K)} & \multicolumn{1}{c}{$\Delta_f H^o$} & \multicolumn{1}{c}{$\Delta_f G^o$} & \multicolumn{1}{c}{$\log K_f$} \\
\cline{2-4}
\cline{6-8}
\multicolumn{1}{c}{(K)} & \multicolumn{3}{c}{J K$^{-1}$ mol$^{-1}$} & & \multicolumn{3}{c}{kJ mol$^{-1}$} & \\
\hline
 100 &    53.997 &   272.972 &   448.402 & &   $-$17.543 &  1307.526 &  1283.819 & $-$Infinity \\
 200 &    90.784 &   322.103 &   373.308 & &   $-$10.241 &  1306.182 &  1260.592 & $-$Infinity \\
 298.15 &   115.657 &   363.434 &   363.434 & &     0.000 &  1305.020 &  1238.479 &  $-$216.972 \\
 300 &   116.011 &   364.151 &   363.438 & &     0.214 &  1304.998 &  1238.064 &  $-$215.562 \\
 400 &   130.568 &   399.737 &   368.190 & &    12.619 &  1303.882 &  1215.920 &  $-$158.780 \\
 500 &   139.042 &   429.883 &   377.593 & &    26.145 &  1302.466 &  1194.085 &  $-$124.743 \\
\hline
\end{tabular}
\end{table*}

\begin{table*}
\caption{Thermochemical properties of Ti$_4$C$_4$ at 1 bar.} \label{table:ti4c4}
\centering
\small
\begin{tabular}{rrrrrrrrrr}
\hline \hline
\multicolumn{1}{c}{$T$} & \multicolumn{1}{c}{$C_P^o$} & \multicolumn{1}{c}{$S^o$} & \multicolumn{1}{c}{$-$[$G^o-H^o$(298.15 K)]/$T$} & & \multicolumn{1}{c}{$H^o-H^o$(298.15 K)} & \multicolumn{1}{c}{$\Delta_f H^o$} & \multicolumn{1}{c}{$\Delta_f G^o$} & \multicolumn{1}{c}{$\log K_f$} \\
\cline{2-4}
\cline{6-8}
\multicolumn{1}{c}{(K)} & \multicolumn{3}{c}{J K$^{-1}$ mol$^{-1}$} & & \multicolumn{3}{c}{kJ mol$^{-1}$} & \\
\hline
 100 &    45.073 &   248.232 &   421.762 & &   $-$17.353 &  1175.377 &  1154.239 & $-$Infinity \\
 200 &    89.948 &   293.078 &   346.183 & &   $-$10.621 &  1173.137 &  1133.969 &  $-$296.156 \\
 298.15 &   123.665 &   335.836 &   335.836 & &     0.000 &  1171.690 &  1115.089 &  $-$195.355 \\
 300 &   124.152 &   336.603 &   335.840 & &     0.229 &  1171.667 &  1114.736 &  $-$194.089 \\
 400 &   144.319 &   375.370 &   340.988 & &    13.753 &  1170.647 &  1095.917 &  $-$143.109 \\
 500 &   156.171 &   408.981 &   351.305 & &    28.838 &  1169.464 &  1077.366 &  $-$112.549 \\
\hline
\end{tabular}
\end{table*}

\begin{table*}
\caption{Thermochemical properties of Ti$_3$C$_8$ at 1 bar.} \label{table:ti3c8}
\centering
\small
\begin{tabular}{rrrrrrrrrr}
\hline \hline
\multicolumn{1}{c}{$T$} & \multicolumn{1}{c}{$C_P^o$} & \multicolumn{1}{c}{$S^o$} & \multicolumn{1}{c}{$-$[$G^o-H^o$(298.15 K)]/$T$} & & \multicolumn{1}{c}{$H^o-H^o$(298.15 K)} & \multicolumn{1}{c}{$\Delta_f H^o$} & \multicolumn{1}{c}{$\Delta_f G^o$} & \multicolumn{1}{c}{$\log K_f$} \\
\cline{2-4}
\cline{6-8}
\multicolumn{1}{c}{(K)} & \multicolumn{3}{c}{J K$^{-1}$ mol$^{-1}$} & & \multicolumn{3}{c}{kJ mol$^{-1}$} & \\
\hline
 100 &    99.795 &   320.515 &   610.135 & &   $-$28.962 &   799.173 &   770.361 & $-$Infinity \\
 200 &   149.339 &   405.885 &   487.680 & &   $-$16.359 &   803.417 &   739.907 &  $-$193.240 \\
 298.15 &   181.112 &   472.012 &   472.012 & &     0.000 &   807.400 &   707.873 &  $-$124.014 \\
 300 &   181.569 &   473.135 &   472.015 & &     0.336 &   807.467 &   707.253 &  $-$123.141 \\
 400 &   200.983 &   528.284 &   479.404 & &    19.552 &   810.660 &   673.336 &   $-$87.927 \\
 500 &   213.659 &   574.614 &   493.938 & &    40.338 &   812.615 &   638.757 &   $-$66.729 \\
\hline
\end{tabular}
\end{table*}

\begin{table*}
\caption{Thermochemical properties of Ti$_4$C$_8$ at 1 bar.} \label{table:ti4c8}
\centering
\small
\begin{tabular}{rrrrrrrrrr}
\hline \hline
\multicolumn{1}{c}{$T$} & \multicolumn{1}{c}{$C_P^o$} & \multicolumn{1}{c}{$S^o$} & \multicolumn{1}{c}{$-$[$G^o-H^o$(298.15 K)]/$T$} & & \multicolumn{1}{c}{$H^o-H^o$(298.15 K)} & \multicolumn{1}{c}{$\Delta_f H^o$} & \multicolumn{1}{c}{$\Delta_f G^o$} & \multicolumn{1}{c}{$\log K_f$} \\
\cline{2-4}
\cline{6-8}
\multicolumn{1}{c}{(K)} & \multicolumn{3}{c}{J K$^{-1}$ mol$^{-1}$} & & \multicolumn{3}{c}{kJ mol$^{-1}$} & \\
\hline
 100 &    92.081 &   304.405 &   616.355 & &   $-$31.195 &   544.209 &   517.834 &  $-$270.484 \\
 200 &   163.550 &   391.968 &   482.688 & &   $-$18.144 &   546.984 &   490.503 &  $-$128.104 \\
 298.15 &   201.652 &   465.272 &   465.272 & &     0.000 &   550.400 &   462.053 &   $-$80.948 \\
 300 &   202.157 &   466.522 &   465.275 & &     0.374 &   550.458 &   461.503 &   $-$80.353 \\
 400 &   223.064 &   527.842 &   473.495 & &    21.739 &   553.187 &   431.409 &   $-$56.335 \\
 500 &   236.467 &   579.185 &   489.639 & &    44.773 &   554.649 &   400.772 &   $-$41.868 \\
\hline
\end{tabular}
\end{table*}

\begin{table*}
\caption{Thermochemical properties of Ti$_6$C$_{13}$ at 1 bar.} \label{table:ti6c13}
\centering
\small
\begin{tabular}{rrrrrrrrrr}
\hline \hline
\multicolumn{1}{c}{$T$} & \multicolumn{1}{c}{$C_P^o$} & \multicolumn{1}{c}{$S^o$} & \multicolumn{1}{c}{$-$[$G^o-H^o$(298.15 K)]/$T$} & & \multicolumn{1}{c}{$H^o-H^o$(298.15 K)} & \multicolumn{1}{c}{$\Delta_f H^o$} & \multicolumn{1}{c}{$\Delta_f G^o$} & \multicolumn{1}{c}{$\log K_f$} \\
\cline{2-4}
\cline{6-8}
\multicolumn{1}{c}{(K)} & \multicolumn{3}{c}{J K$^{-1}$ mol$^{-1}$} & & \multicolumn{3}{c}{kJ mol$^{-1}$} & \\
\hline
 100 &   142.231 &   347.170 &   842.360 & &   $-$49.519 &   461.748 &   433.225 &  $-$226.289 \\
 200 &   259.543 &   485.256 &   629.836 & &   $-$28.916 &   466.611 &   403.045 &  $-$105.263 \\
 298.15 &   322.656 &   602.049 &   602.049 & &     0.000 &   472.770 &   370.542 &   $-$64.916 \\
 300 &   323.511 &   604.049 &   602.056 & &     0.598 &   472.878 &   369.903 &   $-$64.405 \\
 400 &   359.039 &   702.485 &   615.235 & &    34.900 &   478.203 &   334.732 &   $-$43.711 \\
 500 &   381.637 &   785.250 &   641.186 & &    72.032 &   481.651 &   298.431 &   $-$31.176 \\
\hline
\end{tabular}
\end{table*}

\begin{table*}
\caption{Thermochemical properties of Ti$_7$C$_{13}$ at 1 bar.} \label{table:ti7c13}
\centering
\small
\begin{tabular}{rrrrrrrrrr}
\hline \hline
\multicolumn{1}{c}{$T$} & \multicolumn{1}{c}{$C_P^o$} & \multicolumn{1}{c}{$S^o$} & \multicolumn{1}{c}{$-$[$G^o-H^o$(298.15 K)]/$T$} & & \multicolumn{1}{c}{$H^o-H^o$(298.15 K)} & \multicolumn{1}{c}{$\Delta_f H^o$} & \multicolumn{1}{c}{$\Delta_f G^o$} & \multicolumn{1}{c}{$\log K_f$} \\
\cline{2-4}
\cline{6-8}
\multicolumn{1}{c}{(K)} & \multicolumn{3}{c}{J K$^{-1}$ mol$^{-1}$} & & \multicolumn{3}{c}{kJ mol$^{-1}$} & \\
\hline
 100 &   163.124 &   367.228 &   904.358 & &   $-$53.713 &   643.823 &   614.120 &  $-$320.777 \\
 200 &   280.673 &   520.499 &   675.099 & &   $-$30.920 &   648.959 &   582.590 &  $-$152.154 \\
 298.15 &   342.862 &   645.457 &   645.457 & &     0.000 &   654.770 &   548.771 &   $-$96.140 \\
 300 &   343.727 &   647.582 &   645.462 & &     0.636 &   654.869 &   548.109 &   $-$95.432 \\
 400 &   380.433 &   751.979 &   659.449 & &    37.012 &   659.655 &   511.755 &   $-$66.827 \\
 500 &   404.363 &   839.670 &   686.960 & &    76.355 &   662.573 &   474.410 &   $-$49.560 \\
\hline
\end{tabular}
\end{table*}

\begin{table*}
\caption{Thermochemical properties of Ti$_8$C$_{12}$ at 1 bar.} \label{table:ti8c12}
\centering
\small
\begin{tabular}{rrrrrrrrrr}
\hline \hline
\multicolumn{1}{c}{$T$} & \multicolumn{1}{c}{$C_P^o$} & \multicolumn{1}{c}{$S^o$} & \multicolumn{1}{c}{$-$[$G^o-H^o$(298.15 K)]/$T$} & & \multicolumn{1}{c}{$H^o-H^o$(298.15 K)} & \multicolumn{1}{c}{$\Delta_f H^o$} & \multicolumn{1}{c}{$\Delta_f G^o$} & \multicolumn{1}{c}{$\log K_f$} \\
\cline{2-4}
\cline{6-8}
\multicolumn{1}{c}{(K)} & \multicolumn{3}{c}{J K$^{-1}$ mol$^{-1}$} & & \multicolumn{3}{c}{kJ mol$^{-1}$} & \\
\hline
 100 &   138.419 &   352.081 &   850.201 & &   $-$49.812 &  $-$346.678 &  $-$374.135 &   195.424 \\
 200 &   260.531 &   488.293 &   635.548 & &   $-$29.451 &  $-$345.565 &  $-$401.864 &   104.954 \\
 298.15 &   332.379 &   607.145 &   607.145 & &     0.000 &  $-$342.910 &  $-$430.027 &    75.338 \\
 300 &   333.382 &   609.206 &   607.153 & &     0.616 &  $-$342.862 &  $-$430.573 &    74.968 \\
 400 &   375.669 &   711.471 &   620.806 & &    36.266 &  $-$340.392 &  $-$460.204 &    60.095 \\
 500 &   402.608 &   798.457 &   647.863 & &    75.297 &  $-$339.201 &  $-$490.322 &    51.223 \\
\hline
\end{tabular}
\end{table*}

\begin{table*}
\caption{Thermochemical properties of Ti$_9$C$_{15}$ at 1 bar.} \label{table:ti9c15}
\centering
\small
\begin{tabular}{rrrrrrrrrr}
\hline \hline
\multicolumn{1}{c}{$T$} & \multicolumn{1}{c}{$C_P^o$} & \multicolumn{1}{c}{$S^o$} & \multicolumn{1}{c}{$-$[$G^o-H^o$(298.15 K)]/$T$} & & \multicolumn{1}{c}{$H^o-H^o$(298.15 K)} & \multicolumn{1}{c}{$\Delta_f H^o$} & \multicolumn{1}{c}{$\Delta_f G^o$} & \multicolumn{1}{c}{$\log K_f$} \\
\cline{2-4}
\cline{6-8}
\multicolumn{1}{c}{(K)} & \multicolumn{3}{c}{J K$^{-1}$ mol$^{-1}$} & & \multicolumn{3}{c}{kJ mol$^{-1}$} & \\
\hline
 100 &   187.713 &   387.111 &  1038.781 & &   $-$65.167 &    37.229 &     7.381 &    $-$3.855 \\
 200 &   342.131 &   570.092 &   759.447 & &   $-$37.871 &    42.382 &   $-$24.182 &     6.316 \\
 298.15 &   420.650 &   723.112 &   723.112 & &     0.000 &    49.110 &   $-$58.278 &    10.210 \\
 300 &   421.709 &   725.719 &   723.119 & &     0.780 &    49.227 &   $-$58.950 &    10.264 \\
 400 &   465.730 &   853.694 &   740.269 & &    45.370 &    54.955 &   $-$95.922 &    12.526 \\
 500 &   493.570 &   960.889 &   773.969 & &    93.460 &    58.486 &  $-$134.091 &    14.008 \\
\hline
\end{tabular}
\end{table*}

\begin{table*}
\caption{Thermochemical properties of Ti$_{13}$C$_{22}$ at 1 bar.} \label{table:ti13c22}
\centering
\small
\begin{tabular}{rrrrrrrrr}
\hline \hline
\multicolumn{1}{c}{$T$} & \multicolumn{1}{c}{$C_P^o$} & \multicolumn{1}{c}{$S^o$} & \multicolumn{1}{c}{$-$[$G^o-H^o$(298.15 K)]/$T$} & & \multicolumn{1}{c}{$H^o-H^o$(298.15 K)} & \multicolumn{1}{c}{$\Delta_f H^o$} & \multicolumn{1}{c}{$\Delta_f G^o$} & \multicolumn{1}{c}{$\log K_f$} \\
\cline{2-4}
\cline{6-8}
\multicolumn{1}{c}{(K)} & \multicolumn{3}{c}{J K$^{-1}$ mol$^{-1}$} & & \multicolumn{3}{c}{kJ mol$^{-1}$} & \\
\hline
 100 &   254.409 &   428.612 &  1353.172 & &   $-$92.456 &  $-$619.327 &  $-$649.355 &  Infinity \\
 200 &   487.167 &   685.061 &   955.591 & &   $-$54.106 &  $-$613.070 &  $-$681.331 &   177.942 \\
 298.15 &   601.555 &   903.651 &   903.651 & &     0.000 &  $-$604.170 &  $-$716.723 &   125.564 \\
 300 &   603.058 &   907.380 &   903.663 & &     1.115 &  $-$604.018 &  $-$717.430 &   124.913 \\
 400 &   664.119 &  1090.172 &   928.172 & &    64.800 &  $-$596.808 &  $-$756.403 &    98.774 \\
 500 &   701.434 &  1242.769 &   976.259 & &   133.255 &  $-$593.158 &  $-$796.790 &    83.238 \\
\hline
\end{tabular}
\end{table*}

\end{appendix}

\end{document}